\documentclass[aip,pof,reprint,preprintnumbers,floatfix,nofootinbib]{revtex4-1}

\usepackage{amsmath}
\usepackage{amssymb}

\usepackage{graphicx}
\usepackage{subcaption}
\usepackage{dcolumn}
\usepackage{bm}
\usepackage{marvosym}

\usepackage{color}
\definecolor{light-gray}{gray}{0.5}
\definecolor{blue}{rgb}{0.0,0.0,1.0}
\definecolor{green}{rgb}{0.0,0.5,0.0}
\definecolor{red}{rgb}{1.0,0.0,0.0}
\definecolor{cyan}{rgb}{0.0,0.75,0.75}
\definecolor{magenta}{rgb}{0.75,0.0,0.75}
\definecolor{yellow}{rgb}{0.75,0.75,0.0}

\newcommand{\avg}[1]{\langle{#1}\rangle}
\newcommand{\sdot}{\cdot}

\newcommand{\grad}{\bm \nabla}
\newcommand{\pd}{\partial}
\newcommand{\lrbig}[1]{\left( {#1} \right)}

\newcommand{\lt}{\left}
\newcommand{\rt}{\right}

\newcommand{\dd}{\mathrm{d}}

\begin{document}
\title{Structures and dynamics of small scales in decaying magnetohydrodynamic turbulence}
\author{V. Dallas}
\email{vassilios.dallas@lps.ens.fr}
\author{A. Alexakis}
\affiliation{Laboratoire de Physique Statistique, \'Ecole Normale Sup\'erieure, 24 Rue Lhomond, 75231 Paris, France}

\begin{abstract}
The topological and dynamical features of small scales are studied in the context of decaying magnetohydrodynamic turbulent flows using direct numerical simulations. Joint probability density functions (PDFs) of the invariants of gradient quantities related to the velocity and the magnetic fields demonstrate that structures and dynamics at the time of maximum dissipation depend on the large scale initial conditions. 
This is evident in particular from the fact that each flow has a different shape for the joint PDF of the invariants of the velocity gradient in contrast to the universal teardrop shape of hydrodynamic turbulence. The general picture that emerges from the analysis of the invariants is that regions of high vorticity are correlated with regions of high strain rate $\bm S$ also in contrast to hydrodynamic turbulent flows. Magnetic strain dominated regions are also well correlated with region of high current density $\bm j$. Viscous dissipation ($\propto \bm S^2$) as well as Ohmic dissipation ($\propto \bm j^2$) reside in regions where strain and rotation are locally almost in balance. The structures related to the velocity gradient possess different characteristics than those associated with the magnetic field gradient with the latter being locally more quasi-two dimensional.
\end{abstract}

\maketitle

\section{\label{sec:intro}Introduction}

Numerous studies have shown that the key assumptions of small-scale universality, isotropy and locality of interactions are in question in various contexts of magnetohydrodynamic (MHD) turbulence \cite{alexakisetal07,schekochihinetal08,mininni11}. Therefore, the validity of the classical phenomenology of Kolmogorov (K41) \cite{k41a}, which provides to a good approximation the power law of the energy spectrum in hydrodynamic turbulence (besides intermittency corrections), is questionable in MHD turbulence, where several debatable phenomenological theories exist \cite{iroshnikov64,kraichnan65,goldreichsridhar95,zhouetal04,boldyrev06,ngbhattacharjee97,galtieretal00}. 
In summary the power law scaling exponents obtained in the various MHD turbulence phenomenologies based on weak and strong turbulence arguments both for isotropic and anisotropic fields are $-5/3$, $-3/2$ and $-2$. 

Numerical simulations to date are unable to provide a definitive answer to the scaling of the energy spectrum in MHD turbulence \cite{mullergrappin05,mininnipouquet07}. Recently, large resolution simulations by Lee et al. \cite{leeetal10} (using a code that enforces the symmetries of the Taylor-Green vortex to achieve higher resolution) demonstrated different scaling of the total energy spectra for different initial conditions and thereby suggested that freely decaying MHD turbulent flows are non-universal.

This lack of the detailed knowledge of the energy spectrum in MHD turbulence has many implications because to predict for example heating rates in solar and space physics, the energy dissipation rate is required, which is dependent on the slope of the energy spectrum. This is also the reason why subgrid scale models, required for numerical modelling in astrophysics and geophysics, are less developed in MHD. 

On the other hand, several universal small scale features have been observed in a variety of hydrodynamic turbulent flows since the seminal works by Perry, Chong and Cantwell \cite{chongetal90,cantwell93,perrychong94} on the analysis of the velocity gradient tensor invariants. One of the most important universal results is the well known teardrop shape \cite{tsinober02,davidson04,meneveau11} of the joint probability density function (PDF) of the invariants of the velocity gradient tensor, which describes the topology and dynamics of small scales in hydrodynamic turbulence. Other important directions of research in hydrodynamic turbulence involving the study of such invariants has been the topological classification of the coherent structures \cite{blackburnetal96,chakrabortyetal05} and the use of the invariants in subgrid-scale modelling \cite{bosetal02}.

Therefore, due to the limited information that can be extracted just from the slopes of the energy spectra and the fact that small scale universality is one of the key assumptions in inertial range phenomenologies, we try to gain insight in this paper on the universal or/and non-universal features of MHD turbulence by studying the structures and dynamics of small scales through joint PDFs of the invariants of the velocity gradient, magnetic field gradient and related tensors.
Through this analysis we also attempt to provide a classification of the structures in MHD turbulence. This investigation was carried out using Direct Numerical Simulation (DNS) data of incompressible, homogeneous, decaying MHD turbulence with no imposed symmetries and no magnetic flux either in or out of our periodic boxes. 

The paper is organised as follows. The numerical method, the initial conditions and the parameters of our DNS of decaying MHD turbulent flows are provided in section \ref{sec:dns}. In section \ref{sec:spectra} we present the energy spectra of our flows.
Before presenting our results we give an outline for the classification of fluid flow topology in section \ref{sec:defs}. Then, in sections \ref{sec:invu} and \ref{sec:invb} we unravel our joint PDF analysis for the invariants of gradient quantities related to the velocity and magnetic field, respectively, delineating the structure and dynamics of the examined MHD flows. At the end, in section \ref{sec:end}, we summarise our results.

\section{\label{sec:dns}DNS of decaying MHD turbulence}
\subsection{Governing equations \& numerical method}
We consider the three-dimensional, incompressible MHD equations of fluid velocity $\bm u$ and magnetic induction $\bm b$ to be
\begin{align}
 \pd_t \bm u - \nu \bm\Delta \bm u &= (\bm u \times \bm \omega) + (\bm j \times \bm b) - \grad P 
 \label{eq:ns} \\
 \pd_t \bm b - \kappa \bm\Delta \bm b &= \grad \times (\bm u \times \bm b) 
 \label{eq:induction} \\
 \grad \sdot \bm u &= \grad \sdot \bm b = 0
 \label{eq:incomp}
\end{align}
with $\nu$ the kinematic viscosity, $\kappa$ the magnetic diffusivity,  $\bm \omega \equiv \grad \times \bm u$ the vorticity, $\bm j \equiv \grad \times \bm b$ the current density of the magnetic field and $P = p/\rho + \tfrac{1}{2}\bm u^2$ the fluid pressure, composed by the plasma pressure $p$, the constant mass density $\rho$ and the hydrodynamic pressure $\tfrac{1}{2}\bm u^2$. Note that magnetic induction has units of Alfv\'en velocity, i.e. $\bm b/\sqrt{\rho \mu_0}$, where $\mu_0 = (\kappa \sigma)^{-1}$ is the permeability of free space with $\sigma$ the electrical conductivity. In ideal MHD, where $\nu = \kappa = 0$, the total energy $E_t \equiv \frac{1}{2}\avg{|\bm u|^2 + |\bm b|^2} = E_u + E_b$, the magnetic helicity $H_b \equiv \avg{\bm u \sdot \bm b}$ and the cross helicity $H_c \equiv \avg{\bm a \sdot \bm b}$ are conserved, where the angle brackets $\avg{.}$ in this study denote spatial averages. Here, $\bm a$ is the magnetic 
potential, which is defined as $\bm a \equiv - \bm\Delta^{-1}(\grad \times \bm b)$, since one can define $\bm b \equiv \grad \times \bm a$ with $\grad \sdot \bm a = 0$. 

Our numerical method is pseudo-spectral \cite{gottlieborszag77}, where each component of $\bm u$ and $\bm b$ is represented as truncated Galerkin expansions in terms of the Fourier basis. The non-linear terms are initially computed in physical space and then transformed to spectral space using fast Fourier transforms \cite{fftw98}. Aliasing errors are removed using the 2/3 dealiasing rule, i.e. wavenumbers $k \in [1,N/3]$, where $N$ is the number of grid points in each Cartesian coordinate of our periodic box with period $2\pi$. The non-linear terms along with the pressure term are computed in such a way that $\bm u$ and $\bm b$ are projected on to a divergence-free space so that Eqs. \eqref{eq:incomp} are satisfied. The temporal integration of Eqs. \eqref{eq:ns} and \eqref{eq:induction} is performed using a second-order Runge-Kutta method. The code is parallelised using message passing interface (MPI) with one-dimensional domain decomposition \cite{mpicode05}.

\subsection{Initial conditions \& numerical parameters}
The initial conditions that we consider in this study are the three different cases studied in \cite{leeetal10}. In particular, the initial velocity field is the Taylor-Green (TG) vortex \cite{taylorgreen37} defined as 
\begin{equation}
 \bm u_{TG}(\bm x) = u_0 (\sin x \cos y \cos z, -\cos x \sin y \cos z, 0)
\end{equation}
and the initial conditions of the magnetic field are generalisation of the TG vortex symmetries. In detail, the insulating case (run ``I'' hereafter) is
\begin{equation}\label{eq:icase}
 \bm b_I(\bm x) = b_0^I (\cos x \sin y \sin z, \sin x \cos y \sin z, -2 \sin x \sin y \cos z)
\end{equation}
where $\bm j_I = \grad \times \bm b_I$ is parallel to the faces of a subvolume $[0,\pi]^3$, which can thereby be considered as electrical insulators. Note that in this case the magnetic field $\bm b_I = -(b_0^I/u_0) \grad \times \bm u_{TG}$ and the magnetic as well as cross helicity are globally restricted to vanish for all times due to the TG symmetries. 
The conducting case (run ``C'' hereafter) takes the following form
\begin{equation}\label{eq:ccase}
 \bm b_C(\bm x) = b_0^C (\sin 2x \cos 2y \cos 2z, \cos 2x \sin 2y \cos 2z, -2 \cos 2x \cos 2y \sin 2z)
\end{equation}
with $\bm j_C = \grad \times \bm b_C$ perpendicular to the faces of a subvolume $[0,\pi]^3$, which can consequently be considered as electrically conductive. In this configuration, $H_b = 0$ for all times but $H_c \neq 0$ although negligible 
(i.e. $H_c\ell/E_t < 0.04$ at its maximum over time, where $\ell$ is a typical length scale). The final case that is considered by Lee et al. \cite{leeetal10} is an alternative (run ``A'' hereafter) to the insulating initial conditions above (see Eq. \eqref{eq:icase}), namely
\begin{equation}\label{eq:acase}
 \bm b_A(\bm x) = b_0^A (\cos 2x \sin 2y \sin 2z, -\sin 2x \cos 2y \sin 2z, 0)
\end{equation}
for which again $H_b = H_c = 0$ for all times, at least up to the peak of dissipation.

The above TG fields exhibit several intrinsic symmetries within a cubic box of size $[0,2\pi]^3$, where periodic boundary conditions are applied. These are mirror (anti)symmetries about the planes $x=0$, $x=\pi$, $y=0$, $y=\pi$, $z=0$ and $z=\pi$ as well as rotational (anti)symmetries of angle $N\pi$ about the axes $(x,y,z)=(\tfrac{\pi}{2},y,\tfrac{\pi}{2})$ and $(x,\tfrac{\pi}{2},\tfrac{\pi}{2})$ and of angle $N\pi/2$ about the axis $(\tfrac{\pi}{2},\tfrac{\pi}{2},z)$ for $N \in \mathbb{Z}$. The above mentioned planes that possess mirror symmetries form the insulating and conducting walls of $[0,\pi]^3$ sub-boxes, also called impermeable boxes \cite{brachetetal83}, for the corresponding initial conditions.

It is important to mention that Lee et al. \cite{leeetal10} imposed numerically these symmetries in order to gain substantial savings in computational resources. Unlike \cite{leeetal10}, our computations were performed without imposing any symmetry constrains, allowing thus the turbulence to evolve freely with the view that the initial TG vortex symmetries will break at high enough Reynolds numbers. However, even for our highest resolution simulations with Taylor Reynolds number of the order of 100 the TG symmetries are not broken within the time interval of reaching the peak of dissipation. They seem to be a strong property of the MHD equations, preserved by time evolution of the solutions (see also \cite{leeetal08}).

Due to the fact that there are special global restrictions on these TG flows, we further consider a run with random initial conditions (run ``R'' hereafter) for comparison, ensuring that $H_b = H_c = 0$ and kinetic helicity $H_u \equiv \avg{\bm u \sdot \bm \omega} = 0$ at time $t=0$. During the time evolution magnetic and cross helicity remain zero for all times relative to the total energy. However, the kinetic helicity reaches an approximate value of $H_u\ell/E_t < 0.2$ at its absolute maximum over time but when dissipation is maximum $H_u\ell/E_t < 0.04$. 

We report results based on the analysis of decaying MHD turbulence simulated with $N = 1024^3$ grid points. In order to obtain the broadest inertial range, runs I, A and C are initialised at the largest scales and 
run R at wavenumbers $k = 1$ and 2, adding extra randomness. At time $t=0$ the fields are normalised such that the kinetic and magnetic energies are in equipartition, i.e. $E_u(t=0) = E_b(t=0) = 0.125$. Note that all flows have unit magnetic Prandtl number (i.e. $\nu = \kappa$). The numerical parameters of our computations are provided in Table \ref{tbl:dnsparam}.

\begin{table}[!ht]
  \caption{Numerical parameters of the DNS. The values presented are taken at the peak of total dissipation. Note that $k_{max} = N/3$, using the $2/3$ dealiasing rule.}
  \label{tbl:dnsparam}
   \begin{ruledtabular}
    \begin{tabular}{*{10}{c}}
      \textbf{Run} & \textbf{N} & $\bm{\nu}$ & $\bm{Re_{\lambda_t}}$ & $\bm{L_t}$ & $\bm{\lambda_t}$ & $\bm{\eta_t}$ & $\bm{u'}$ & $\bm{b'}$ & $\bm {k_{max}\eta_t}$ \\
       & & $(\times 10^{-4})$ & & $(\times 10^{-1})$ & $(\times 10^{-1})$ & $(\times 10^{-3})$ & & & \\
     \hline
      R  & 1024 & 5.5 & 140.7 & 8.33 & 2.15 & 7.80 & 0.36 & 0.48 & 2.66 \\
      I  & 1024 & 4.5 & 121.8 & 6.84 & 2.03 & 6.54 & 0.27 & 0.62 & 2.23 \\
      C  & 1024 & 4.5 & 138.0 & 6.23 & 1.35 & 5.82 & 0.46 & 0.35 & 1.97 \\
      A  & 1024 & 4.5 & 115.1 & 3.76 & 1.40 & 5.77 & 0.37 & 0.46 & 1.99 \\
    \end{tabular}
  \end{ruledtabular}
\end{table}

The rms velocity $u'$ is defined as
\begin{equation}
  u' \equiv \lt( \frac{2}{3}\int E_u(k)dk \rt)^{1/2}.
\end{equation}
and similarly for $b'$, the rms of the magnetic field.
The integral length scales are then defined as the total, kinetic and magnetic integral length scale, respectively
\begin{equation}
 L_{t,u,b} \equiv \frac{3\pi}{4}\frac{\int k^{-1}E_{t,u,b}(k)dk}{\int E_{t,u,b}(k)dk}
\end{equation}
and likewise for the Taylor scales
\begin{equation}
 \lambda_{t,u,b} \equiv \lrbig{5\frac{\int E_{t,u,b}(k)dk}{\int k^2E_{t,u,b}(k)dk}}^{1/2}.
\end{equation}
In Table \ref{tbl:dnsparam}, we report the total integral and Taylor length scales as well as the Reynolds number based on $\lambda_t$ given by $Re_{\lambda_t} \equiv u' \lambda_t / \nu$. Finally, the smallest length scale in our flows is defined based on K41 scaling $\eta_t \equiv (\nu^3 / \epsilon_t)^{1/4}$, where $\epsilon_t = \nu \avg{|\bm \omega|^2} + \kappa \avg{|\bm j|^2}$ is the total dissipation. The time we address in this study is the moment which the dissipation reaches its maximum value and therefore the highest scale separation occurs $\eta_t \ll \ell \ll L_t$, where $\ell$ is a typical length scale in the inertial range. Therefore, the values provided in Table \ref{tbl:dnsparam} correspond to that moment.

\section{\label{sec:spectra}Energy spectra}
\begin{figure}[!ht]
  \begin{subfigure}{0.35\textwidth}
   \includegraphics[width=\textwidth]{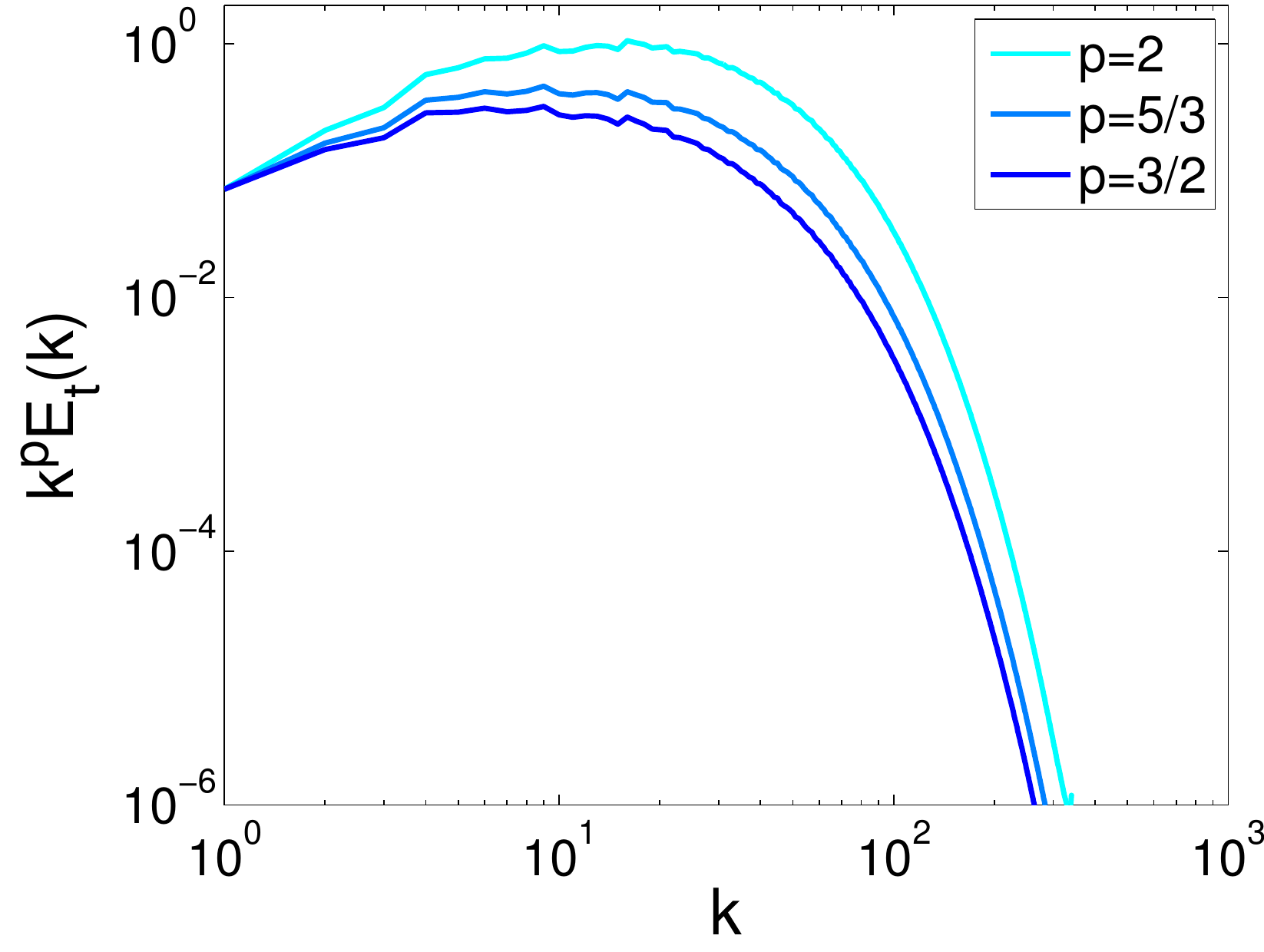} 
   \caption{Run R}
  \end{subfigure}
  \begin{subfigure}{0.35\textwidth}
   \includegraphics[width=\textwidth]{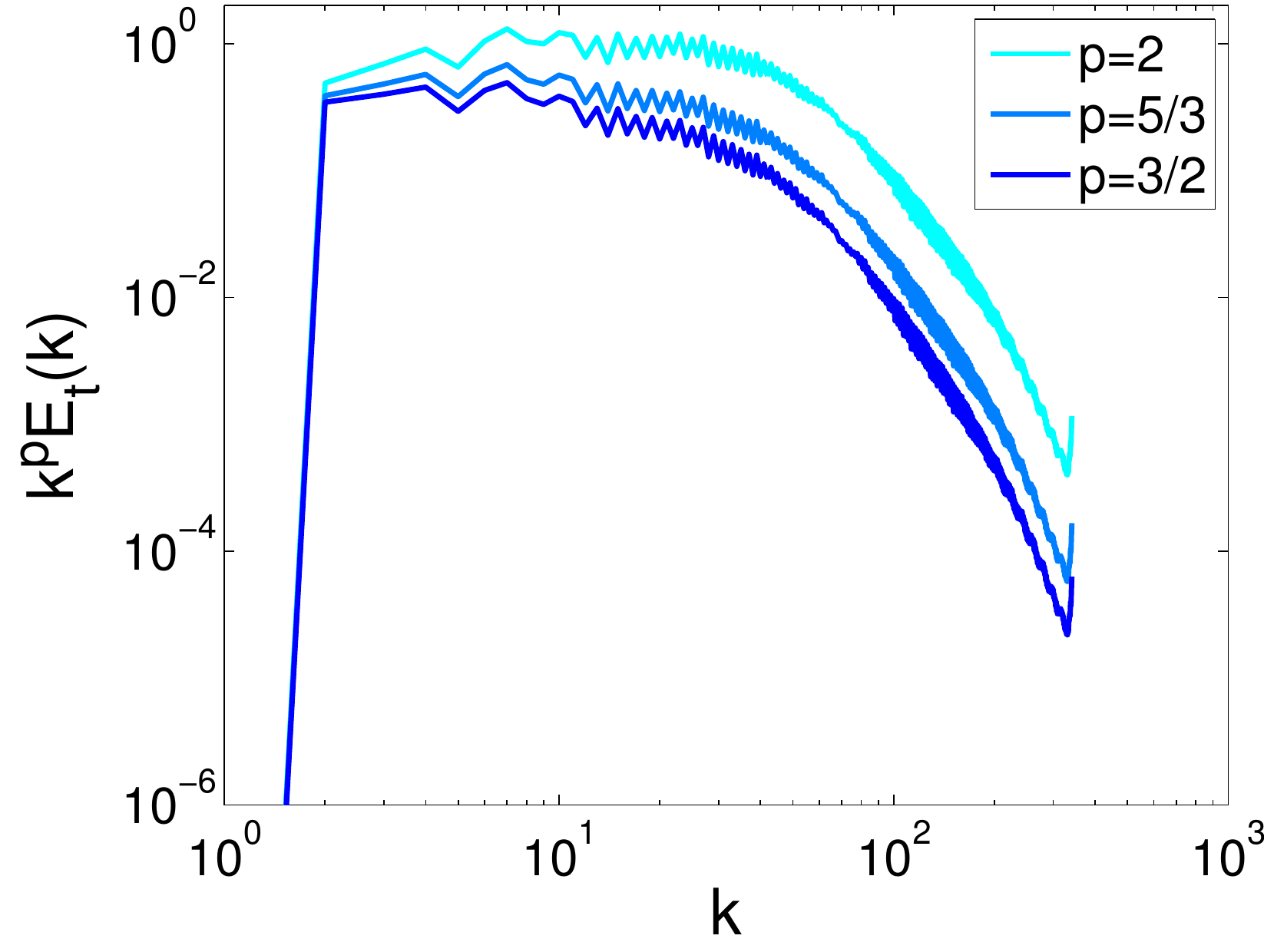}
   \caption{Run I}
  \end{subfigure} \\
  \begin{subfigure}{0.35\textwidth}
   \includegraphics[width=\textwidth]{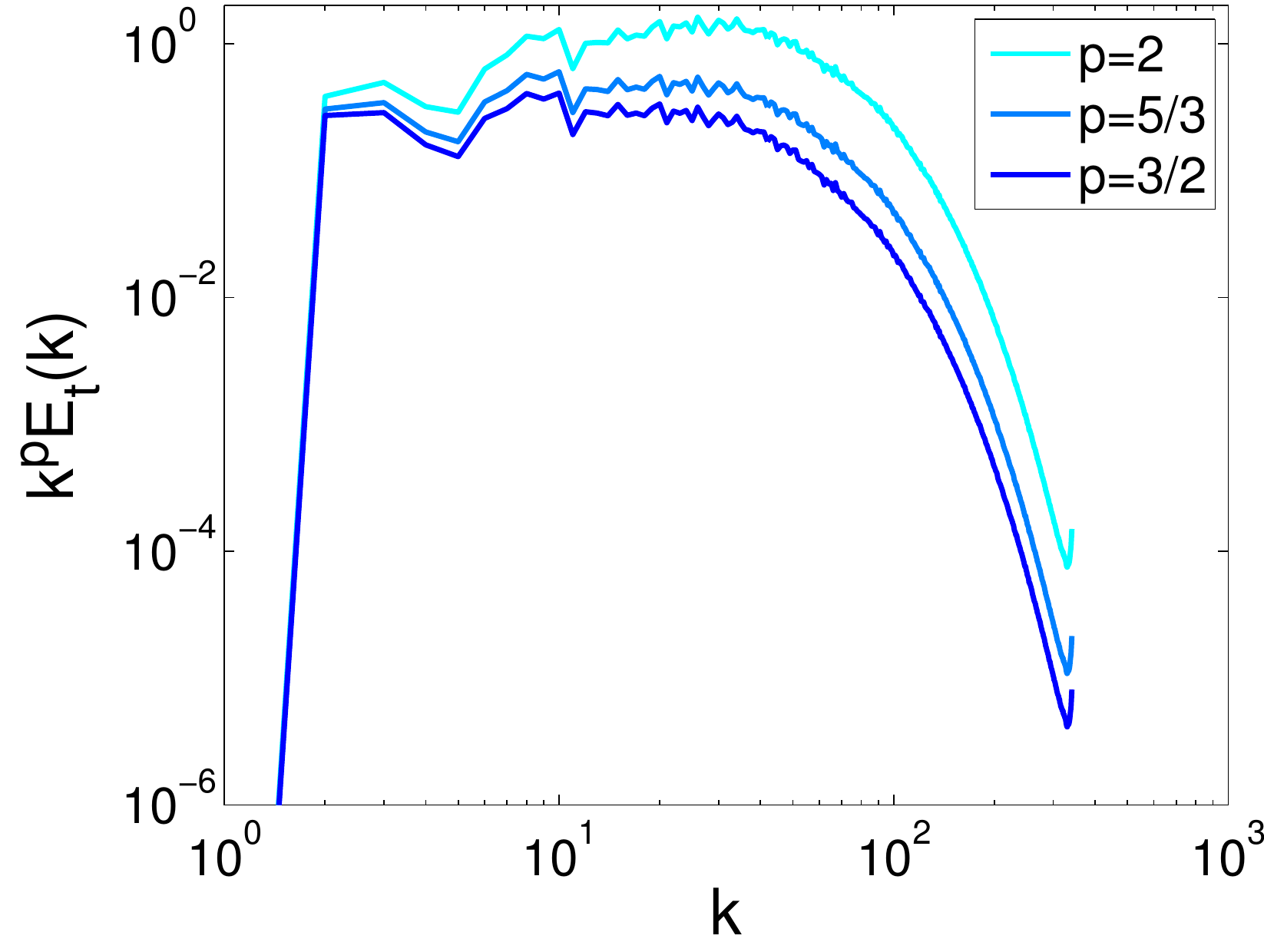}
   \caption{Run C}
  \end{subfigure}
  \begin{subfigure}{0.35\textwidth}
   \includegraphics[width=\textwidth]{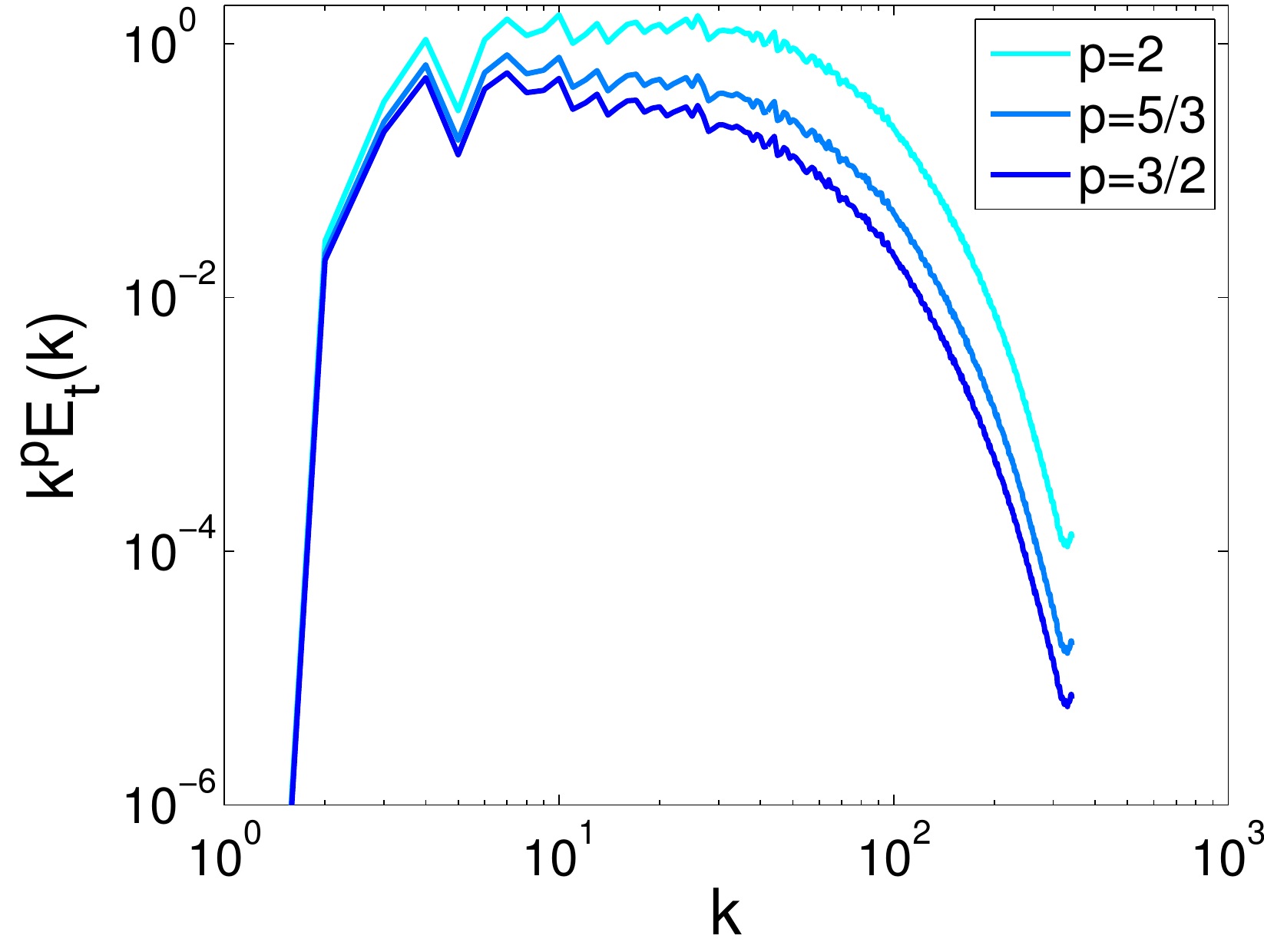}
   \caption{Run A}
  \end{subfigure}
  \caption{(Color online) Three-dimensional compensated total energy spectra $k^pE_t(k)$ with scaling exponents $p = 2,\; 5/3,\; 3/2$ for a) run R, b) run I, c) run C and  d) run A of Table \ref{tbl:dnsparam}.}
  \label{fig:et_spectra}
 \end{figure}

Figure \ref{fig:et_spectra} presents the three-dimensional compensated total energy spectra $k^pE_t(k)$ that we obtain at the peak of dissipation for all the runs of Table \ref{tbl:dnsparam}. The spectra are compensated with the scaling exponents $p = 2,\; 5/3$ and $3/2$. The small peaks at high wavenumbers show the quality of our simulations.

According to Lee et al. \cite{leeetal10}, the energy spectrum of the magnetically dominated flow I, i.e. $E_b/E_u > 1$ (see also $u'$ and $b'$ values in Table \ref{tbl:dnsparam}), is close to a $k^{-2}$ power law (see Fig. \ref{fig:et_spectra}b), which is the weak turbulence (WT) theory expectation \cite{ngbhattacharjee97,galtieretal00}. Here, we would like to emphasize, however, that the WT scaling ($E_\perp \propto k_\perp^{-2}$) is for an anisotropic energy spectrum, where perpendicular denotes the direction relative to an imposed large scale mean magnetic field $\bm B_0$ that does not exist in this flow. In fact, it is argued that in MHD turbulence there is no prescribed cascade in the parallel direction \cite{biskamp03}. This is based on the idea that small-scale turbulent fluctuations become anisotropic, as it is easier to shuffle strong magnetic field lines than to bend them due to the preventing action of the Lorentz force $\bm j \times \bm b$. 
caused by a large-scale field $\bm B_0$. 

Furthermore, Lee et al. \cite{leeetal10} argues that the kinetic energy dominated flow C, i.e. $E_b/E_u < 1$ (see also Table \ref{tbl:dnsparam}), is compatible with a $k^{-3/2}$ slope (Fig. \ref{fig:et_spectra}c) and the less magnetically dominated flow A is near a $k^{-5/3}$ scaling (Fig. \ref{fig:et_spectra}d). In addition, we report that the power law of the total energy spectrum for our also magnetically dominated run R (see Table \ref{tbl:dnsparam}) seems to be between $k^{-5/3}$ and $k^{-3/2}$. The difference between these two power laws is subtle enough that any type of contamination, such as intermittency or any dissipative small-scale effects, will blur the results. However, even a $k^{-2}$ spectrum which is slightly more transparent in these high Reynolds numbers can be misinterpreted. For example, in contrast to \cite{leeetal10}, we claim that the total energy spectrum of run A (Fig. \ref{fig:et_spectra}d) scales like $E_t \propto k^{-2}$ but we leave this to the readers' judgement.

Therefore, the following questions are raised: How can we circumvent this ambiguity of the results? Is there a dependence of small scales on the large scale initial conditions and thereby non-universality in decaying MHD turbulence? Since limited information can be extracted just from the slopes of the spectra, we try to answer these questions by examining the topology of the small scales through the invariants of related gradient statistics, which some have shown universal characteristics for hydrodynamic turbulent flows.

\section{\label{sec:defs}Classification of the fluid flow topology}
An approach that provides a well-defined and unambiguous language to describe eddying motions and flow patterns is the framework of critical point concepts from bifurcation theory \cite{glendinning94}, which was studied extensively in the context of hydrodynamic turbulent flows by Perry, Chong, Cantwell and co-workers \cite{chongetal90,soriaetal94,chongetal98,ooietal99}. 
Here we provide a brief outline on the background material related to the geometric invariants of second-order tensorial quantities in turbulence before going to consider various statistics of these invariants in the following sections. Extensive reviews on the subject can be found in \cite{cantwell02,tsinober02,davidson04} and references therein. 

Geometric invariants remain unchanged under the full group of rotations (i.e. rotations plus reflections) \cite{tsinober02}, being independent of the frame of reference. Any traceless second-order tensor $\bm M$ has the following characteristic polynomial
\begin{equation}
 \label{eq:cubic}
 \det[\bm M - \lambda_i \bm I] = 0 \Rightarrow 
 \lambda_i^3 + P \lambda_i^2 + Q \lambda_i + R = 0
\end{equation}
where $\lambda_i$ are the eigenvalues of $\bm M$ and its invariants are
\begin{align}
  \label{eq:invI}
 P &= -tr(\bm M) = - (\lambda_1 + \lambda_2 + \lambda_3) = 0 \\
  \label{eq:invII}
 Q &=  \frac{1}{2}[P^2 - tr(\bm M^2)] = \lambda_1\lambda_2 + \lambda_2\lambda_3 + \lambda_3\lambda_1 \\
  \label{eq:invIII}
 R &= -\det(\bm M) = - \lambda_1\lambda_2\lambda_3
\end{align}
The value of the discriminant for $P = 0$ is
\begin{equation}
 D = \tfrac{27}{4}R^2 + Q^3
\end{equation}
and provides a general classification for the solutions of the cubic equation \eqref{eq:cubic} dividing the (Q,R) space into the following regions
 \begin{enumerate}
  \item $D > 0$: 1 real \& 2 complex-conjugate eigenvalues
  \item $D = 0$: 3 real eigenvalues of which 2 are equal
  \item $D < 0$: 3 real distinct eigenvalues
 \end{enumerate}
which correspond to various local flow topologies. In this study, the first invariant is $P = 0$ from definition \eqref{eq:invI} since the vector fields that we consider are solenoidal.

%





%
\section{\label{sec:invu}Invariants of the velocity gradient, the strain rate and rotation rate tensors}
\subsection{Joint PDFs of the velocity gradient invariants}
\begin{figure}[!ht]
  \includegraphics[width=0.35\textwidth]{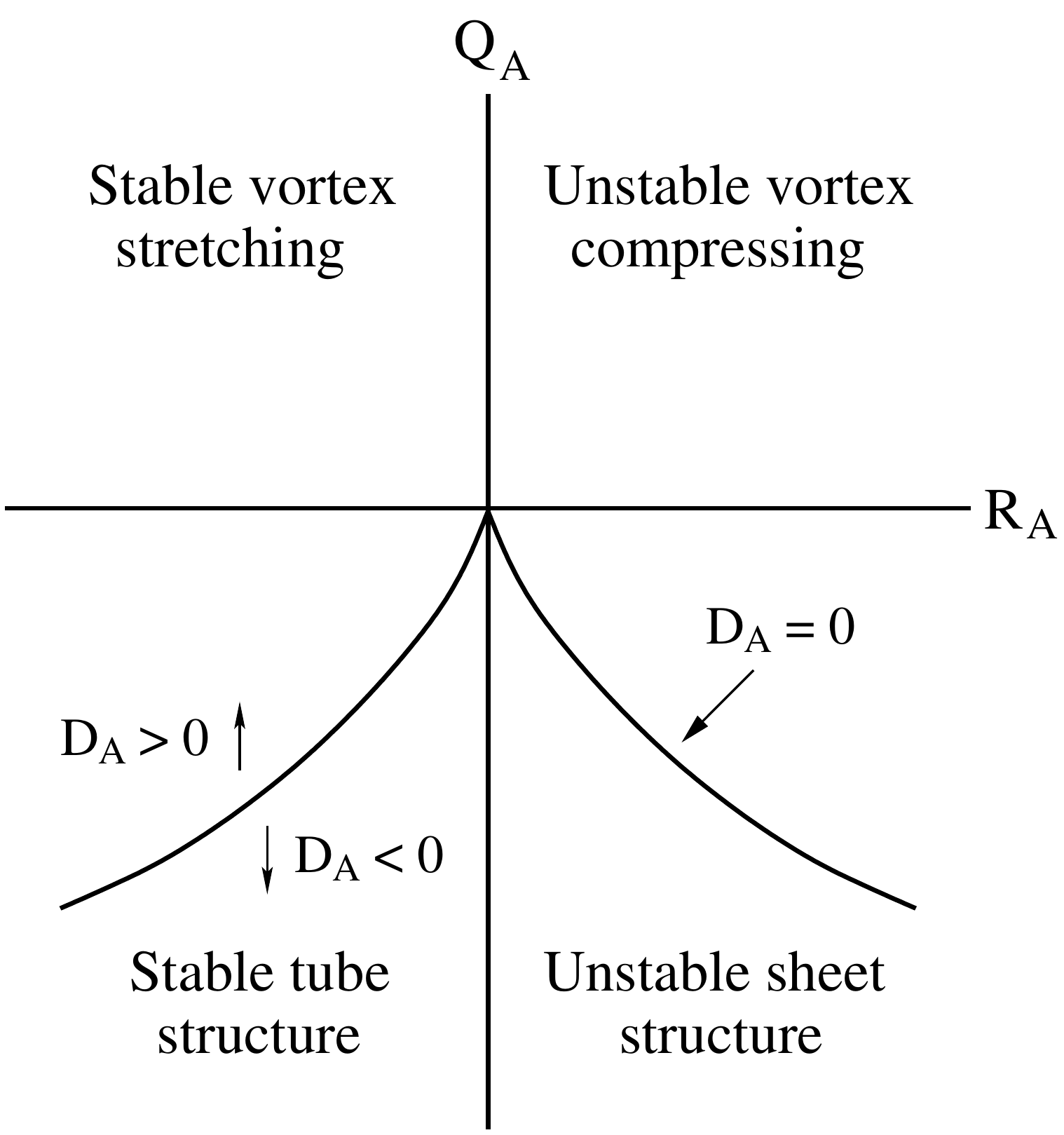}
  \caption{Diagram of the ($R_A,Q_A$) invariant map indicating the local flow topologies related to each zone.}
  \label{fig:QaRa_map}
 \end{figure}

The velocity gradient tensor $\bm A = \grad \bm u$ can be decomposed into a symmetric and skew-symmetric component,
\begin{equation}
 \bm A = \bm S + \bm \Omega = S_{ij} - \tfrac{1}{2}\epsilon_{ijk}\omega_k
\end{equation}
where $\bm S = \tfrac{1}{2}(\grad \bm u + \grad \bm u^T)$ and $\bm \Omega = \tfrac{1}{2}(\grad \bm u - \grad \bm u^T)$ are the strain rate and rotation rate tensors, respectively. According to Eqs. \eqref{eq:invII} and \eqref{eq:invIII}, the second and third invariants of $\bm A$ are
\begin{equation}
 \label{eq:Qa}
 Q_A = \tfrac{1}{4}[\bm\omega^2 - 2tr(\bm S^2)]
\end{equation}
and
\begin{equation}
 \label{eq:Ra}
 R_A = -\tfrac{1}{3}[tr(\bm S^3) + \tfrac{3}{4}\omega_i\omega_jS_{ij}],
\end{equation}
respectively. Here we are interested in the joint probability density function (PDF) of these invariants. A diagram of this joint PDF called the ($R_A,Q_A$) invariant map is presented in Fig. \ref{fig:QaRa_map}, labelling the various topological classifications.

If $Q_A > 0$ then enstrophy $\bm \omega^2$ dominates over $tr(\bm S^2)$ and vice versa if $Q_A < 0$. For positive values of $R_A$ the topologies are unstable, whereas for negative $R_A$ the topologies are stable. Moreover, the $D_A = 0$ line (see Fig. \ref{fig:QaRa_map}), where $D_A = \tfrac{27}{4} R_A ^2 + Q_A^3$ is the discriminant, divides the invariant map into two regions. One where $D_A > 0$ with one real and two complex-conjugate eigenvalues as solutions of Eq. \eqref{eq:cubic} for the velocity gradient tensor and the other where $D_A < 0$ with three real distinct eigenvalues. Note that along the vertical $R_A = 0$ axis one of the eigenvalues is zero and therefore locally the flow topology is invariant in this direction. 

Now, if $Q_A$ is much greater than zero (i.e. $D_A > 0$) then $R_A \approx -\tfrac{1}{4}\omega_i\omega_jS_{ij}$. In this case, for $R_A < 0$ vortex stretching dominates over vortex compression, whereas for $R_A > 0$ vortex compression dominates (see Fig. \ref{fig:QaRa_map}). On the other hand, if $Q_A$ is much less than zero and 
$D_A < 0$ then $R_A \approx -\tfrac{1}{2}tr(\bm S^3)$. In this case, $R_A > 0$ locally is related to a sheetlike structure (or unstable node/saddle/saddle topologies according to the terminology of Chong et al. \cite{chongetal90}) whereas $R_A < 0$ with a tubelike structure (or stable node/saddle/saddle topologies \cite{chongetal90}). This will also become more transparent when we will deal later with the third invariant of the strain rate tensor (see section \ref{sec:QsRs}). 

In hydrodynamic turbulent flows, ranging from atmospheric boundary layers to free shear flows in wind tunnels and even simulations of compressible turbulence, there is the prominent tendency of the joint PDF of ($R_A,Q_A$) to develop an inclined teardrop shape. This shape aligns with the second and fourth quadrants, with a cusp lying along the $R_A > 0$, $D_A = 0$ branch (see for example Fig. 10.1 in \cite{tsinober02}) and is considered to be a universal feature. Therefore, there is a preference for vortex stretching and sheetlike structures. In many visualisations of enstrophy in hydrodynamic turbulent flows the dominant structures seem to be tubelike structures but between these vortex-tubes there are sheetlike structures, where most of the dissipation is located \cite{tanakakida93,tsinober02}.

Before analysing the results, we would like to note that the aspect ratio of the axes of all joint PDFs, that are reported in this paper, has been kept the same but the abscissa and the ordinate are different to reflect the change in magnitude of the plotted quantities in the four flows that we consider. In addition, the points near the origin correspond to low gradient values associated with the large scale motions, whereas points far away characterize the high-gradient small scales. 
All the joint PDFs were computed at the instant of maximum dissipation.

\begin{figure}[!ht]
  \begin{subfigure}{0.35\textwidth}
   \includegraphics[width=\textwidth]{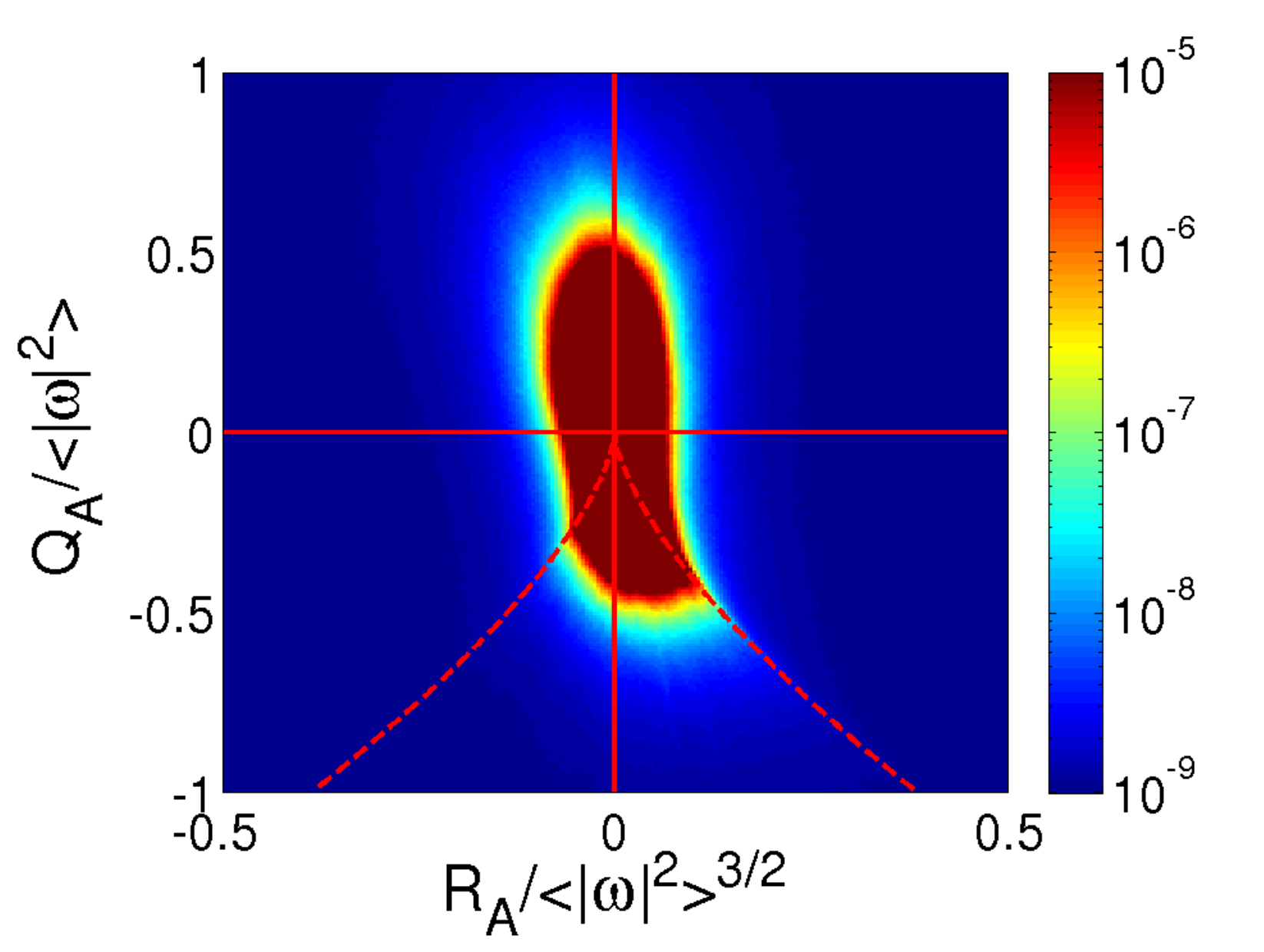}
   \caption{Run R}
  \end{subfigure}  
  \begin{subfigure}{0.35\textwidth}
   \includegraphics[width=\textwidth]{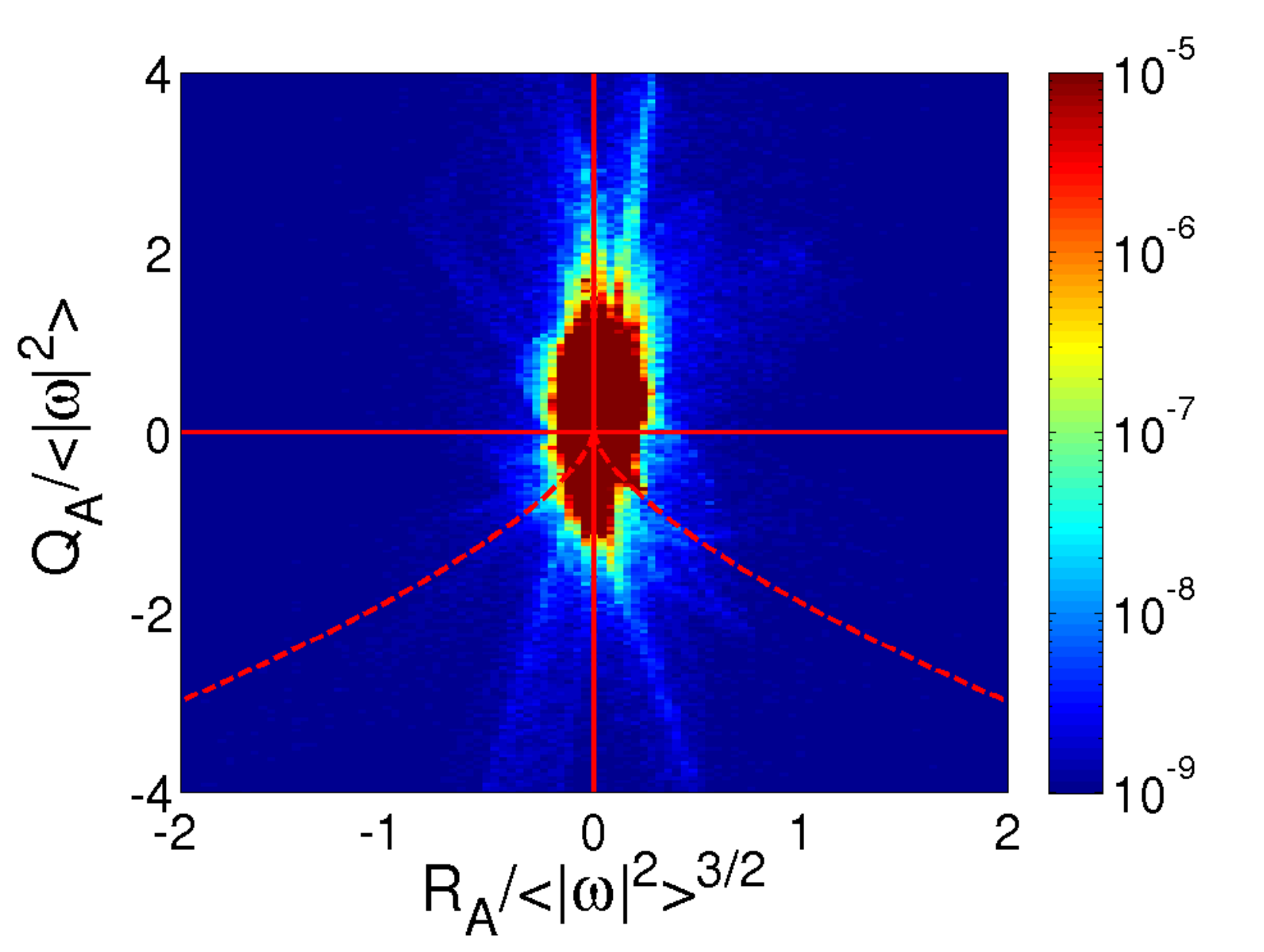}
   \caption{Run I}
  \end{subfigure} \\
  \begin{subfigure}{0.35\textwidth}
   \includegraphics[width=\textwidth]{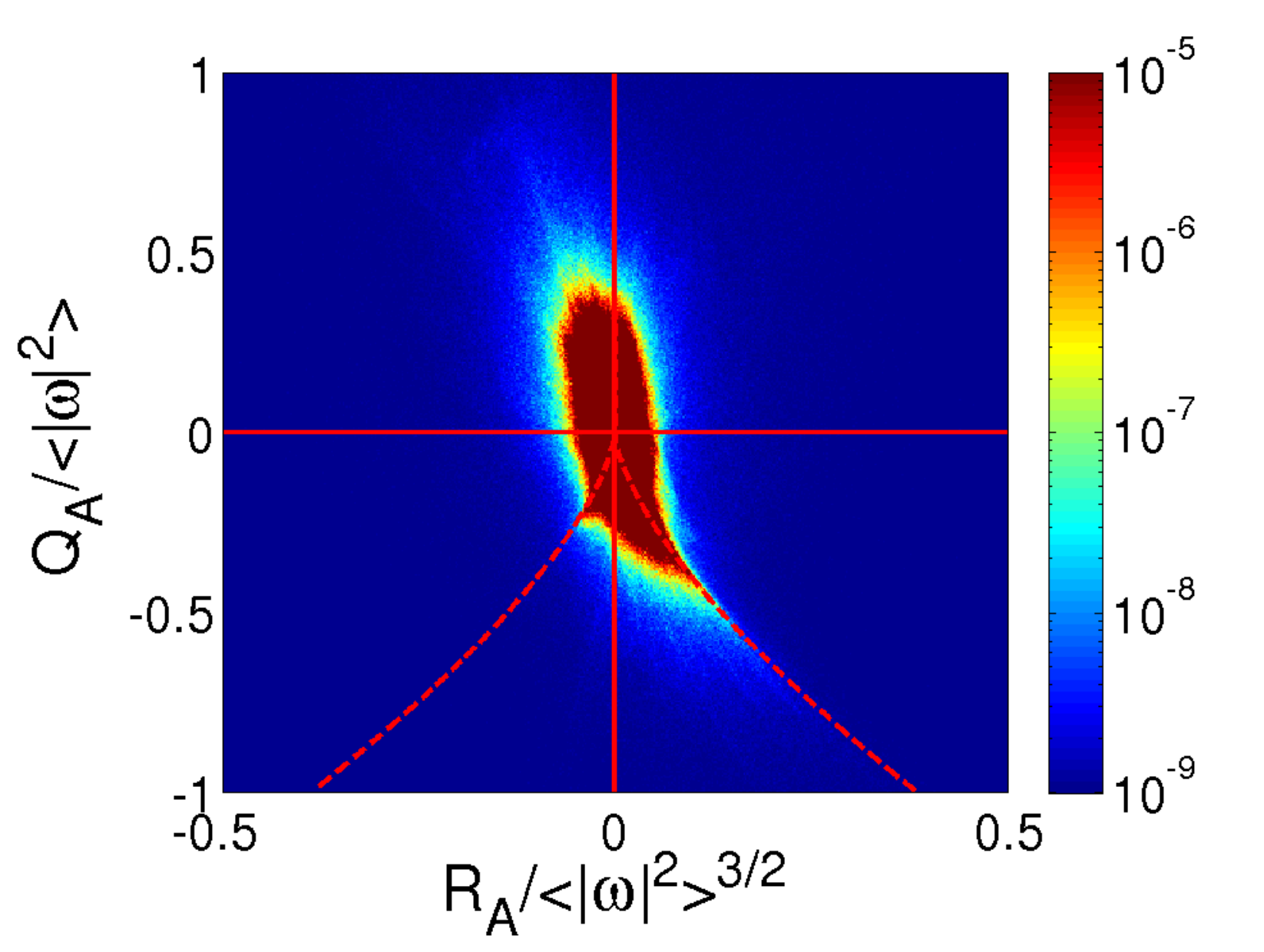}
   \caption{Run C}
  \end{subfigure}
  \begin{subfigure}{0.35\textwidth}
   \includegraphics[width=\textwidth]{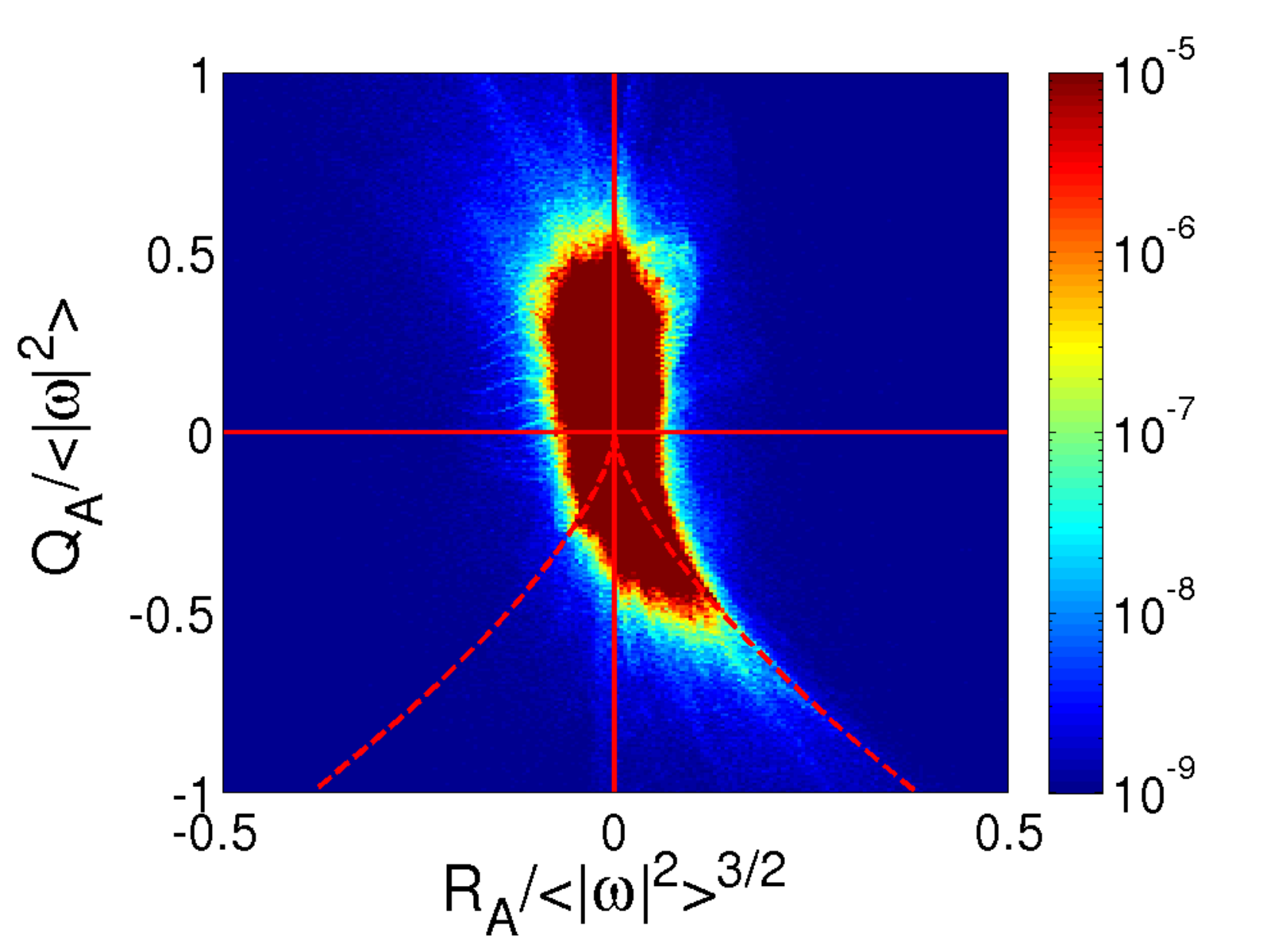}
   \caption{Run A}
  \end{subfigure}
  \caption{(Color online) Joint PDFs of the second invariant $Q_A$ and the third invariant $R_A$ of the velocity gradient tensor normalised appropriately by powers of the mean enstrophy for a) run R, b) run I, c) run C and  d) run A of Table \ref{tbl:dnsparam}. The line $D_A = \tfrac{27}{4} R_A ^2 + Q_A^3 = 0$ is plotted for reference.}
  \label{fig:QaRa}
 \end{figure}

In Fig. \ref{fig:QaRa} we present the joint PDFs of $R_A$ versus $Q_A$ for all the decaying MHD runs of Table \ref{tbl:dnsparam}.
The most important outcome from the plots in Fig. \ref{fig:QaRa} is that the shape of the joint PDF of $R_A$ with $Q_A$ is not universal in decaying MHD turbulence and small-scales seem to depend on the large-scale initial conditions. However, we should be cautious here as it is not clear whether the self-preservation of the TG vortex symmetries during the evolution restrict the dynamics in some way. 

On the other hand, it is clear that there is a modest but still present trend of the ($R_A,Q_A$) map to align along the second and fourth quadrants for our simulation with random initial conditions (Fig. \ref{fig:QaRa}a). It is noteworthy that the shape of the joint PDF is more symmetric with respect to the $R_A = 0$ axis in comparison to hydrodynamic flows (see for example \cite{ooietal99}). Run I gives a striking joint PDF (Fig \ref{fig:QaRa}b) with a significant percentage of its points lying in the first quadrant and with high absolute values of $Q_A$ in comparison to the rest of the runs.
Points of the joint PDF in the first quadrant that are far from the origin (see Fig \ref{fig:QaRa}b) are associated with very low rates of kinetic energy dissipation. This suggests that the 
structure is likely to be quite long-lived. Run C seems to resemble more the teardrop shape of hydrodynamic turbulence with the classic narrow cusp in $R_A > 0$, $D_A = 0$ branch (Fig. \ref{fig:QaRa}c). Finally, Fig. \ref{fig:QaRa}d shows the ($R_A,Q_A$) map of run A, which has a shape with features in between the random MHD and hydrodynamic turbulence. In other words, there is a modest tendency of the joint PDF to align with the second quadrant like in the random MHD run (Fig. \ref{fig:QaRa}a) but there is a high correlation between $R_A > 0$ and $Q_A < 0$ values forming a long cusp in analogy to hydrodynamic turbulent flows.

%
%
\subsection{\label{sec:QsRs}Joint PDFs of strain rate invariants}
Setting $\bm \Omega$ to zero or essentially $\bm \omega$ to zero in Eqs. \eqref{eq:Qa} and \eqref{eq:Ra}, we can obtain the invariants of the strain rate tensor, which are
\begin{align}
 Q_s &= -\tfrac{1}{2}tr(\bm S^2) \\
\label{eq:rs}
 R_s &= -\tfrac{1}{3}tr(\bm S^3).
\end{align}

\begin{figure}[!ht]
  \includegraphics[width=0.35\textwidth]{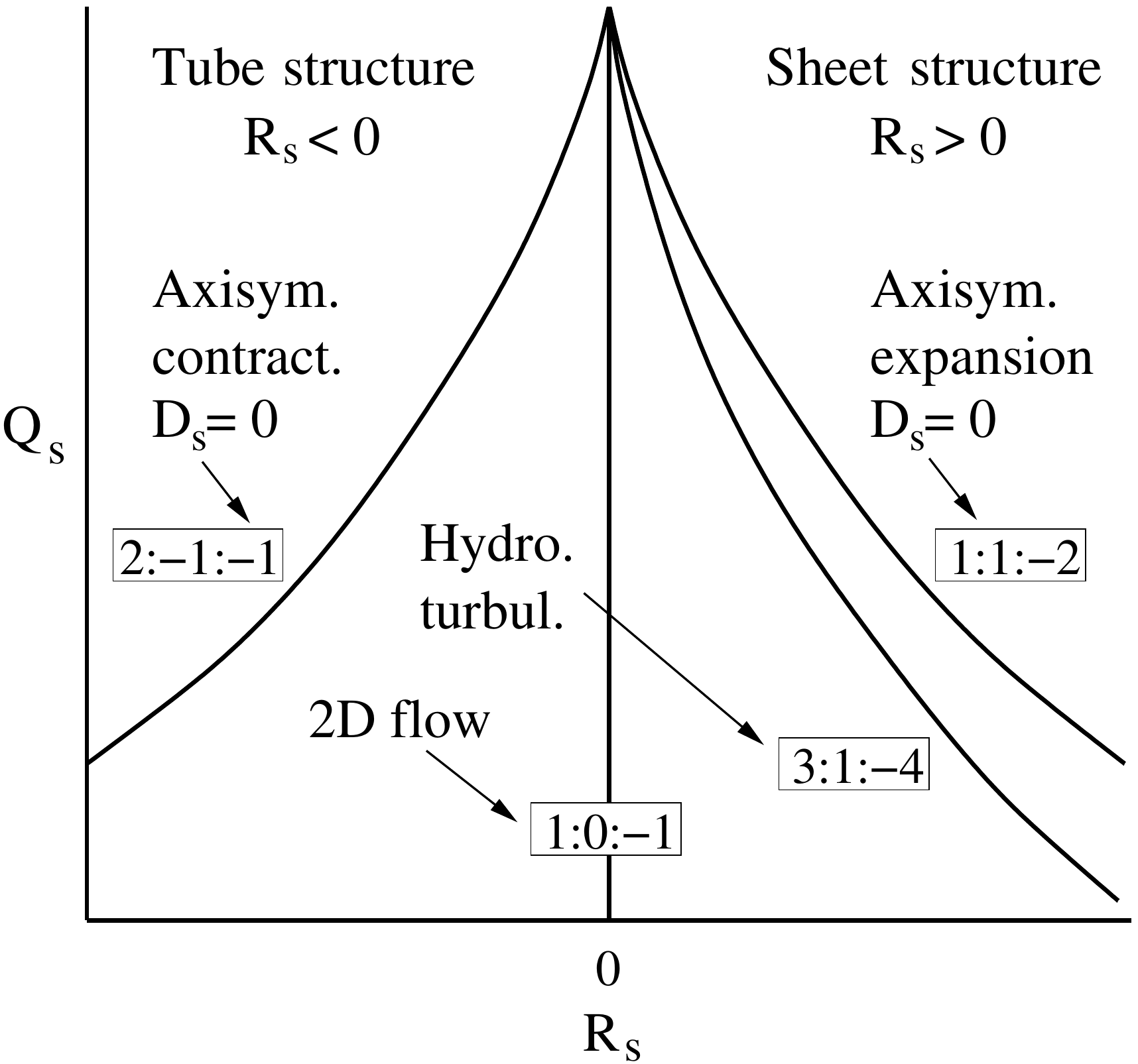}
  \caption{Diagram of the ($R_s,Q_s$) invariant map. Each plotted curve corresponds to the following flow geometries: $\lambda_1:\lambda_2:\lambda_3 = 2:-1:-1$ (axisymmetric contraction), $1:0:-1$ (two-dimensional flow), $3:1:-4$ (biaxial stretching) and $1:1:-2$ (axisymmetric stretching).}
  \label{fig:QsRs_map}
 \end{figure}

\begin{figure}[!ht]
  \begin{subfigure}{0.35\textwidth}
   \includegraphics[width=\textwidth]{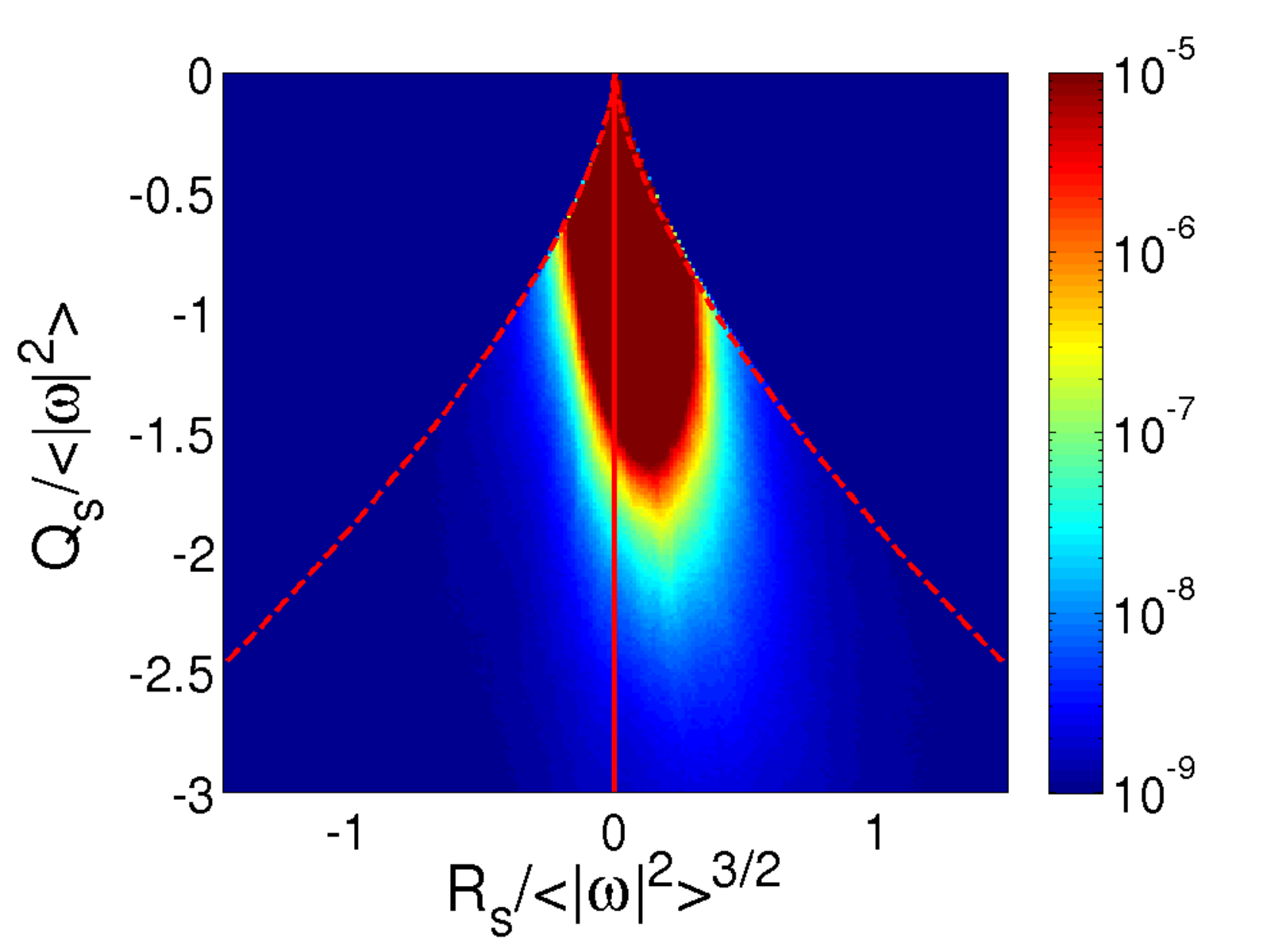}
   \caption{Run R}
  \end{subfigure}
    \begin{subfigure}{0.35\textwidth}
   \includegraphics[width=\textwidth]{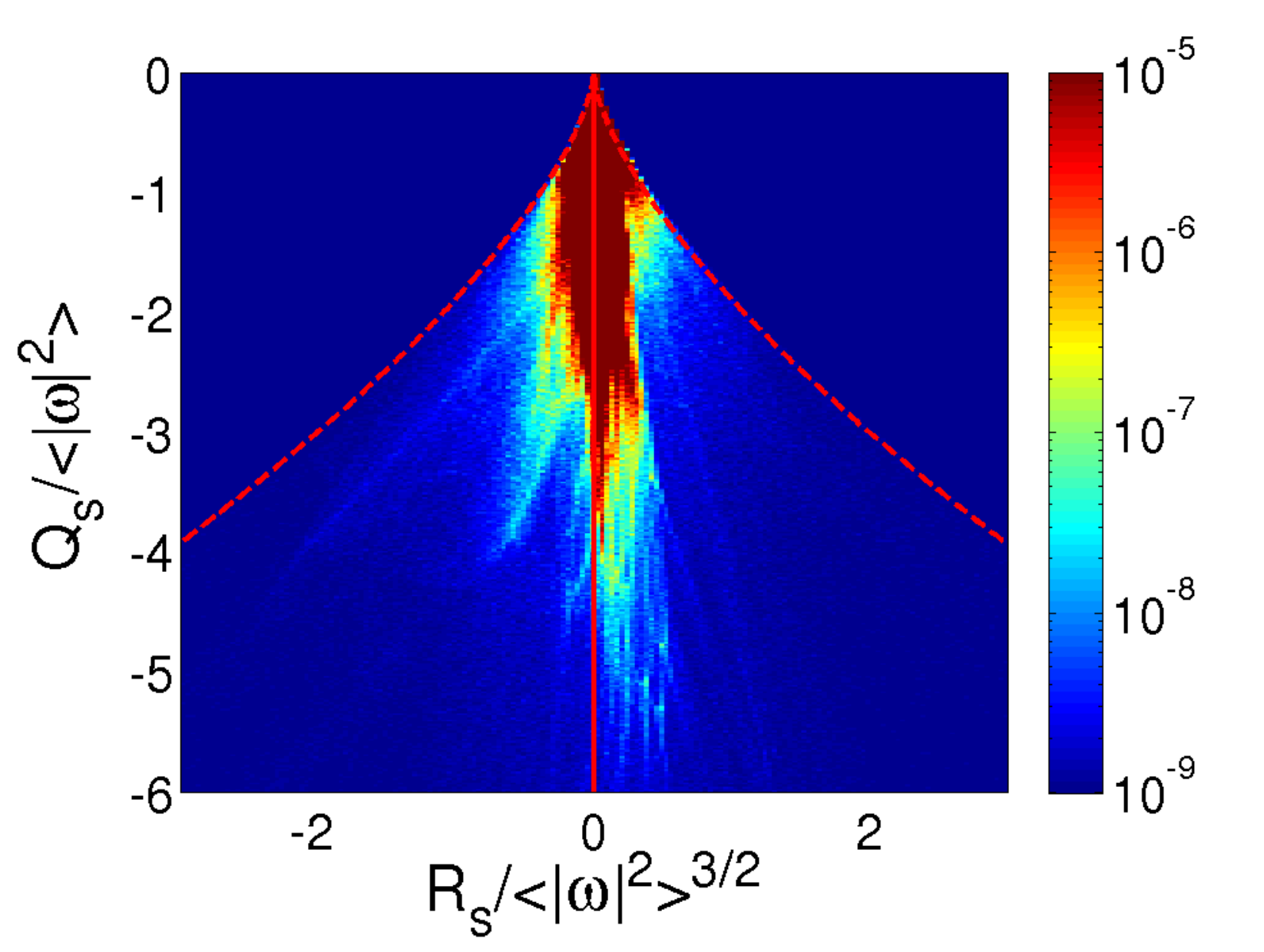}
   \caption{Run I}
  \end{subfigure} \\
    \begin{subfigure}{0.35\textwidth}
   \includegraphics[width=\textwidth]{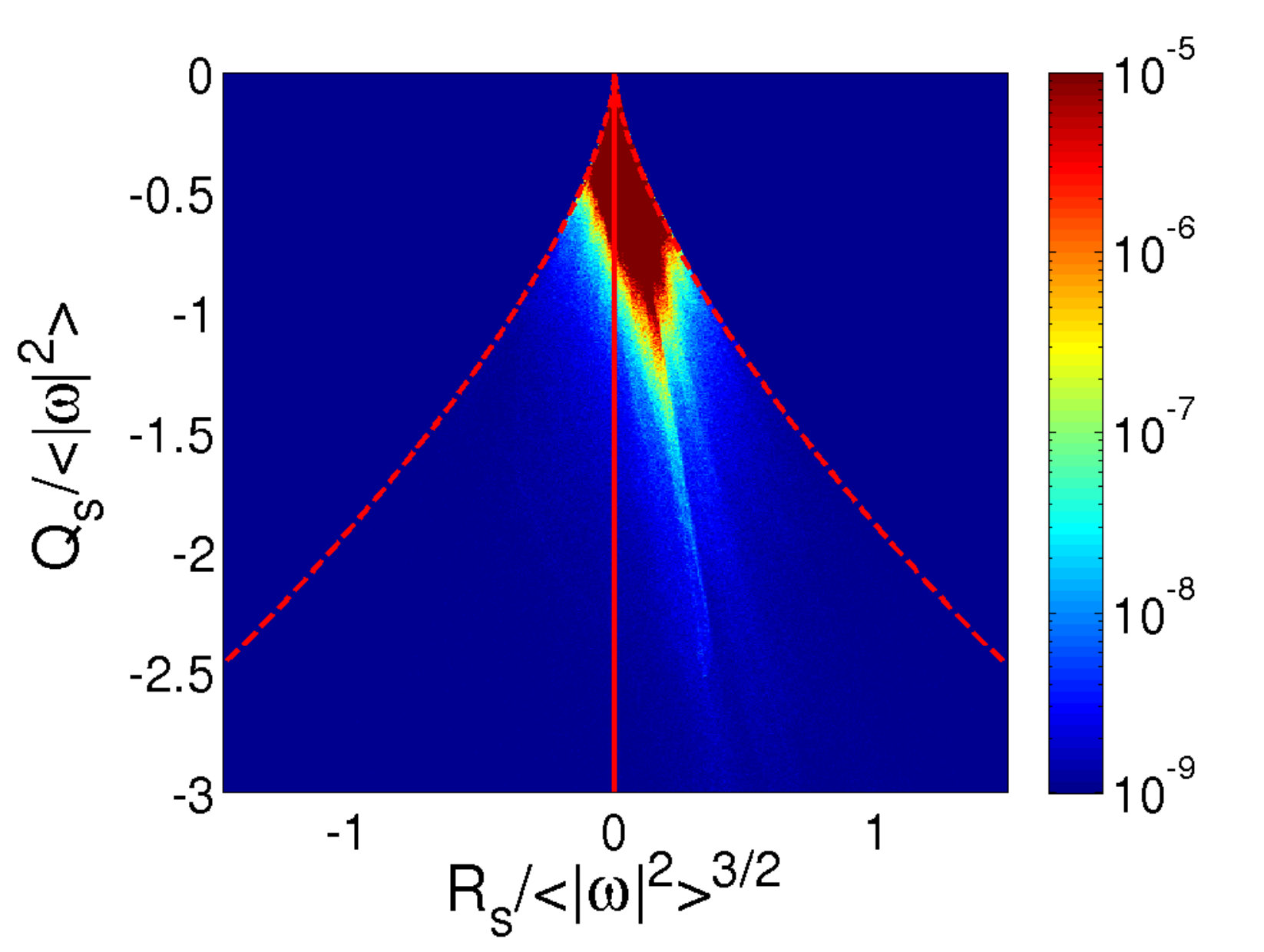}
   \caption{Run C}
  \end{subfigure}
    \begin{subfigure}{0.35\textwidth}
   \includegraphics[width=\textwidth]{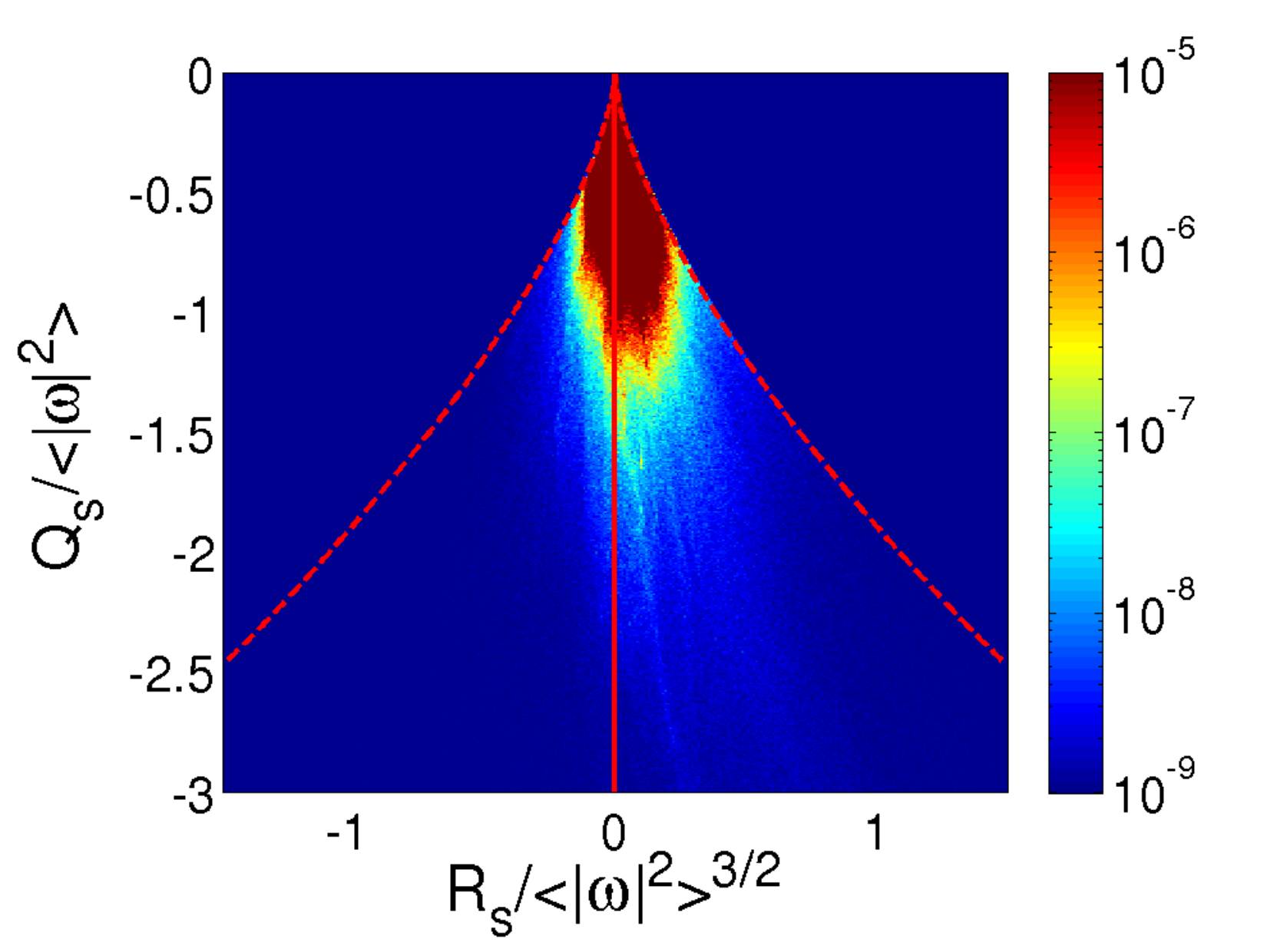}
   \caption{Run A}
  \end{subfigure}
  \caption{(Color online) Joint PDFs of the second invariant $Q_s$ and third invariant $R_s$ of the strain rate tensor normalised appropriately by powers of the mean enstrophy for a) run R, b) run I, c) run C and  d) run A of Table \ref{tbl:dnsparam}. The line $D_s = \tfrac{27}{4} R_s ^2 + Q_s^3 = 0$ is plotted for reference.}
  \label{fig:QsRs}
 \end{figure}

The ($R_s,Q_s$) invariant map features the geometry of the local straining of the fluid elements (see Fig. \ref{fig:QsRs_map}).
The second invariant $Q_s$ is related to viscous dissipation $\epsilon = 2 \nu \bm S^2$ through $Q_s = -\tfrac{1}{4}\epsilon / \nu$ because the strain rate tensor is symmetric, i.e. $\bm S^2 = S_{ij}S_{ji} = tr(\bm S^2)$. So, locations with $Q_s$ much less than zero are highly dissipative regions. Note that $Q_s$ is negative definite. The third invariant $R_s$ has two important physical meanings. First, it is proportional to strain skewness $S_{ij}S_{jk}S_{ki}$, which appears as part of the production term in the evolution equation of $\bm S^2$ (see \cite{tsinober02}). Second, it can be written as a function of the eigenvalues of $S_{ij}$, viz. 
\begin{equation}
 \label{eq:rseig}
 R_s = -\tfrac{1}{3}(\lambda_1^3 + \lambda_2^3 + \lambda_3^3) = -\lambda_1\lambda_2\lambda_3 
\end{equation}
since $tr(\bm S) = \lambda_1 + \lambda_2 + \lambda_3 = 0$ due to incompressibility, with $\lambda_1 \geq \lambda_2 \geq \lambda_3$. Owing to the symmetry of $S_{ij}$ all eigenvalues are real and 
therefore the ($R_s,Q_s$) invariant map is contained only in the region where $D_s = \tfrac{27}{4} R_s ^2 + Q_s^3 \leq 0$ (see Figs. \ref{fig:QsRs_map} and \ref{fig:QsRs}). So, $R_s > 0$ implies production of $\bm S^2$ and hence of viscous dissipation with $\lambda_1,\lambda_2 > 0$ and $\lambda_3 < 0$ related to sheetlike structures. On the contrary, $R_s < 0$ indicates destruction of $\bm S^2$ with $\lambda_1 > 0$ and $\lambda_2,\lambda_3 < 0$ associated with tubelike structures. Note, therefore that $\text{sgn}(R_s) = \text{sgn}(\lambda_2)$. We should point out here that if we define the following ratio $a = \lambda_2 /\lambda_1$ of the eigenvalues of $S_{ij}$, then each value of $a$ corresponds to a line in the ($R_s,Q_s$) plane with the following expression
\begin{equation}
\label{eq:schematic}
 R_s = (-Q_s)^{3/2}a(1+a)(1+a+a^2)^{-3/2}
\end{equation}
where each line is associated with a flow topology (see caption of Fig. \ref{fig:QsRs_map}) \cite{cantwell02,blackburnetal96}.

The ($R_s,Q_s$) invariant map in many hydrodynamic turbulent flows away from boundaries manifests a tendency for the $R_s > 0$ region, implying a predominance of sheetlike structures related to the strain rate (see for example Fig. 8c in \cite{ooietal99}). In particular, numerical and experimental evidences in homogeneous hydrodynamic turbulence propose the ratios of the mean eigenvalues of $S_{ij}$ to be $\avg{\lambda_1}$:$\avg{\lambda_2}$:$\avg{\lambda_3} =$ 3:1:-4 \cite{ashurstetal87,tsinoberetal92} (see the corresponding line in Fig. \ref{fig:QsRs_map}).

The joint PDFs of $R_s$ versus $Q_s$ for the four runs of Table \ref{tbl:dnsparam} are illustrated in Fig. \ref{fig:QsRs}.
Their dependence on initial conditions is clearly depicted. The shape of the ($R_s,Q_s$) map for run R (Fig. \ref{fig:QsRs}a) moves away from the $D_s = 0$ line towards the $R_s = 0$ axis expressing a more quasi two-dimensional (2D) character of the structures related to $S_{ij}$ than in hydrodynamic turbulent flows away from the boundaries. We should point out, however, that this particular shape is reminiscent to the joint PDFs of ($R_s,Q_s$) found in the buffer layer, i.e. a region very close to the wall, of wall-bounded turbulent shear flows (see for example Fig. 6f in \cite{blackburnetal96}). The joint PDF of run I (Fig. \ref{fig:QsRs}b) is aligned along the $R_s = 0$ with some highly dissipative small scales in contrast to the rest of the runs. The local topology in this case seems to have a strong tendency towards quasi two-dimensionality. Part of the shape of this joint PDF can be explained through two-dimensional shearing (or vortex sheet), i.e.
\begin{equation}
 \label{eq:approxA}
 A_{ij} = 
  \begin{pmatrix}
   0 & \pd_yu_x & 0 \\
   0 & 0        & 0 \\
   0 & \pd_yu_z & 0
  \end{pmatrix}
\end{equation}
which gives $Q_s = -\tfrac{1}{4}[(\pd_yu_x)^2+(\pd_yu_z)^2]$ and $R_s = 0$ in analogy to the influence of the wall on the velocity gradient in wall-bounded flows \cite{blackburnetal96}. The ($R_s,Q_s$) invariant map of run C in Fig. \ref{fig:QsRs}c also falls away from the $D_s = 0$ branch with low correlations between $R_s$ and $Q_s$. Finally, the joint PDF of run A (Fig. \ref{fig:QsRs}d) is almost identical in shape but less correlated with respect to Fig. \ref{fig:QsRs}a.

\begin{figure}[!ht]
  \includegraphics[width=0.5\textwidth]{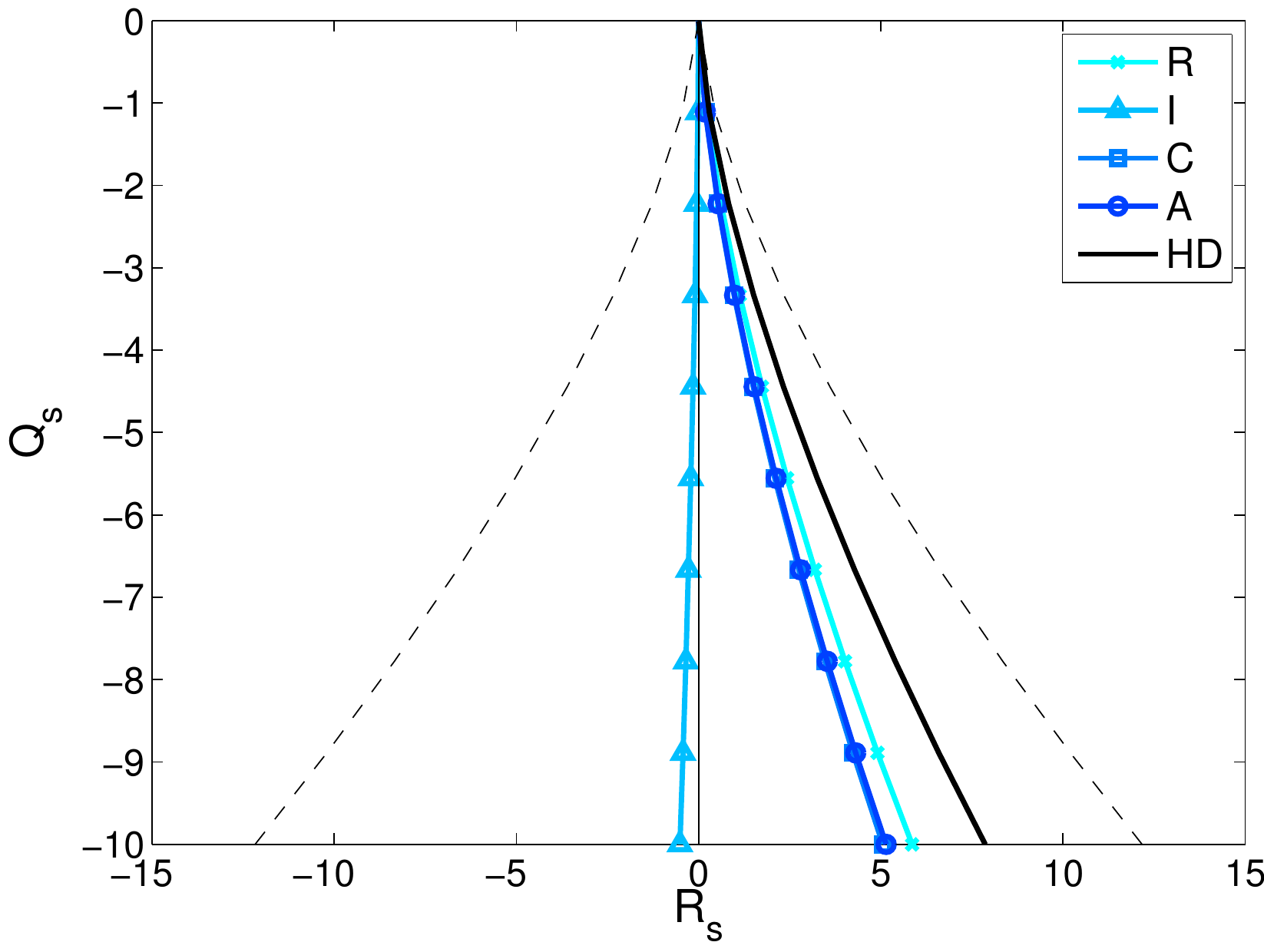}
  \caption{(Color online) Plots of Eq. \eqref{eq:schematic} using the mean eigenvalues of $S_{ij}$ from Table \ref{tbl:Seig}. The dashed line $D_s = \tfrac{27}{4} R_s ^2 + Q_s^3 = 0$ is plotted for reference. With HD we label the curve that corresponds to 3:1:-4, the characteristic eigenvalue ratios for homogeneous hydrodynamic turbulence.}
  \label{fig:Seig}
 \end{figure}

 
We now try to summarise and clarify our arguments by tabulating the mean eigenvalues of the strain rate tensor and their ratios for all our runs in Table \ref{tbl:Seig} but also by plotting the curves that can be constructed from Eq. \eqref{eq:schematic} using the mean eigenvalues of Table \ref{tbl:Seig} (see Fig. \ref{fig:Seig}). 
\begin{table}[!ht]
 \caption{Mean eigenvalues of the strain rate tensor $S_{ij}$ and their ratios.}
 \label{tbl:Seig}
 \begin{ruledtabular}
    \begin{tabular}{*{6}{c}} 
      \textbf{Run} & $\bm{\avg{\lambda_1}}$ & $\bm{\avg{\lambda_2}}$ & $\bm{\avg{\lambda_3}}$ & $\bm{\avg{\lambda_1}:\avg{\lambda_2}:\avg{\lambda_3}}$ \\
     \hline
      R & 0.14 &  0.03 & -0.17 &  5 : 1 : -6 \\
      I & 0.25 & -0.00 & -0.25 &  1 : 0 : -1 \\
      C & 0.25 &  0.04 & -0.29 &  6 : 1 : -7 \\
      A & 0.25 &  0.04 & -0.29 &  6 : 1 : -7 \\
    \end{tabular}
 \end{ruledtabular}
\end{table}
 
In Fig. \ref{fig:Seig}, we plot for reference the curve that corresponds to 3:1:-4, the characteristic eigenvalue ratios for homogeneous hydrodynamic turbulent flows that we denote as ``HD''. In that respect, all the ratios of the mean eigenvalues that we obtain are different than 3:1:-4. However, all the cases represent biaxial expansion apart from run I, which is characterised by quasi two-dimensionality with weak biaxial contraction (see Table \ref{tbl:Seig} and Fig. \ref{fig:Seig}). Figure \ref{fig:Seig} makes clear that on average the flow topologies related to $S_{ij}$ of run C and A are close to run R giving weight to our argument for the similarity of their ($R_s,Q_s$) joint PDFs. The curve for run I also summarises Fig. \ref{fig:QsRs}b by demonstrating that the quasi 2D structures associated to the strain rate tensor are weakly contracted in a average sense.


%
\subsection{Joint PDFs of the second invariants of the strain and rotation rate tensors}
\begin{figure}[!ht]
  \includegraphics[width=0.35\textwidth]{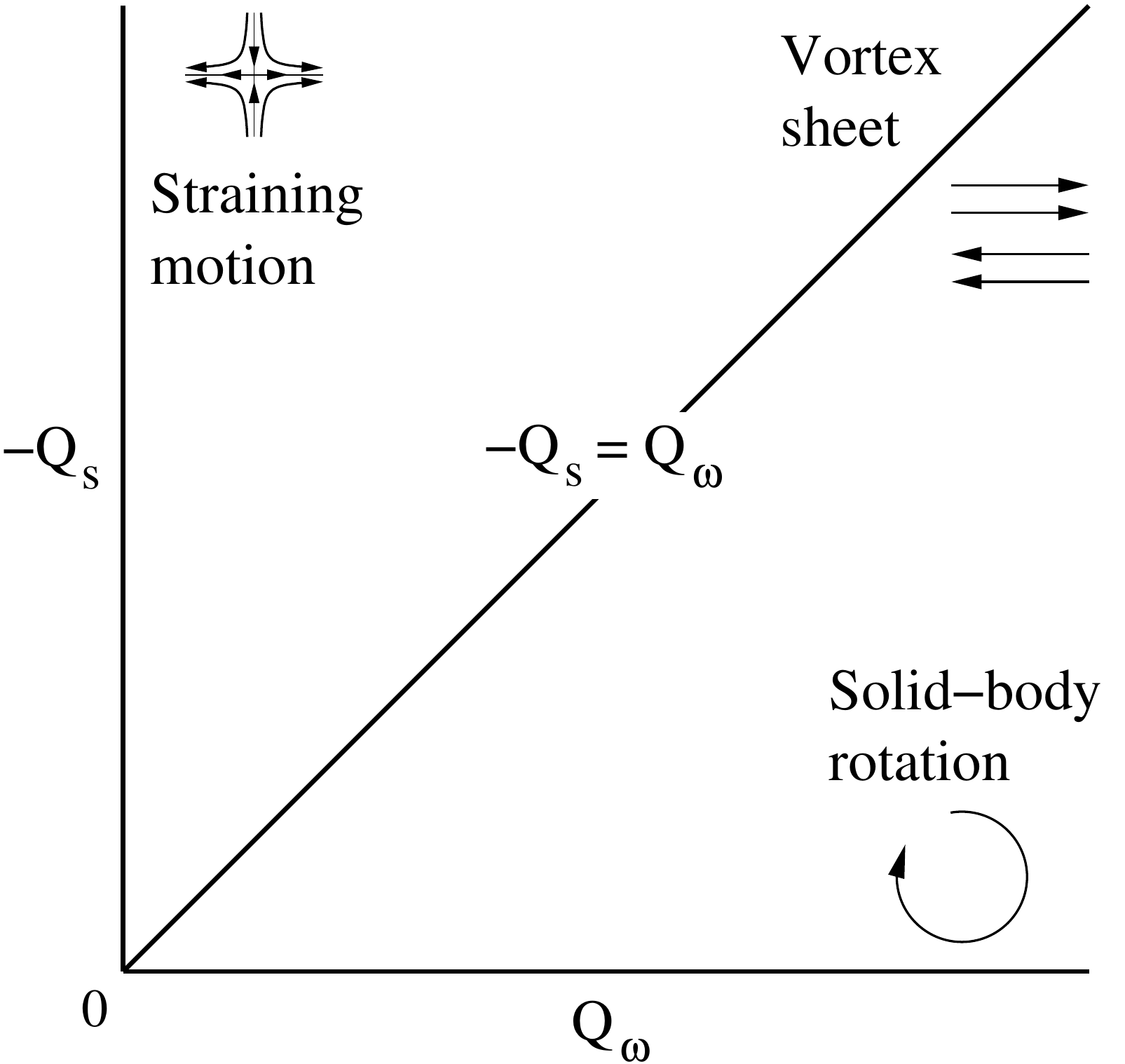}
  \caption{Diagram of the ($Q_\omega,-Q_s$) invariant map pointing out the important regions related to strain and rotation.}
  \label{fig:QwQs_map}
 \end{figure}

Another important joint PDF to analyse is the one of $-Q_s$ versus the second invariant of the rotation rate tensor, $Q_\omega$, which is in fact the only invariant for $\bm \Omega$. To see this, set $\bm S$ to zero in Eqs. \eqref{eq:Qa} and \eqref{eq:Ra}, then
\begin{equation}
 Q_\omega = -\tfrac{1}{2}tr(\bm \Omega^2) = \tfrac{1}{4}\bm \omega^2
\end{equation}
which is positive definite and it is related to the second invariants of $\bm A$ and $\bm S$ through $Q_\omega = Q_A - Q_s$. The ($Q_\omega,-Q_s$) invariant map that is shown schematically in Fig. \ref{fig:QwQs_map} identifies the relative importance of the straining and rotational part of velocity gradient tensor.
A good example that describes simply the physical meanings of Fig. \ref{fig:QwQs_map} is the Burger's vortex tube \cite{saffman95}. As it was mentioned before $Q_s$ characterises the topology associated with viscous dissipation. So, points near the $-Q_s$ axis reflect nearly pure straining motions, i.e. regions of strong dissipation but negligible enstrophy, like outside and away from the Burger's vortex tube. On the other hand, points close to the $Q_\omega$ axis are in nearly pure solid-body rotation, like at the centre of the Burger's vortex tube with high enstrophy but very weak dissipation. Regions with comparable strain rate and rotation map to points close to the $Q_\omega = -Q_s$ line, which correspond to vortex sheets.

Generally, from observations in many hydrodynamic turbulent flows, regions of intense enstrophy tend to be concentrated in tubelike structures, whereas regions of high dissipation are not correlated with regions of concentrated enstrophy \cite{tsinober02}. So, the joint PDF of $Q_\omega$ versus $-Q_s$ is very spread for many hydrodynamic turbulent flows away from walls (see results in \cite{jimenezetal93,blackburnetal96}).

\begin{figure}[!ht]
  \begin{subfigure}{0.35\textwidth}
   \includegraphics[width=\textwidth]{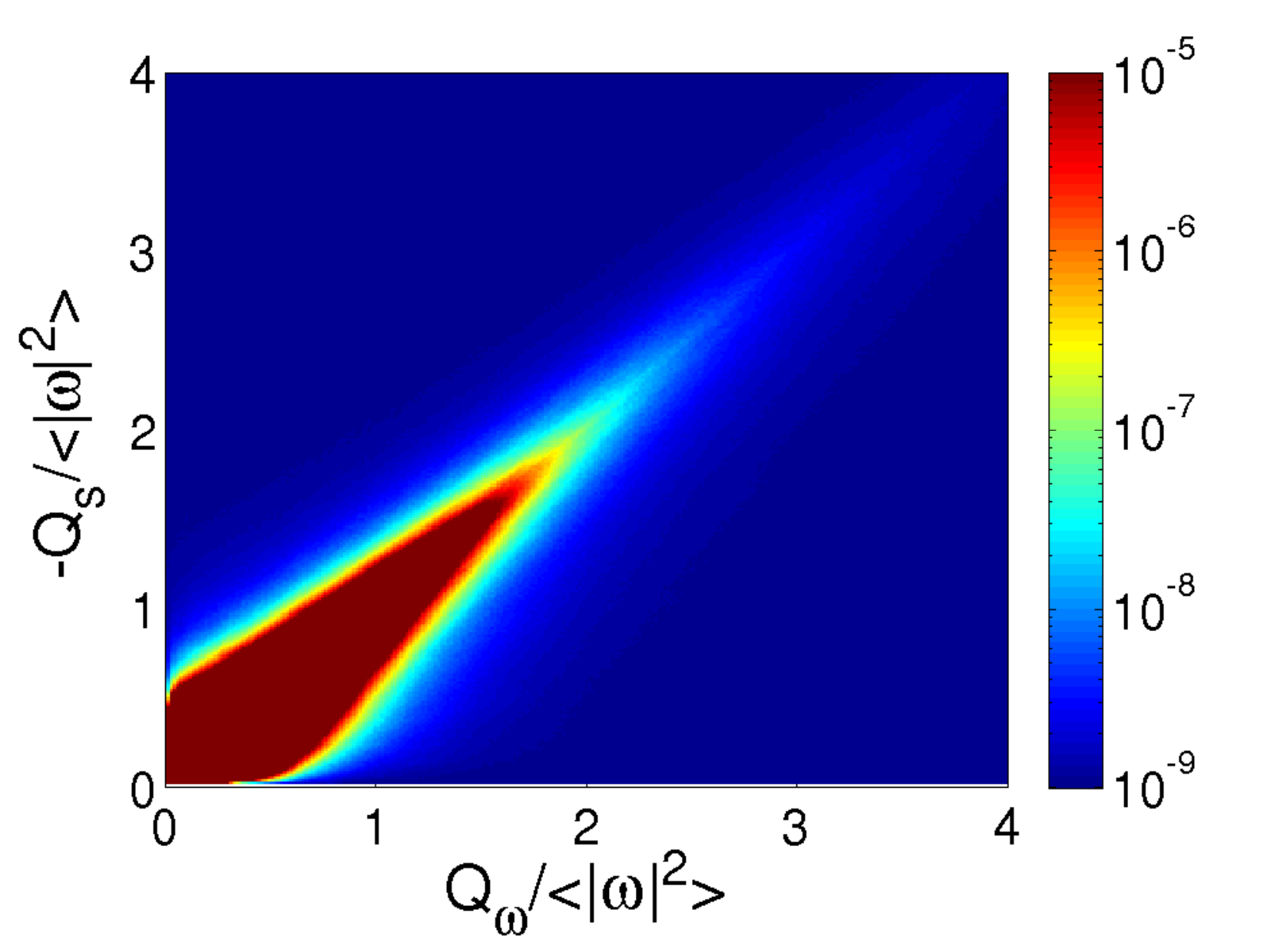}
   \caption{Run R}
  \end{subfigure}
  \begin{subfigure}{0.35\textwidth}
   \includegraphics[width=\textwidth]{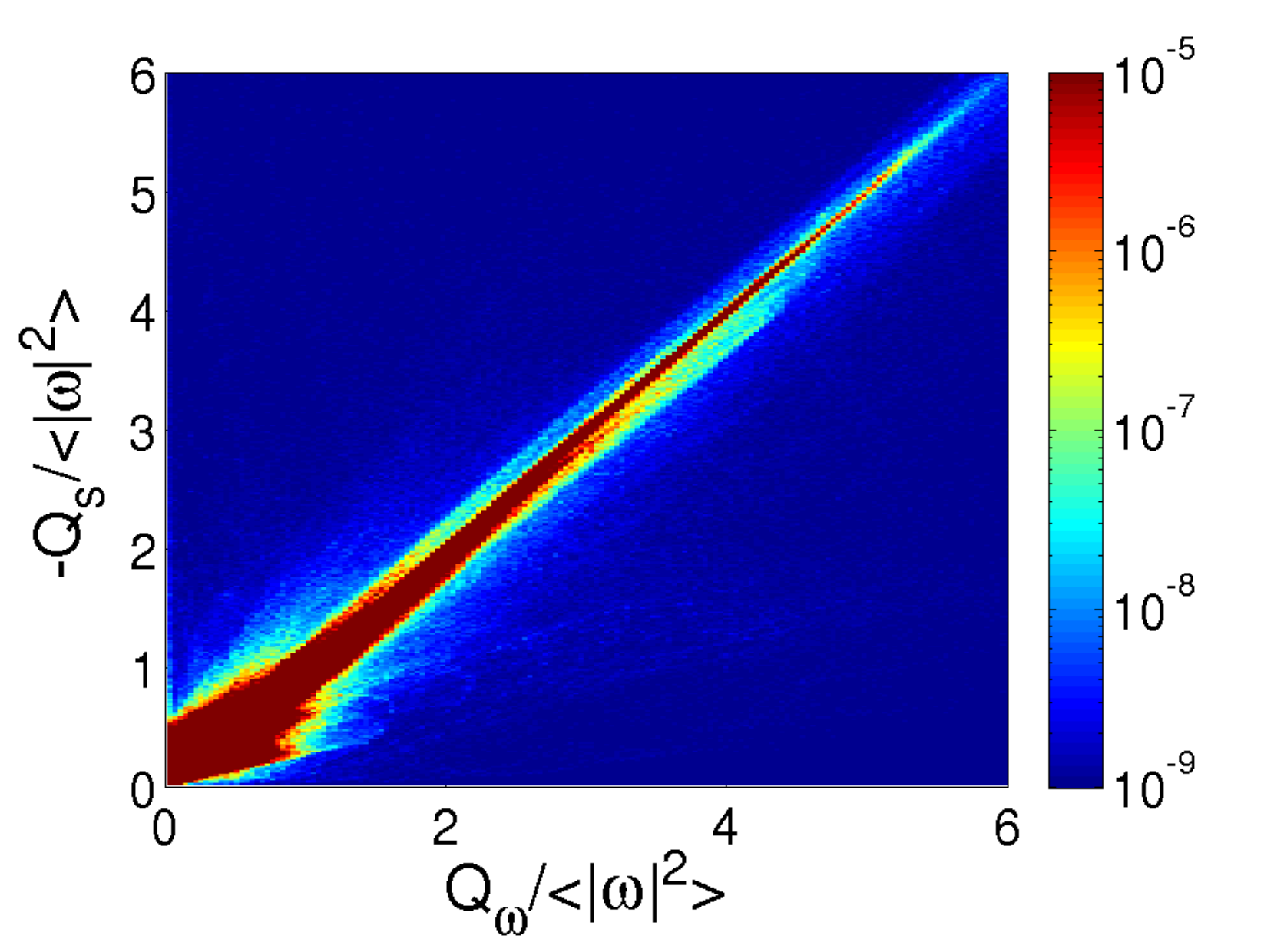}
   \caption{Run I}
  \end{subfigure} \\
  \begin{subfigure}{0.35\textwidth}
   \includegraphics[width=\textwidth]{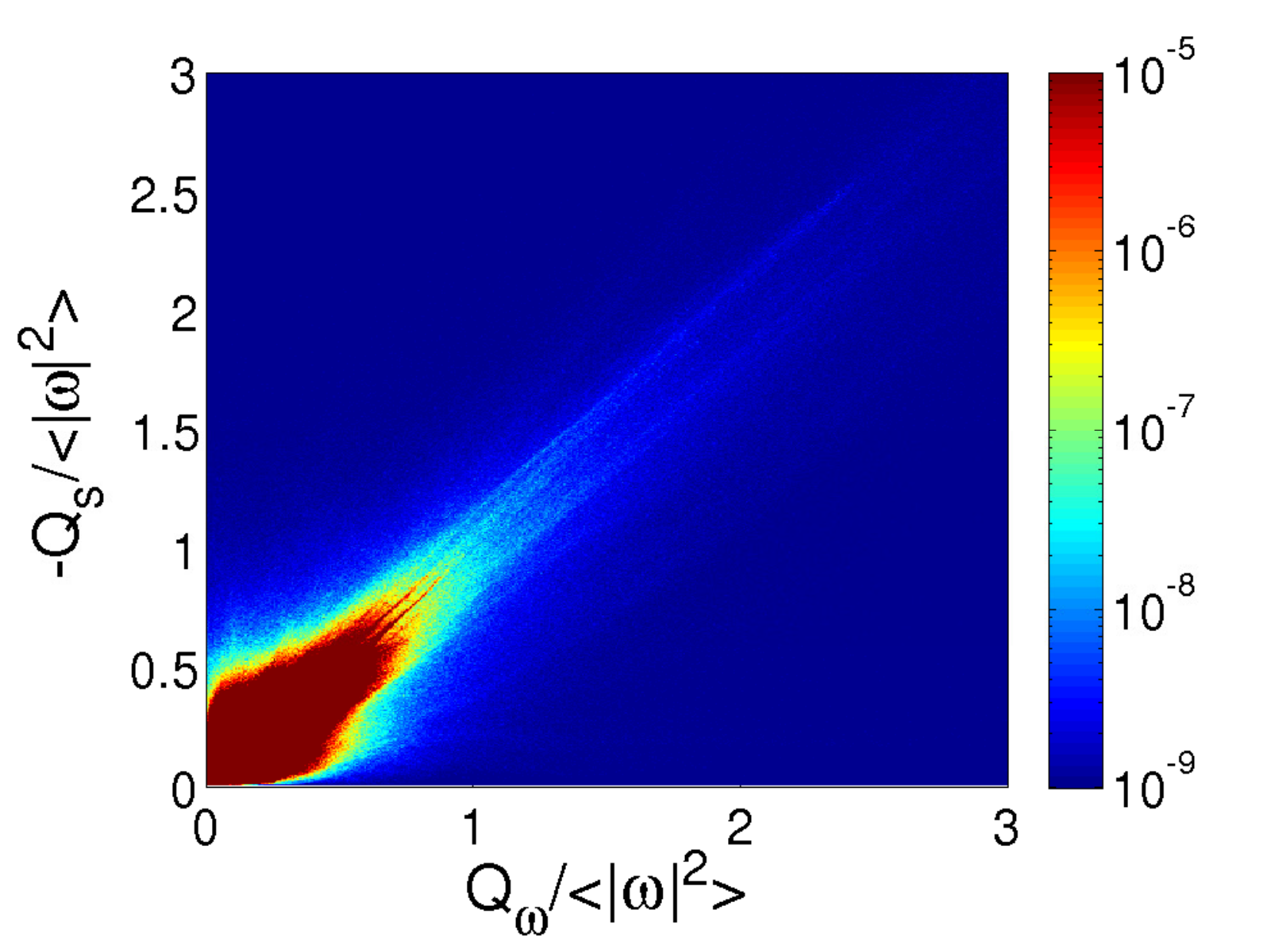}
   \caption{Run C}
  \end{subfigure}  
  \begin{subfigure}{0.35\textwidth}
   \includegraphics[width=\textwidth]{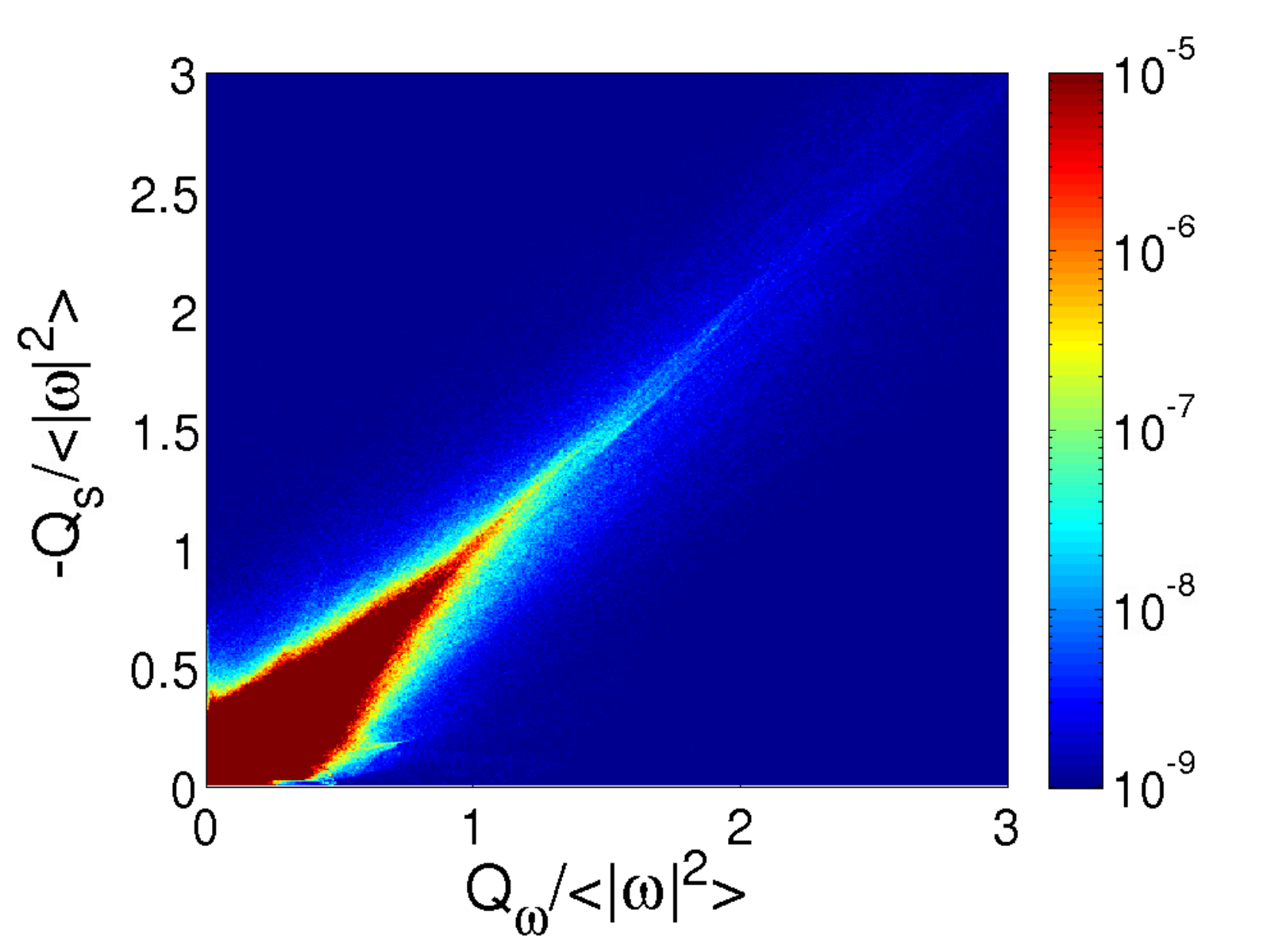}
   \caption{Run A}
  \end{subfigure}
  \caption{(Color online) Joint PDFs of the second invariants of strain rate and rotation rate tensors normalised by the mean enstrophy for a) run R, b) run I, c) run C and  d) run A of Table \ref{tbl:dnsparam}.}
  \label{fig:QsQw}
 \end{figure}

Figure \ref{fig:QsQw} shows the joint PDFs of $Q_\omega$ versus $-Q_s$, normalised with the mean enstrophy, for the four runs of Table \ref{tbl:dnsparam}.
The dependence on initial conditions is pronounced once more in these plots. The ($Q_\omega,-Q_s$) invariant map of run R (Fig. \ref{fig:QsQw}a) is very different to hydrodynamic turbulence away from walls. Here, the joint PDF is concentrated around the $Q_\omega = -Q_s$ line demonstrating stronger correlation between these two variables. This result in conjunction with the outcome from Fig. \ref{fig:Seig} confirms many visualisations of homogeneous MHD turbulent flows \cite{biskamp03}, which illustrate large population of sheetlike rather than tubelike structures. 

The shape of the joint PDF ($Q_\omega,-Q_s$) for run I is even more extreme with a very narrow distribution along the main diagonal (Fig. \ref{fig:QsQw}b), where regions of high dissipation are strongly correlated by high levels of enstrophy particularly for points far from the origin. The high gradients in this flow can be well approximated by Eq. \eqref{eq:approxA} where $Q_\omega = -Q_s = \tfrac{1}{4}[(\pd_yu_x)^2+(\pd_yu_z)^2]$. According to Cantwell \cite{cantwell02} the presence or absence of points very far from the origin, associated with quite long-lived structures, is closely related to the regularity of the initial conditions. He further mentions that such structures are much less prominent in a flow with randomised initial conditions. Here, this is transparent if one compares the run with random initial conditions (Fig. \ref{fig:QsQw}a) with run I (Fig. \ref{fig:QsQw}b). Moreover, it could be argued that the core of the joint PDF ($Q_\omega,-Q_s$) of run I is similar to the joint PDF obtained in 
the buffer layer of wall-bounded flows (see results by \cite{blackburnetal96,chongetal98}). The ($Q_\omega,-Q_s$) map of run A (Fig. \ref{fig:QsQw}d) resembles Fig. \ref{fig:QsQw}a but with weaker correlations between high dissipation and high enstrophy regions. Finally, the joint PDF of Fig. \ref{fig:QsQw}c, which corresponds to run C, presents the weakest correlations between $Q_\omega$ and $-Q_s$ among the four cases with a weak trend of alignment along the main diagonal. 

\subsection{\label{sec:vis} Flow structures and enstrophy dynamics}
\begin{figure}[!ht]
  \begin{subfigure}{0.3\textwidth}
   \includegraphics[width=\textwidth]{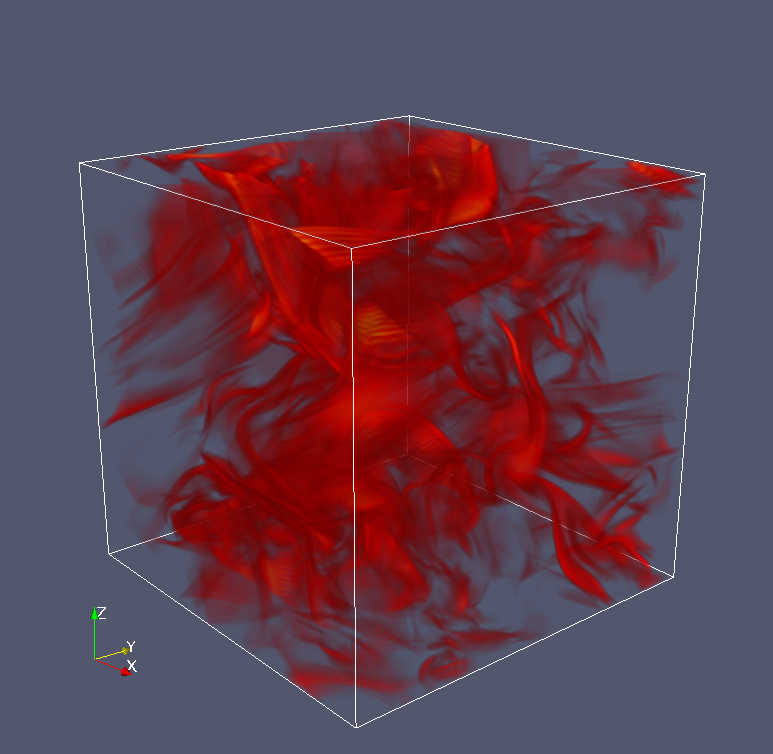}
   \caption{Run R}
  \end{subfigure}
  \begin{subfigure}{0.3\textwidth}
   \includegraphics[width=\textwidth]{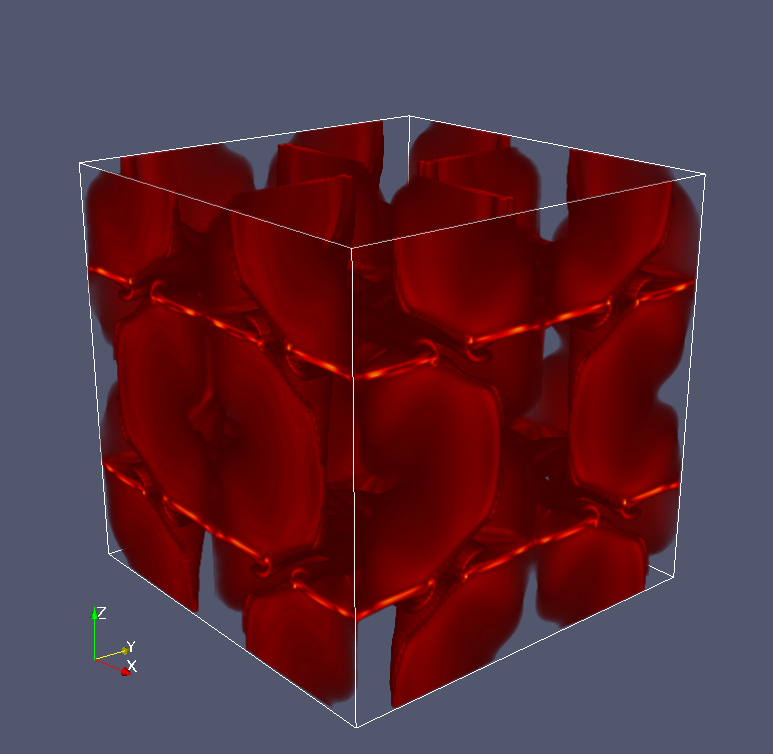}
   \caption{Run I}
  \end{subfigure} \\
  \begin{subfigure}{0.3\textwidth}
   \includegraphics[width=\textwidth]{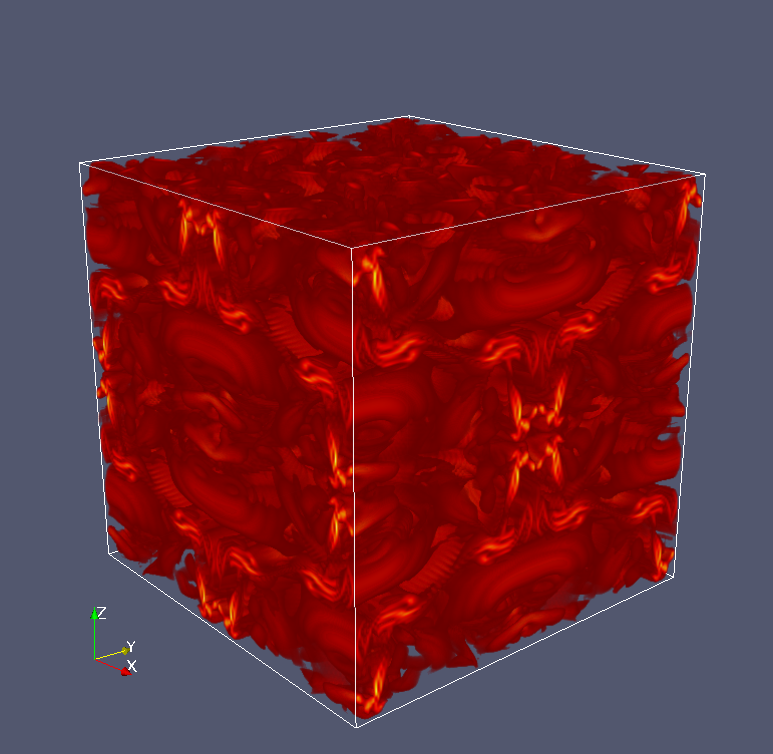}
   \caption{Run C}
  \end{subfigure}
  \begin{subfigure}{0.3\textwidth}
   \includegraphics[width=\textwidth]{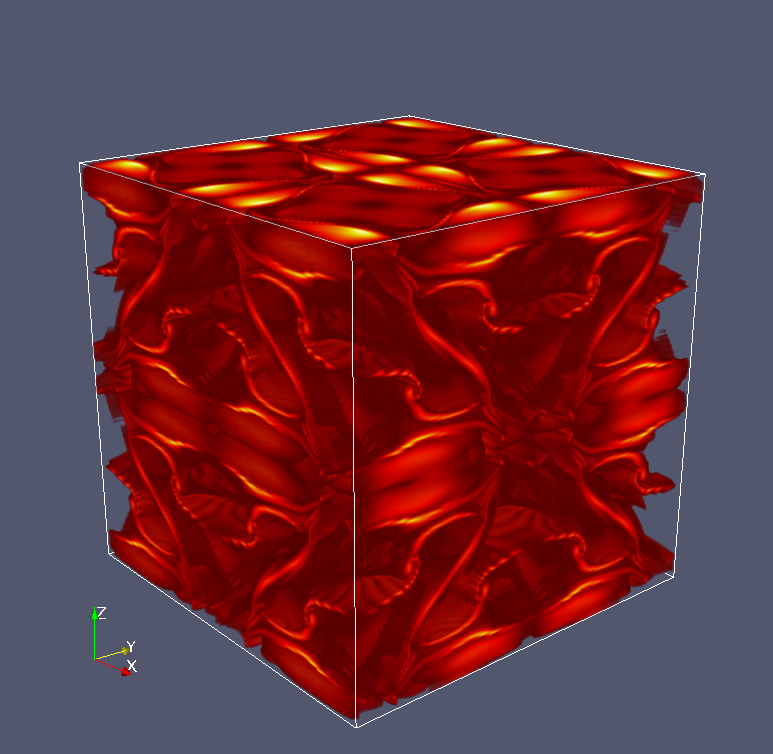}
   \caption{Run A}
  \end{subfigure}
  \caption{(Color online) Vorticity field iso-contours with $|\omega| \geq 3\omega'$ for a) run R, b) run I, c) run C and d) run A of Table \ref{tbl:dnsparam}.}
  \label{fig:vis}
 \end{figure}

Various flow field quantities were viewed interactively using a visualisations software \cite{paraview} to get an idea of the spatial structures in our flows. In order to substantiate our approach, we present indicatively plots of iso-contours of the vorticity field in our $[0,2\pi]^3$ periodic boxes at the moment of maximum dissipation for the four runs of Table \ref{tbl:dnsparam} (see Fig. \ref{fig:vis}).
Figure \ref{fig:vis}a (run R) displays iso-contours of vorticity for $|\omega| \geq 3\omega'$ where $\omega' \equiv (|\bm \omega|^2)^{1/2}$. The predominant structures in this plot are randomly oriented sheetlike structures in support of our joint PDF analysis. In comparison to the randomly oriented structures of run R, the TG vortex symmetries become apparent in Figs. \ref{fig:vis}b, c and d revealing their preservation in time. Remember that we did not impose any symmetries during the evolution of our runs. According to the above analysis, the peculiar run I should be prevailed by quasi two-dimensional sheetlike structures, which are shown in Fig. \ref{fig:vis}b. These flat structures are formed on the insulating faces of the $[0,\pi]^3$ boxes and on their mid-planes in the vertical direction, i.e. $z = \pi/2$. The structures of run A (Fig. \ref{fig:vis}d) are also sheetlike but more randomly oriented in contrast to run I. In the end, run C is a more complicated TG flow.
This is demonstrated in Fig. \ref{fig:vis}c for $|\omega| \geq 3\omega'$. 
It is interesting that the initial conditions of the TG velocity with the TG magnetic fields for the insulating runs I and A create less randomness in the flow fields, which are mainly dominated by quasi 2D sheetlike structures in contrast to run C.

According to Jim\'enez et al. \cite{jimenezetal93}, in hydrodynamic turbulent flows away from walls, it is qualitative clear that there is no other way of production of enstrophy other than straining of weak vorticity to form stronger vortex regions. Then, strain itself is induced by vorticity and the process may become non-linear. This mechanism is called self-amplification of velocity derivatives \cite{tsinober02,sagautcambon08}. 

In order to have an initial picture of this mechanism and in particular of the formation of the vorticity fields in our MHD flows, we examine the rate of vortex stretching
\begin{equation}
 \label{eq:sigma}
 \Sigma = \frac{\bm \omega \sdot \bm S \sdot \bm \omega}{|\bm \omega|^2} = \frac{R_s - R_A}{Q_\omega}
\end{equation}
which is essentially the part of the strain that is aligned with the local vorticity and it is the term that stretches or compresses the vortex lines in the evolution equation of the enstrophy
\begin{equation}
 \label{eq:enstrophy}
 \dd_t (\tfrac{1}{2} \bm \omega^2) = \bm \omega \sdot \bm S \sdot \bm \omega 
                                   + \nu \bm \omega \sdot \bm \Delta \bm \omega 
                                   + \bm \omega \sdot \grad \times (\bm j \times \bm b).
\end{equation}
Notice that $\Sigma$ can be written as a function of the invariants $R_A$, $R_s$ and $Q_\omega$ (see Eq. \eqref{eq:sigma}). 

\begin{figure}[!ht]
 \begin{subfigure}{0.35\textwidth}
  \includegraphics[width=\textwidth]{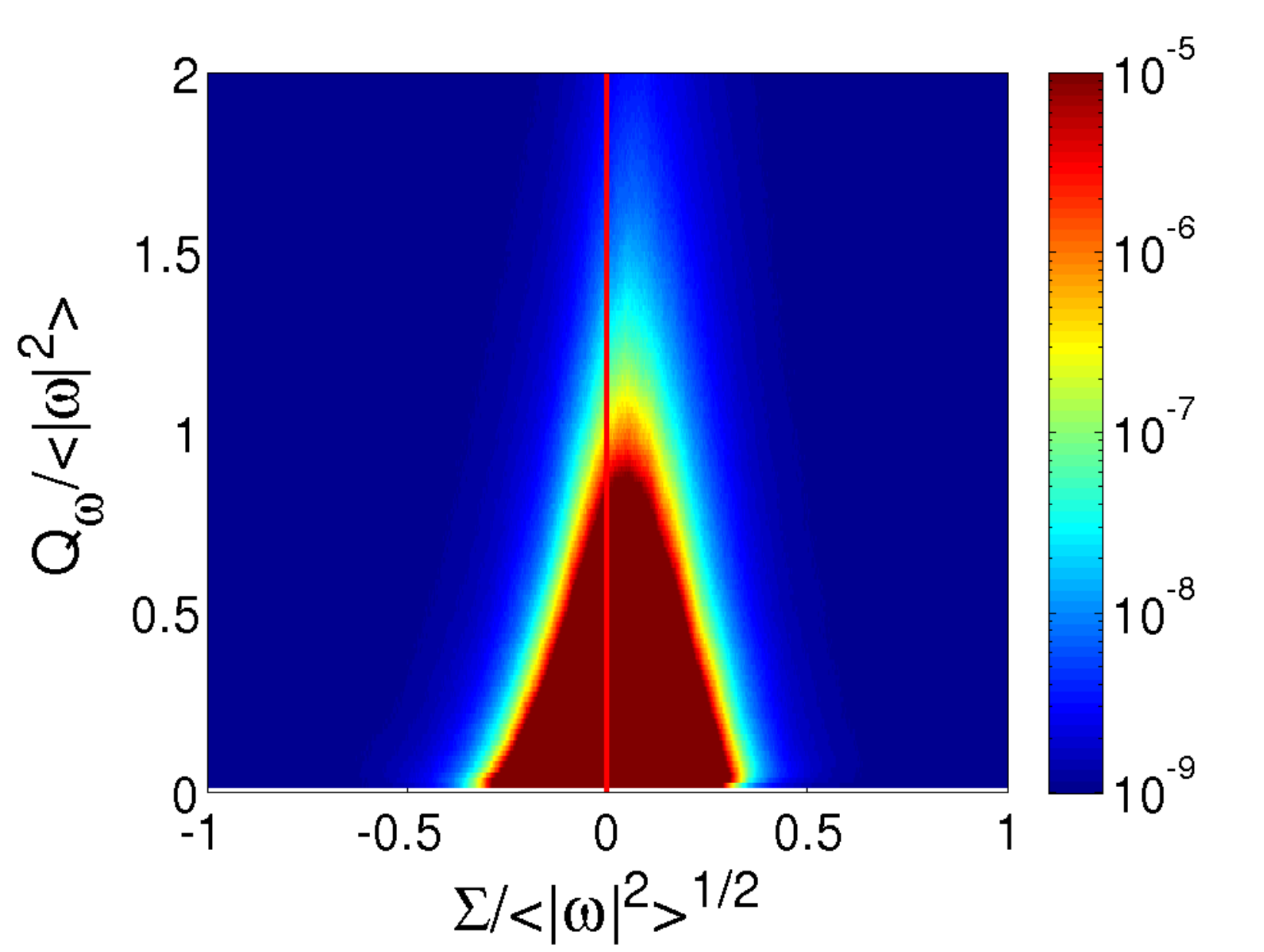}
  \caption{Run R}
 \end{subfigure}
 \begin{subfigure}{0.35\textwidth}
  \includegraphics[width=\textwidth]{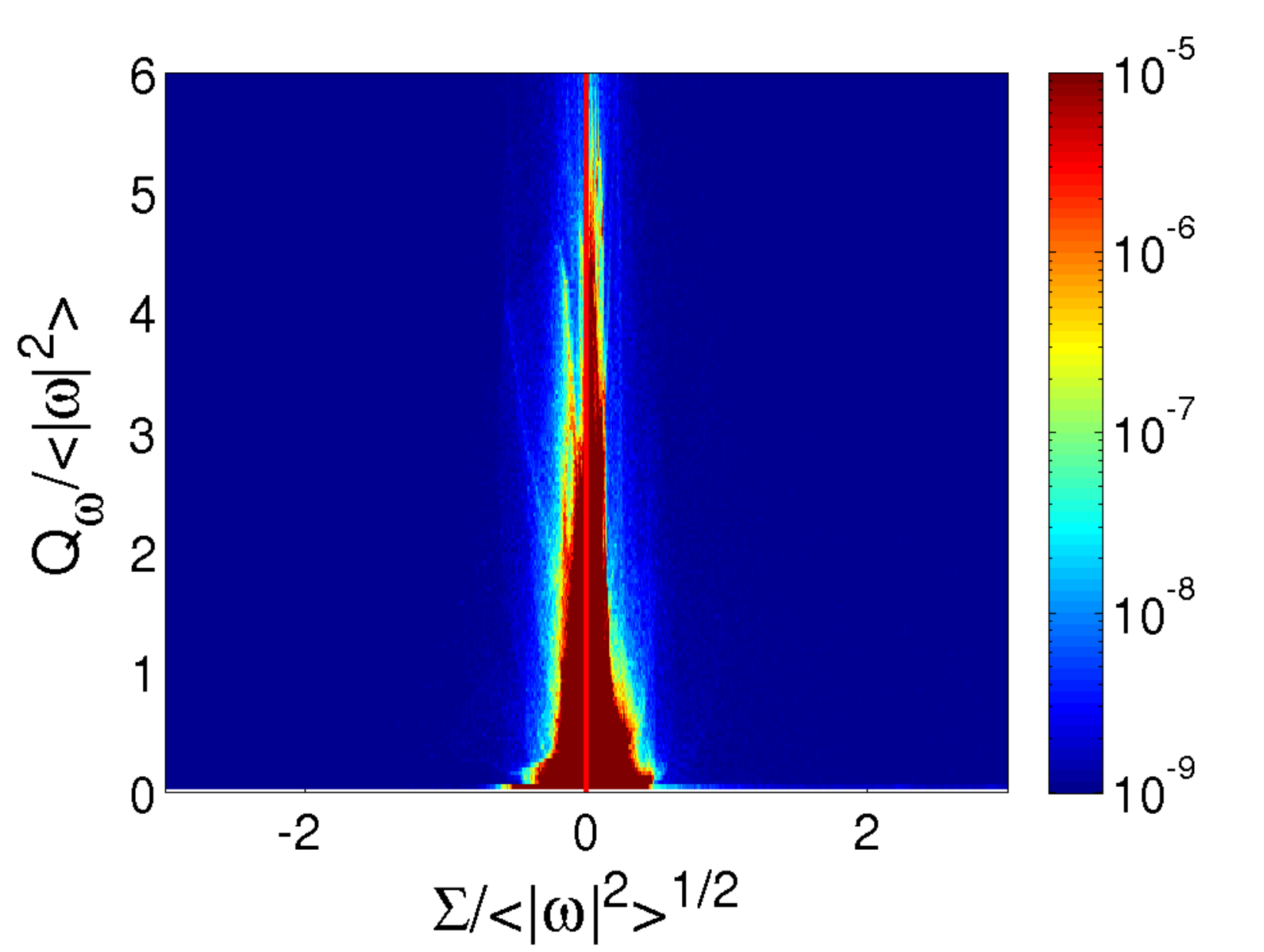}
  \caption{Run I}
 \end{subfigure} \\
 \begin{subfigure}{0.35\textwidth}
  \includegraphics[width=\textwidth]{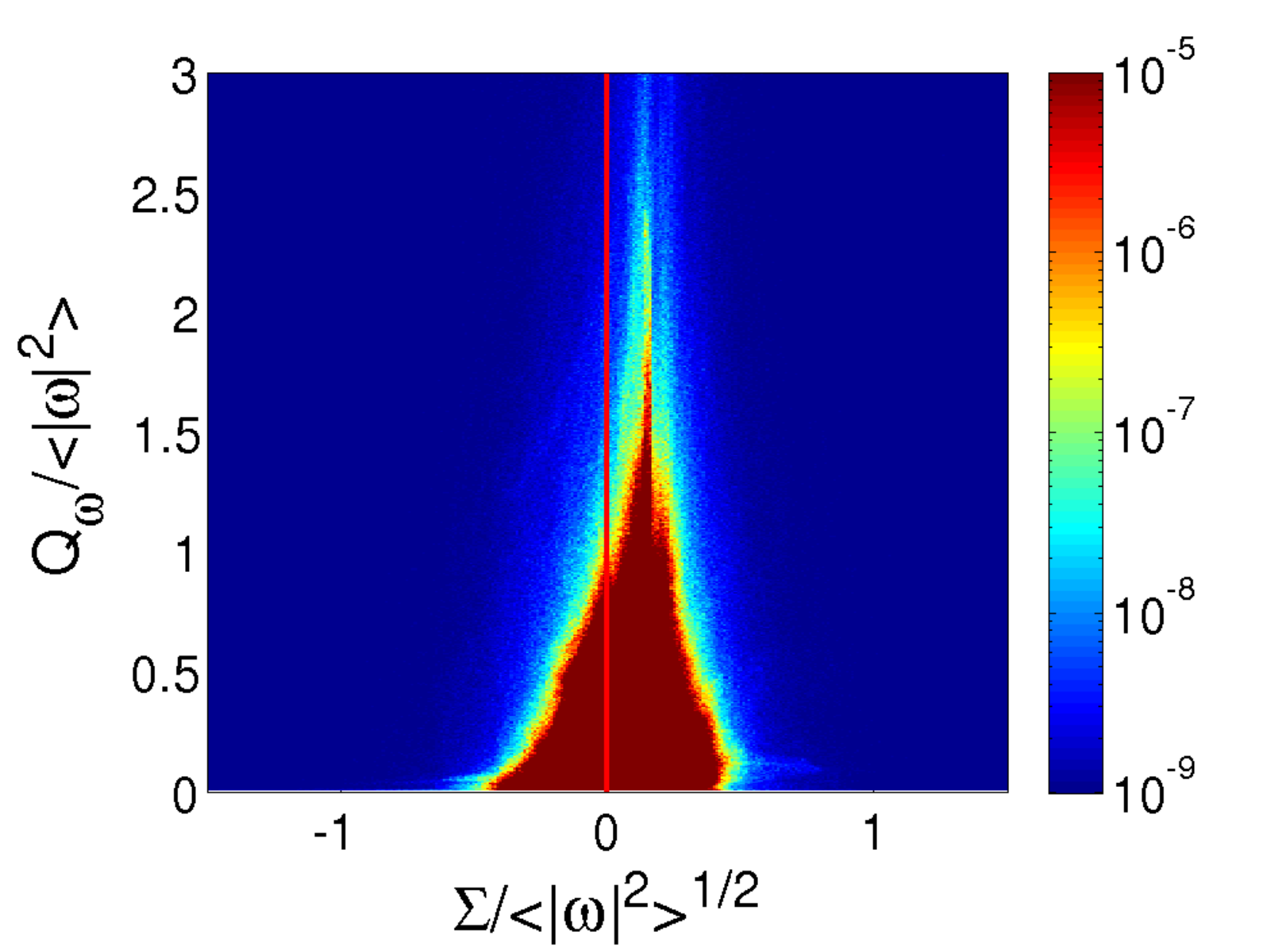}
  \caption{Run C}
 \end{subfigure}
 \begin{subfigure}{0.35\textwidth}
  \includegraphics[width=\textwidth]{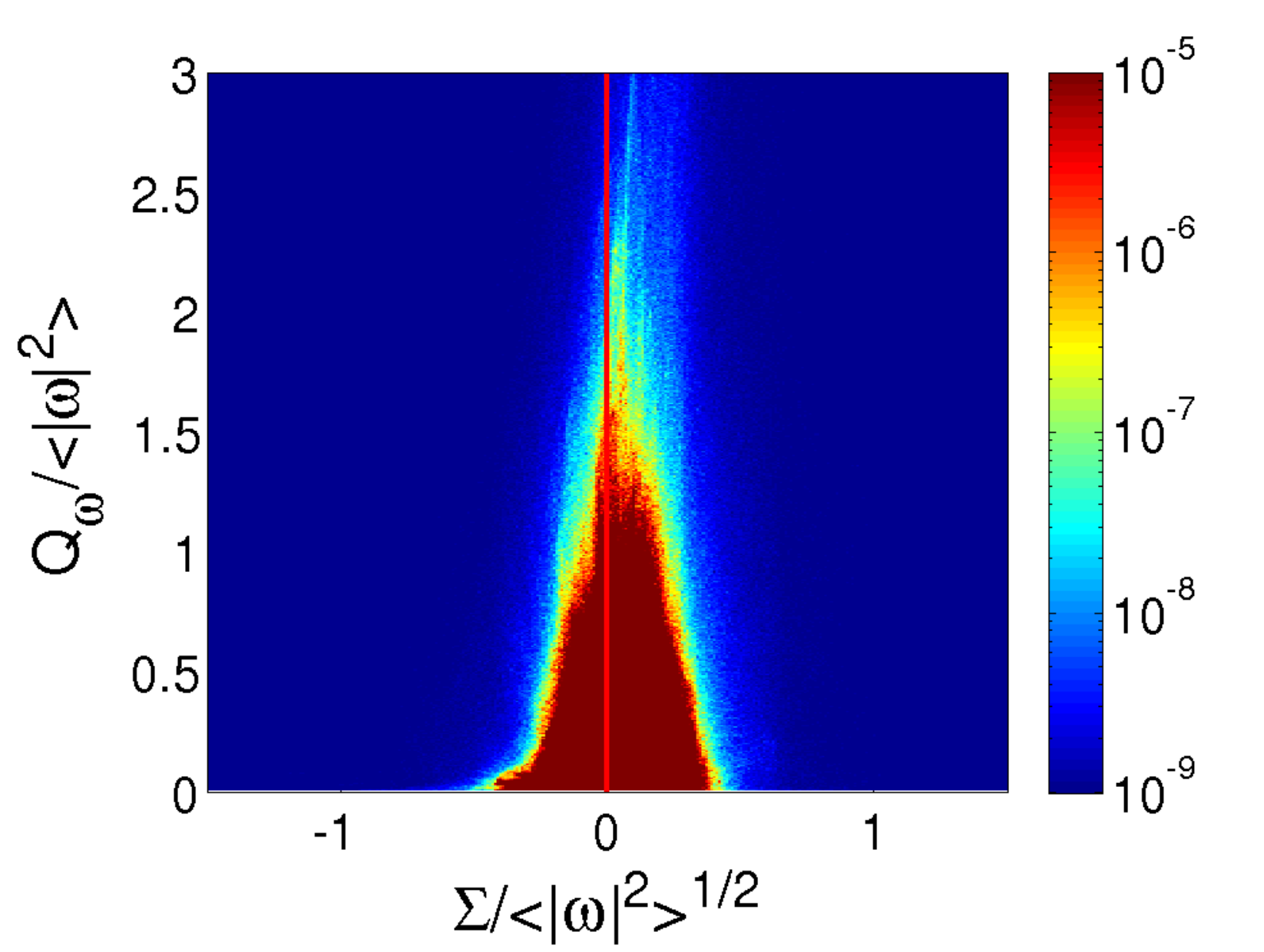}
  \caption{Run A}
 \end{subfigure}
  \caption{(Color online) Joint PDFs of the second invariant of the rotation rate tensor $Q_\omega$ and the vortex stretching rate $\Sigma$ normalised appropriately by powers of the mean enstrophy for a) run R, b) run I, c) run C and  d) run A of Table \ref{tbl:dnsparam}.}
  \label{fig:QwSigma}
 \end{figure}

Figure \ref{fig:QwSigma} shows joint PDFs of essentially the enstrophy (i.e. $Q_\omega$) with the rate of vortex stretching $\Sigma$ appropriately normalised for all the flows of Table \ref{tbl:dnsparam}.
Various common features can be observed in Fig. \ref{fig:QwSigma}. To be more specific, the highest values of enstrophy are associated with positive but low values of $\Sigma$, i.e. stretching of vorticity, whereas high rates of stretching as well as compression correlate with regions of low $Q_\omega$. So, there is little evidence of self-stretching by structures in the flow which have large enstrophy in analogy to hydrodynamic turbulence \cite{jimenezetal93,ooietal99}. Another common feature in all the plots of Fig. \ref{fig:QwSigma} is the tilt towards positive values, i.e. vorticity vectors are being more stretched than compressed. 


On the other hand, quantitative differences are evident, such as
the asymmetry of the ($\Sigma,Q_\omega$) joint PDFs, which seems to be different for each flow. In other words, the joint PDF of run C (Fig. \ref{fig:QwSigma}c) is shifted more towards $\Sigma > 0$ values, akin to hydrodynamic turbulence (see results in \cite{jimenezetal93,ooietal99}), in comparison to run A (Fig. \ref{fig:QwSigma}d) which is closer to the joint PDF of run R (Fig. \ref{fig:QwSigma}a). Another quantitative difference between the four flows is the very high values of enstrophy ($Q_\omega \simeq 6\avg{|\bm \omega|^2}$) that are obtained in run I (Fig. \ref{fig:QwSigma}b) for values of vortex stretching rate of the same order for all the flows (i.e. $\Sigma < 0.3\avg{|\bm \omega|^2}^{1/2}$).

\begin{figure}[!ht]
  \begin{subfigure}{0.35\textwidth}
   \includegraphics[width=\textwidth]{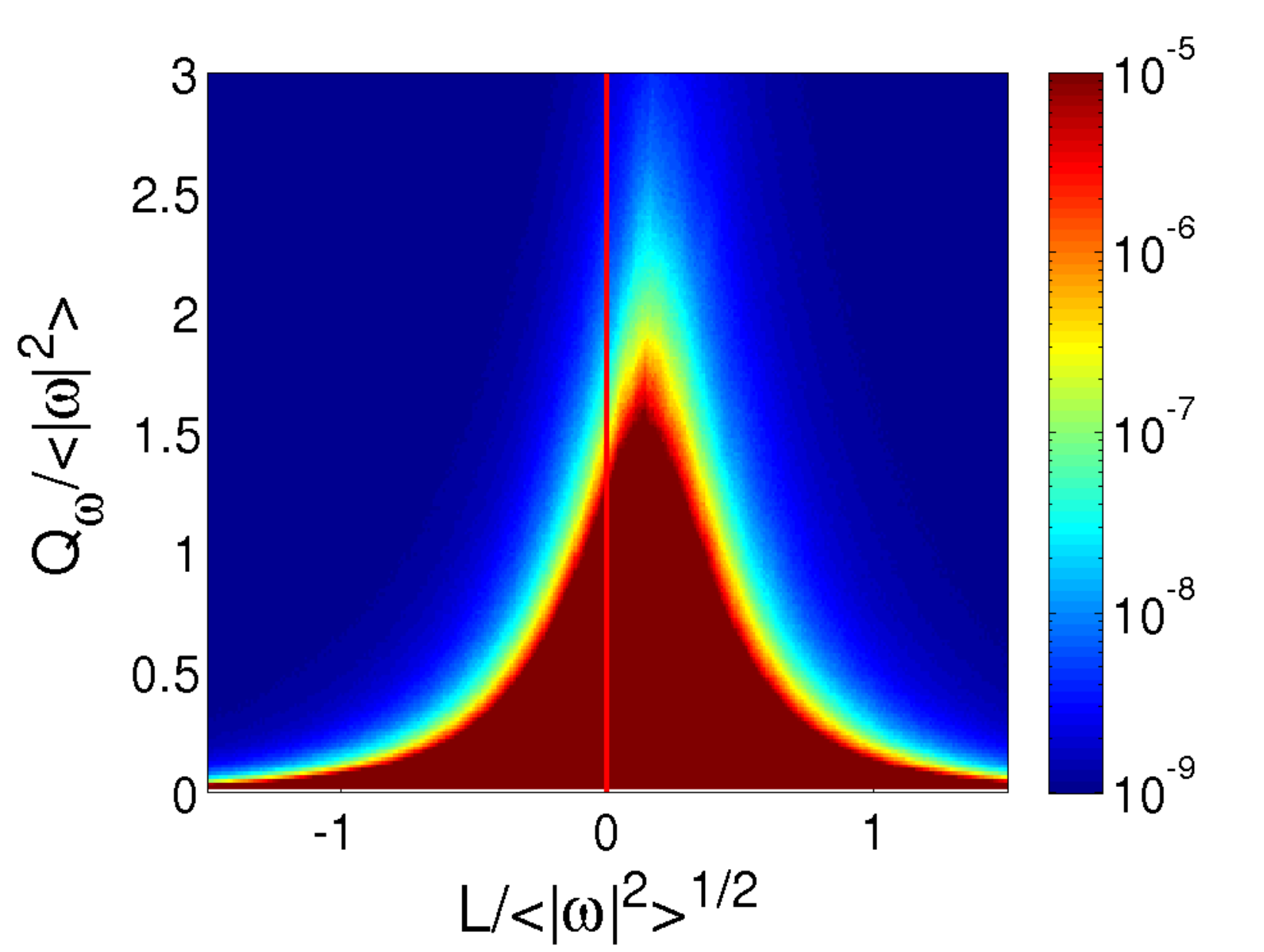}
   \caption{Run R}
  \end{subfigure}
  \begin{subfigure}{0.35\textwidth}
   \includegraphics[width=\textwidth]{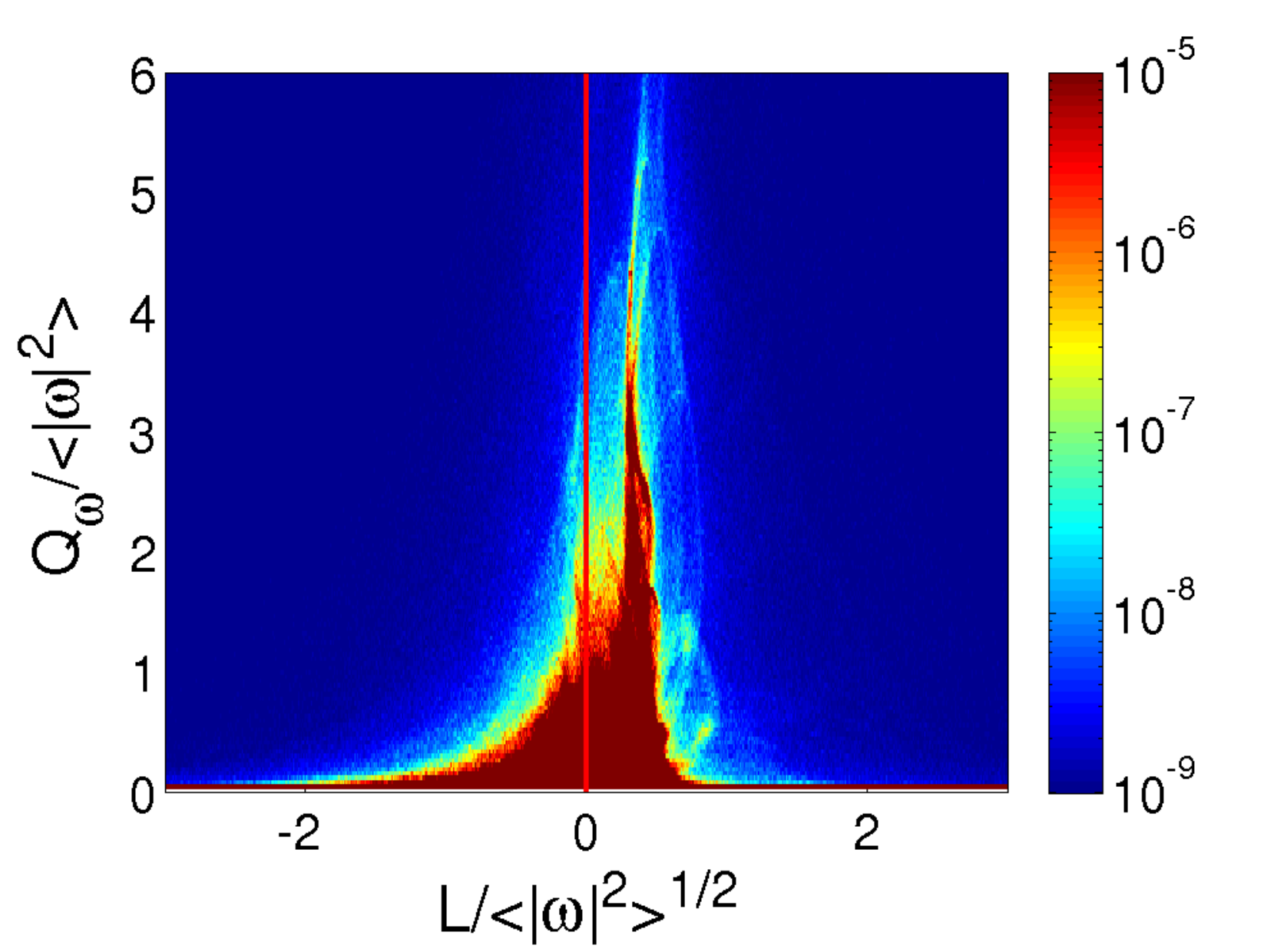}
   \caption{Run I}
  \end{subfigure} \\
  \begin{subfigure}{0.35\textwidth}
   \includegraphics[width=\textwidth]{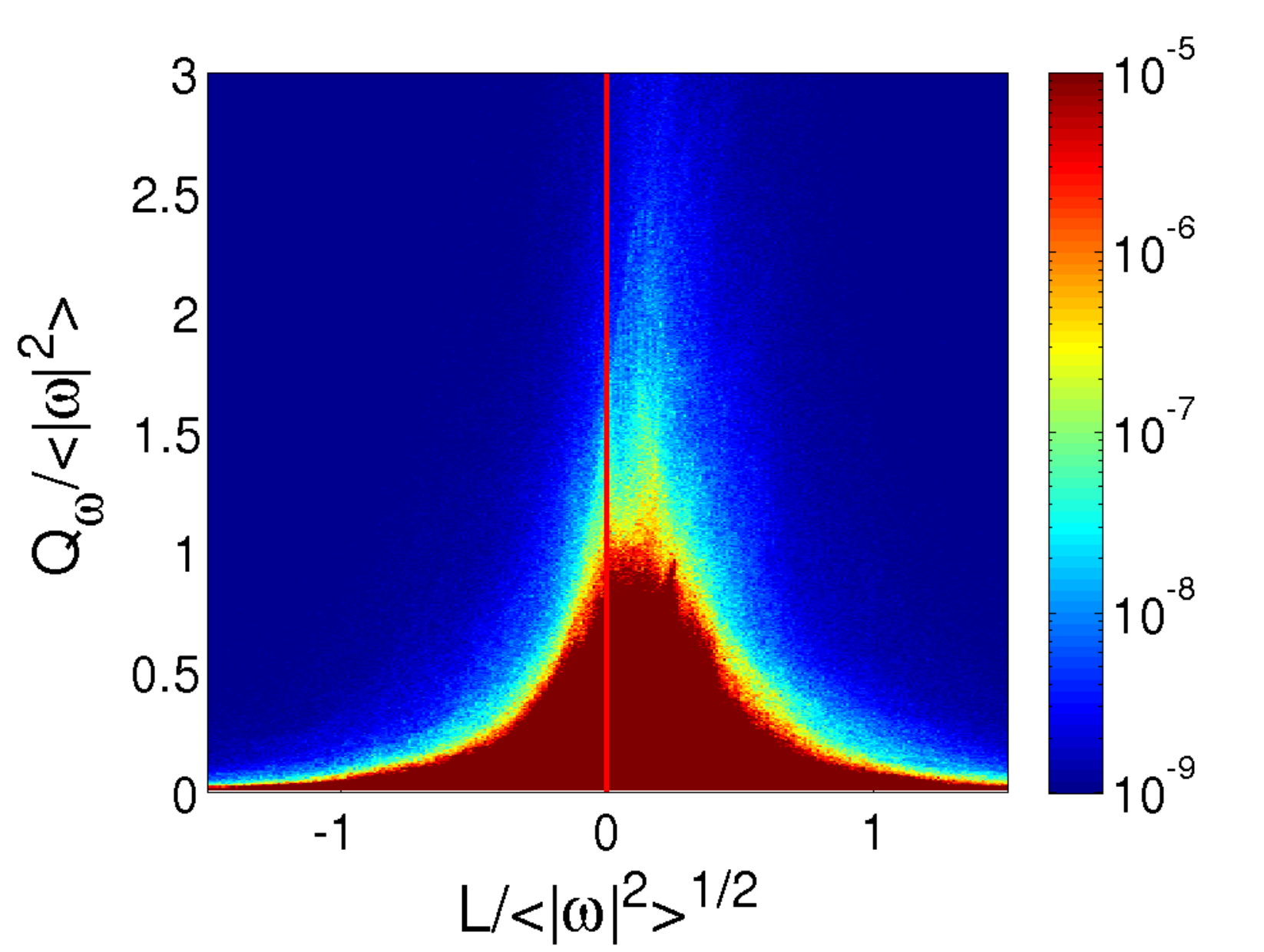}
   \caption{Run C}
  \end{subfigure}
  \begin{subfigure}{0.35\textwidth}
   \includegraphics[width=\textwidth]{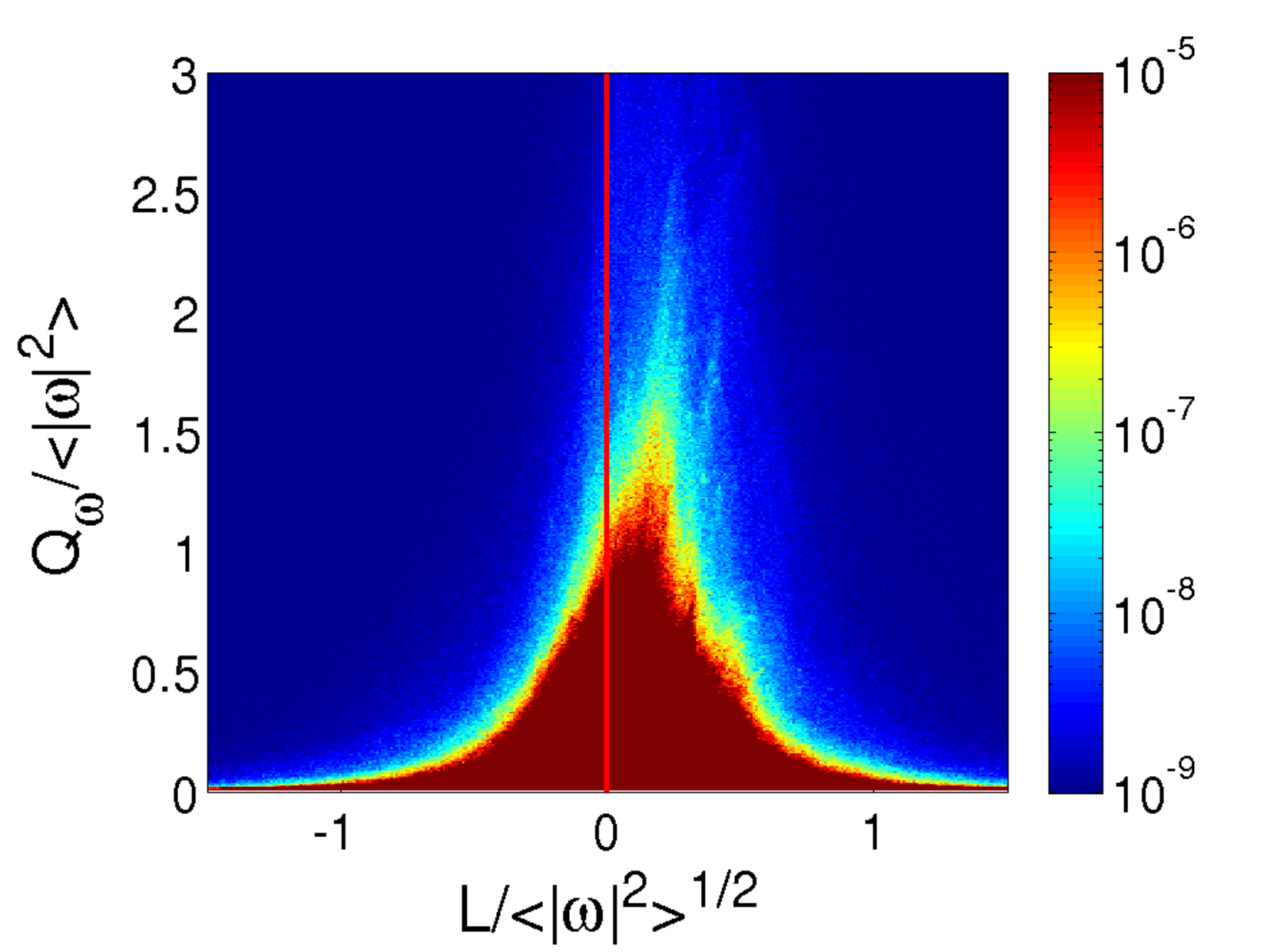}
   \caption{Run A}
  \end{subfigure}
 \caption{(Color online) Joint PDFs of the second invariant of the rotation rate tensor $Q_\omega$ and $L$ normalised appropriately by powers of the mean enstrophy for a) run R, b) run I, c) run C and d) run A of Table \ref{tbl:dnsparam}.}
 \label{fig:QwL}
\end{figure}

Another important mechanism for amplification or reduction of enstrophy that exists only in MHD turbulence is that due to the Lorentz force. This process essentially manifests from the last term of Eq. \eqref{eq:enstrophy}, which we write here as
\begin{equation}
 L = \frac{\bm \omega \sdot \grad \times (\bm j \times \bm b)}{|\bm \omega|^2},
\end{equation}
so that it is comparable with $\Sigma$ (see Eq. \eqref{eq:sigma}). In order to shed light on the dynamics of this term with respect to the enstrophy, we consider in Fig. \ref{fig:QwL} the joint PDFs between $L$ and $Q_\omega$, normalised appropriately, for the four runs of Table \ref{tbl:dnsparam}.

It is characteristic for all the plots of Fig. \ref{fig:QwL} that there is a preference for $L > 0$ for most of the local topology in the flow. It is also common in all the four cases that the highest values of $Q_\omega$ are associated with regions of low but positive $L$, whereas high values of $|L|$ are related to regions of low enstrophy in a similar fashion to the self-amplification mechanism. Once more, the quantitative differences between the plots of Fig. \ref{fig:QwL} are evident with the most notable being the joint PDF of run I (Fig. \ref{fig:QwL}b) with the highest values of $Q_\omega$ in terms of $L$.

A comparison between the two mechanisms of amplification and reduction of enstrophy reveals the cause of high and low enstrophy regions. For the MHD flow with random initial conditions, $L$ is more correlated with regions of higher enstrophy than $\Sigma$ but the opposite is true for the TG flows. On the other hand, the lowest enstrophy regions present correlations with higher absolute values of $L$ than $\Sigma$ for all the runs.

\section{\label{sec:invb} Invariants of the magnetic field gradient, the magnetic strain rate and the current density rate tensors}
In this section we try to classify the topology related to the magnetic field by extending the above joint PDF analysis for the invariants of magnetic field gradient tensor as well as for the invariants of its symmetric and skew-symmetric components.
\subsection{Joint PDFs of the magnetic field gradient invariants}
The magnetic field gradient $\bm X = \grad \bm b$ can be also decomposed into its symmetric and skew-symmetric component
\begin{equation}
 \bm X = \bm K + \bm J = K_{\alpha\beta} - \tfrac{1}{2}\epsilon_{\alpha\beta\gamma}j_\gamma
\end{equation}
where $\bm K = \tfrac{1}{2}(\grad \bm b + \grad \bm b^T)$ and $\bm J = \tfrac{1}{2}(\grad \bm b - \grad \bm b^T)$ are the magnetic strain rate and current density rate tensors, respectively. The skew-symmetric part of $\bm X$ is related to the electric current through Ampere's law $\grad \times \bm b = \mu_0 \bm j$ where $\mu_0 = (\kappa\sigma)^{-1}$ is the permeability of free-space and $\sigma$ is the electrical conductivity. When the magnetic field lines are bended, current is produced providing a Lorentz force that inhibits the bending of the field lines. On the other hand, the symmetric part of $\bm X$ characterises the force-free regions in the magnetic field, where $\bm j = 0$ and therefore $\bm j \times \bm b = 0$. An important relation one can easily derive by taking the divergence of Eq. \eqref{eq:ns} and using the fact that our fields $\bm u$ and $\bm b$ are solenoidal is the following Poisson equation
 \begin{align}
  \grad^2 P &= \grad \sdot [(\bm u \times \bm \omega) + (\bm j \times \bm b)] \nonumber \\
            &= (\Omega_{\alpha\beta}^2 - S_{\alpha\beta}^2) +
               (K_{\alpha\beta}^2 - J_{\alpha\beta}^2).
 \end{align}
What is interesting in this expression is the interchange between the symmetric and skew-symmetric tensors of $\grad \bm u$ and $\grad \bm b$ related to $\grad^2 P$. It is also appealing that the viscous dissipation is related to the symmetric part of the velocity gradient, whereas the Ohmic dissipation to the skew-symmetric part of the magnetic field gradient.

Now, we consider the joint PDF of the second and third invariants of $\bm X$, which are defined according to Eqs. \eqref{eq:invII} and \eqref{eq:invIII} as follows
\begin{equation}
 \label{eq:Qx}
 Q_X = \tfrac{1}{4}[\bm j^2 - 2tr(\bm K^2)]
\end{equation}
\begin{equation}
 \label{eq:Rx}
 R_X = -\tfrac{1}{3}[tr(\bm K^3) + \tfrac{3}{4}j_\alpha j_\beta K_{\alpha\beta}].
\end{equation}
\begin{figure}[!ht]
 \begin{subfigure}{0.35\textwidth}
  \includegraphics[width=\textwidth]{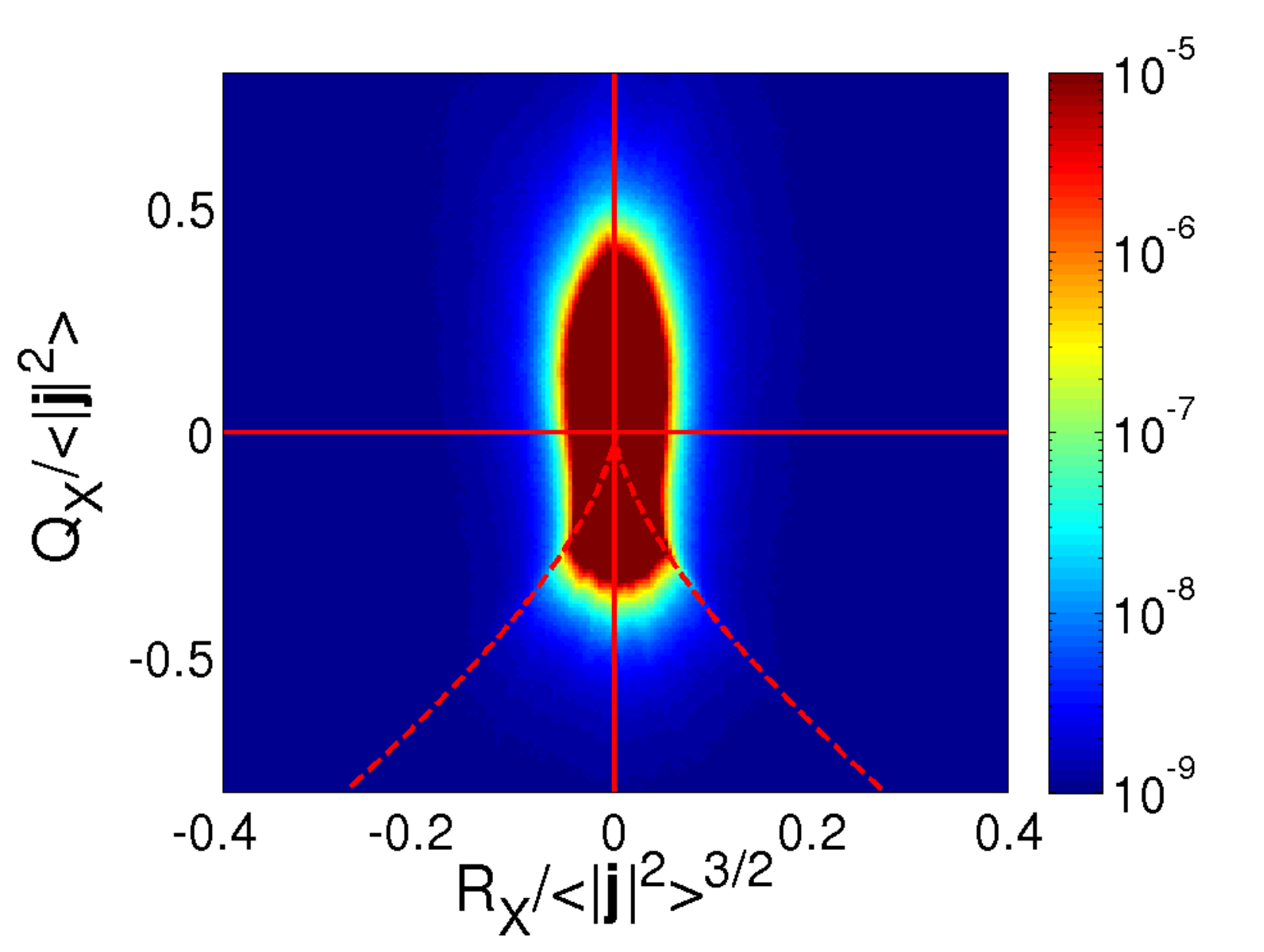}
  \caption{Run R}
 \end{subfigure}
 \begin{subfigure}{0.35\textwidth}
  \includegraphics[width=\textwidth]{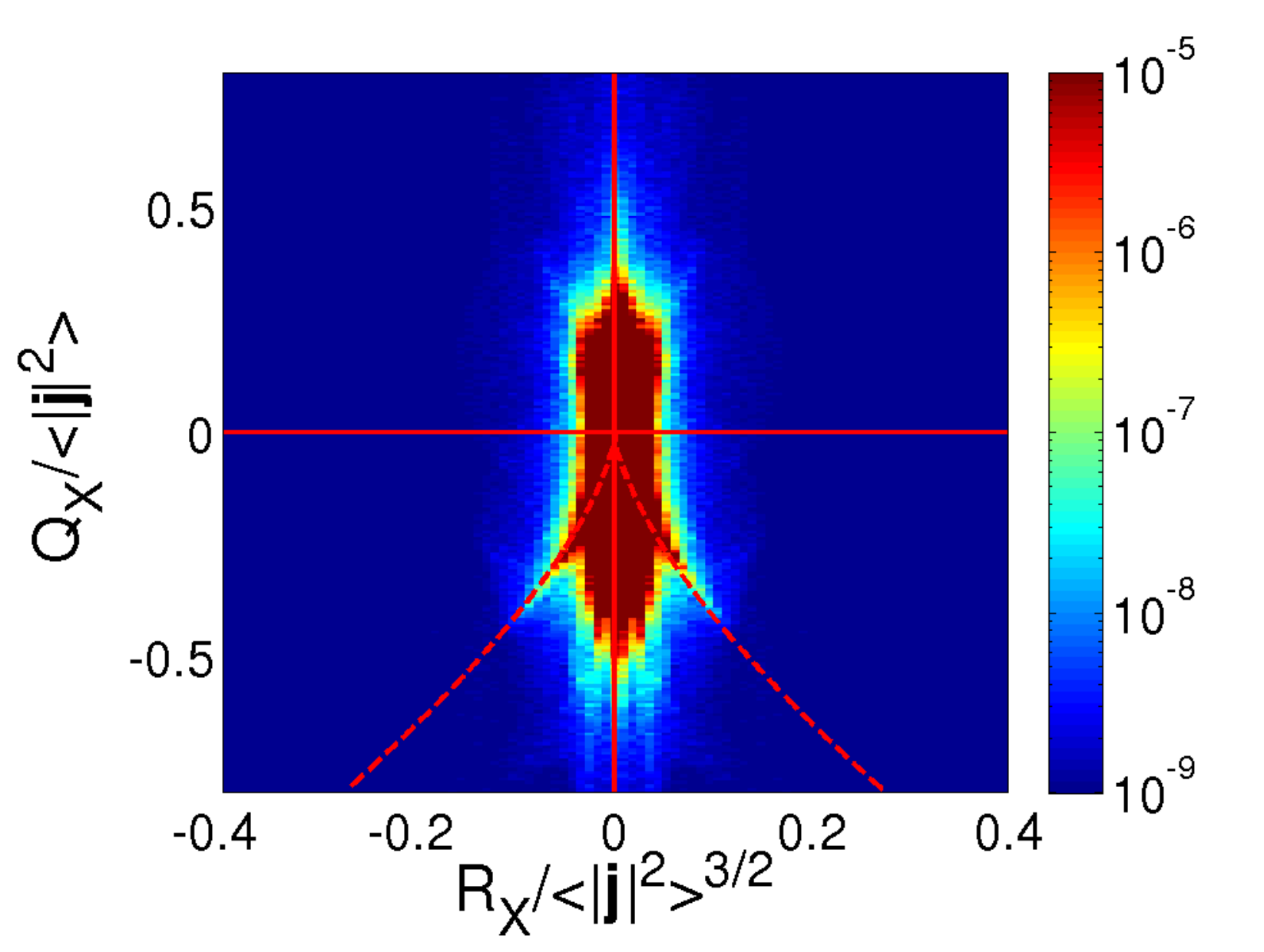}
  \caption{Run I}
 \end{subfigure} \\
 \begin{subfigure}{0.35\textwidth}
  \includegraphics[width=\textwidth]{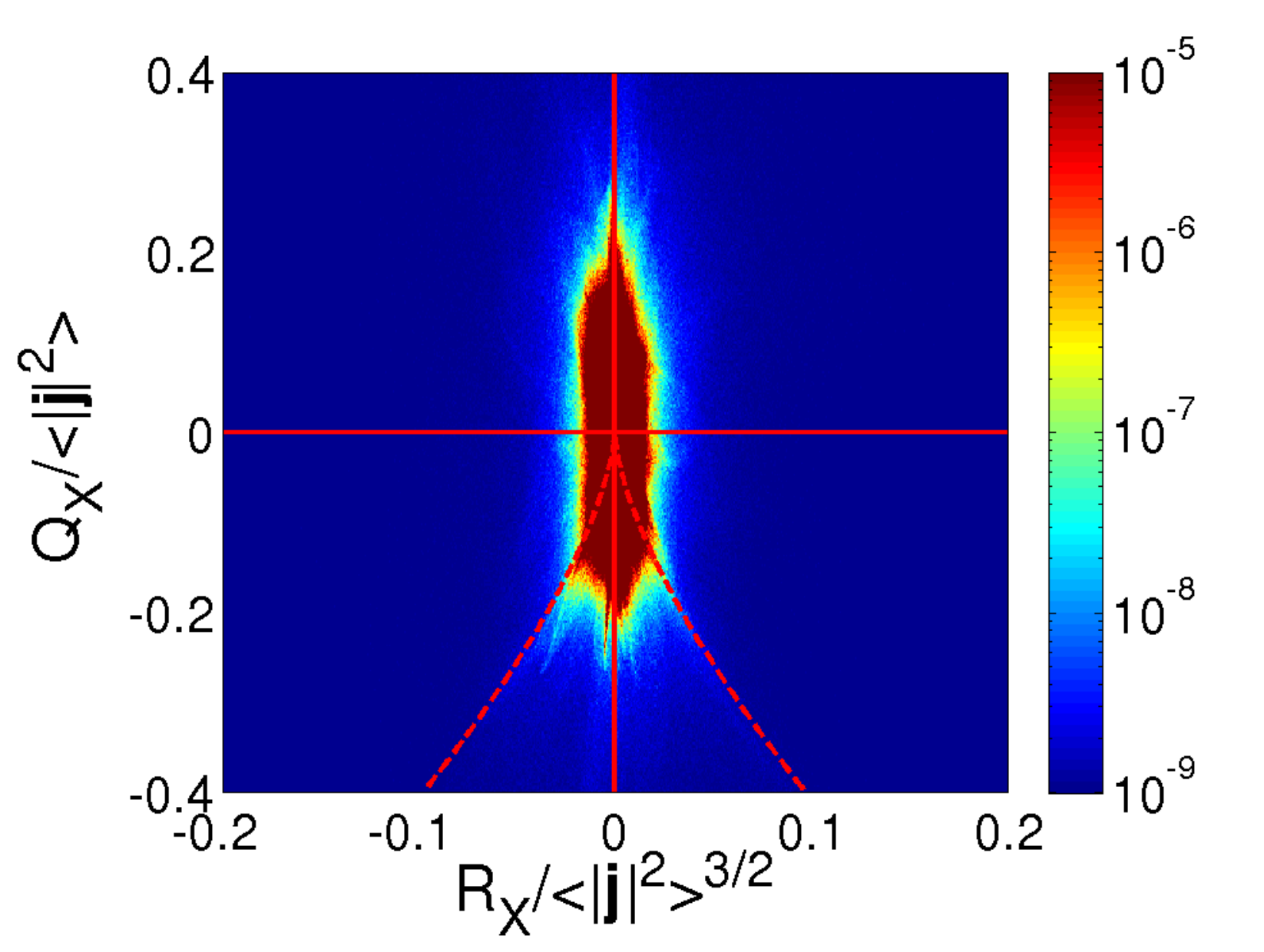}
  \caption{Run C}
 \end{subfigure}
 \begin{subfigure}{0.35\textwidth}
  \includegraphics[width=\textwidth]{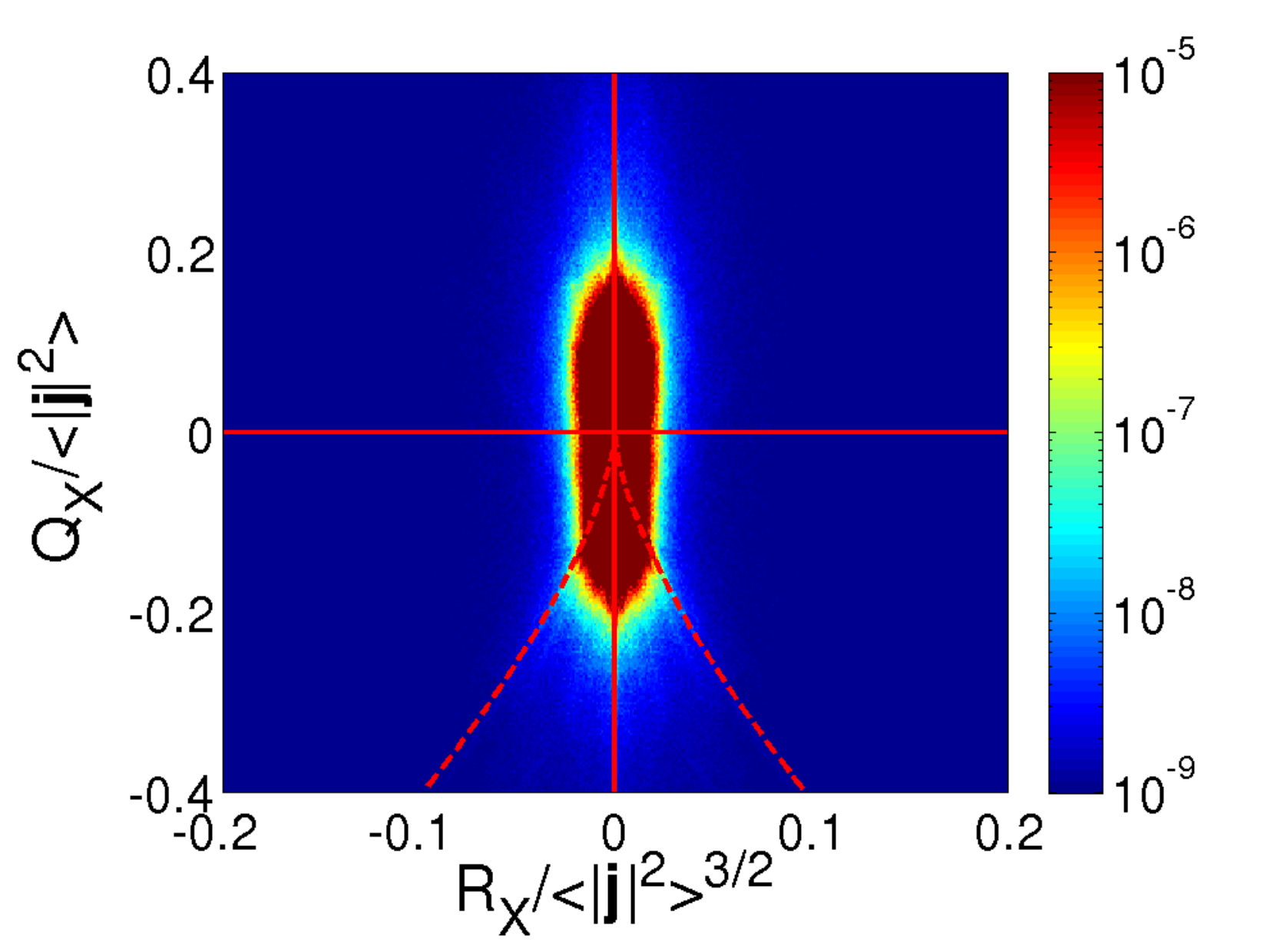}
  \caption{Run A}
 \end{subfigure}
  \caption{(Color online) Joint PDFs of the second invariant $Q_X$ and the third invariant $R_X$ of the magnetic field gradient tensor normalised appropriately by powers of the mean squared current density for a) run R, b) run I, c) run C and  d) run A of Table \ref{tbl:dnsparam}. The line $D_X = \tfrac{27}{4} R_X ^2 + Q_X^3 = 0$ is plotted for reference.}
  \label{fig:QxRx}
\end{figure}
For the classification of the magnetic field structures the $D_X = \tfrac{27}{4} R_X ^2 + Q_X^3 = 0$ line was included in the plots of Fig. \ref{fig:QxRx}. The topological classification emerging from the joint PDFs of $R_X$ and $Q_X$ can be interpreted in analogy to the invariant map of the velocity gradient (Fig. \ref{fig:QaRa_map}). Note, however, that the individual terms of the third invariant in Eq. \eqref{eq:Rx} do not appear in any evolution equation. Thus, $R_X$ does not have a physical meaning here but it is mathematically important for the classification of the magnetic field structures, in terms of the eigenvalues of $X_{\alpha\beta}$ associated with these structures.

In contrast to the invariants of the velocity gradient, the ($R_X,Q_X$) invariant map does not show a particular tendency towards any quadrant (see Fig. \ref{fig:QxRx}). For all the runs the core shape of the joint PDF is symmetric along the $R_X = 0$ axes, meaning that there is a balance between stable and unstable structures. The small scales, on the other hand, are slightly different especially for run I (Fig. \ref{fig:QxRx}b) and run C (Fig. \ref{fig:QxRx}c). Moreover, the joint PDF for runs C and A (see Fig. \ref{fig:QxRx}c and d, respectively) diminish towards the origin of the axes. In general, one could claim that this symmetric shape seems to be a general characteristic for the magnetic field gradient for all initial conditions with some small deviations, which might be due to the TG vortex symmetries.

\subsection{Joint PDFs of the magnetic strain rate invariants}
\begin{figure}[!ht]
 \begin{subfigure}{0.35\textwidth}
  \includegraphics[width=\textwidth]{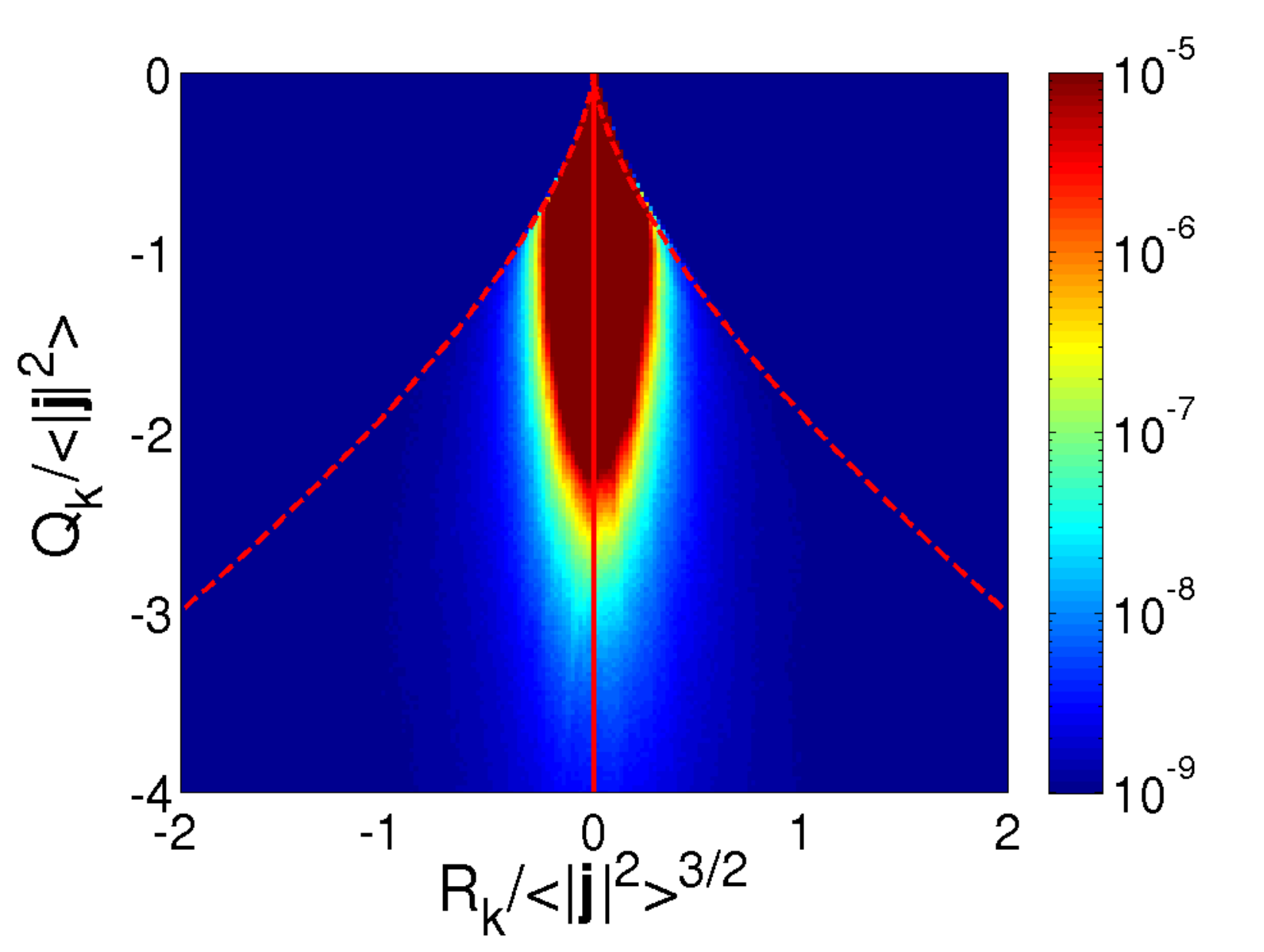}
  \caption{Run R}
 \end{subfigure}
 \begin{subfigure}{0.35\textwidth}
  \includegraphics[width=\textwidth]{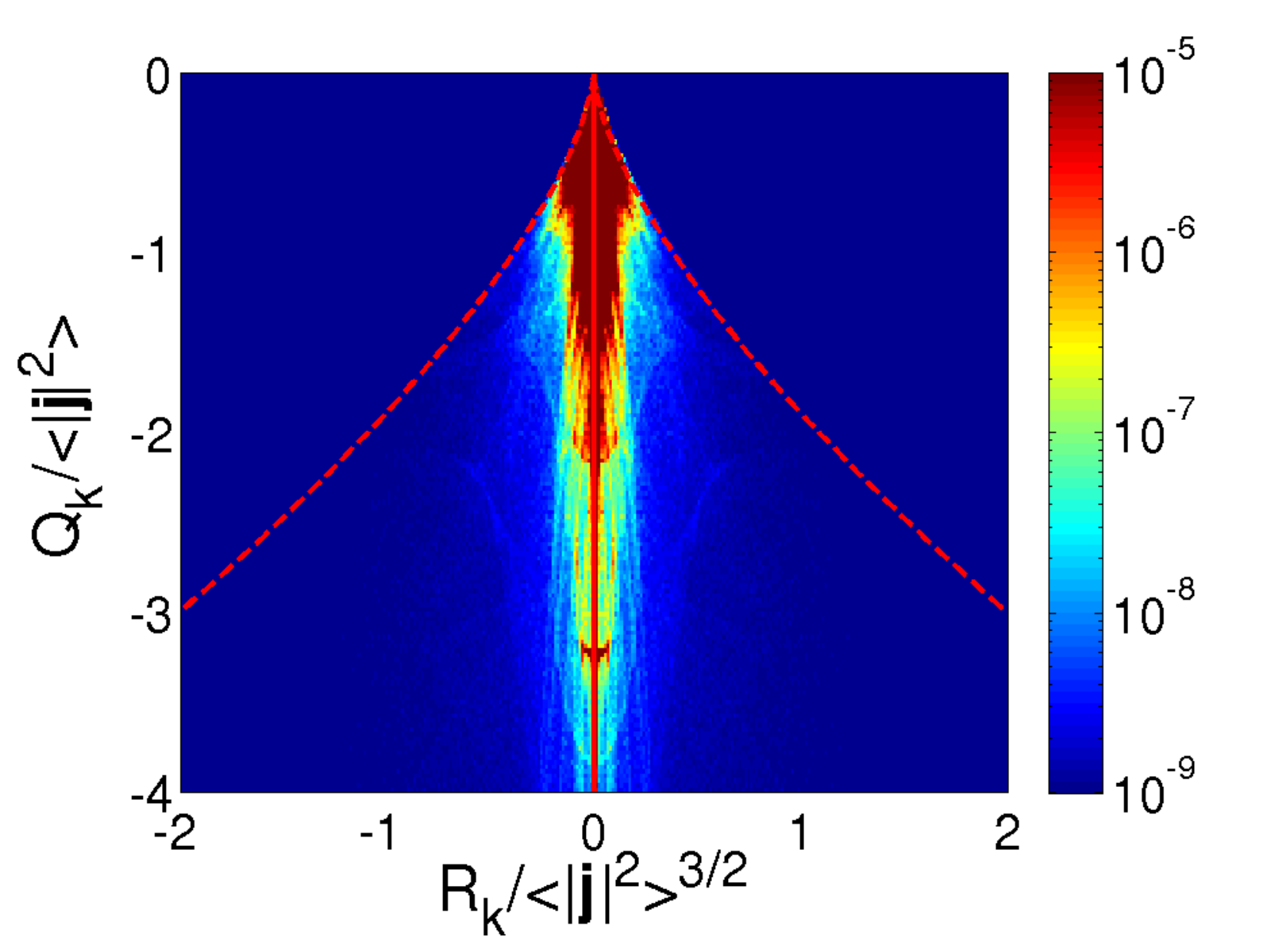}
  \caption{Run I}
 \end{subfigure} \\
 \begin{subfigure}{0.35\textwidth}
  \includegraphics[width=\textwidth]{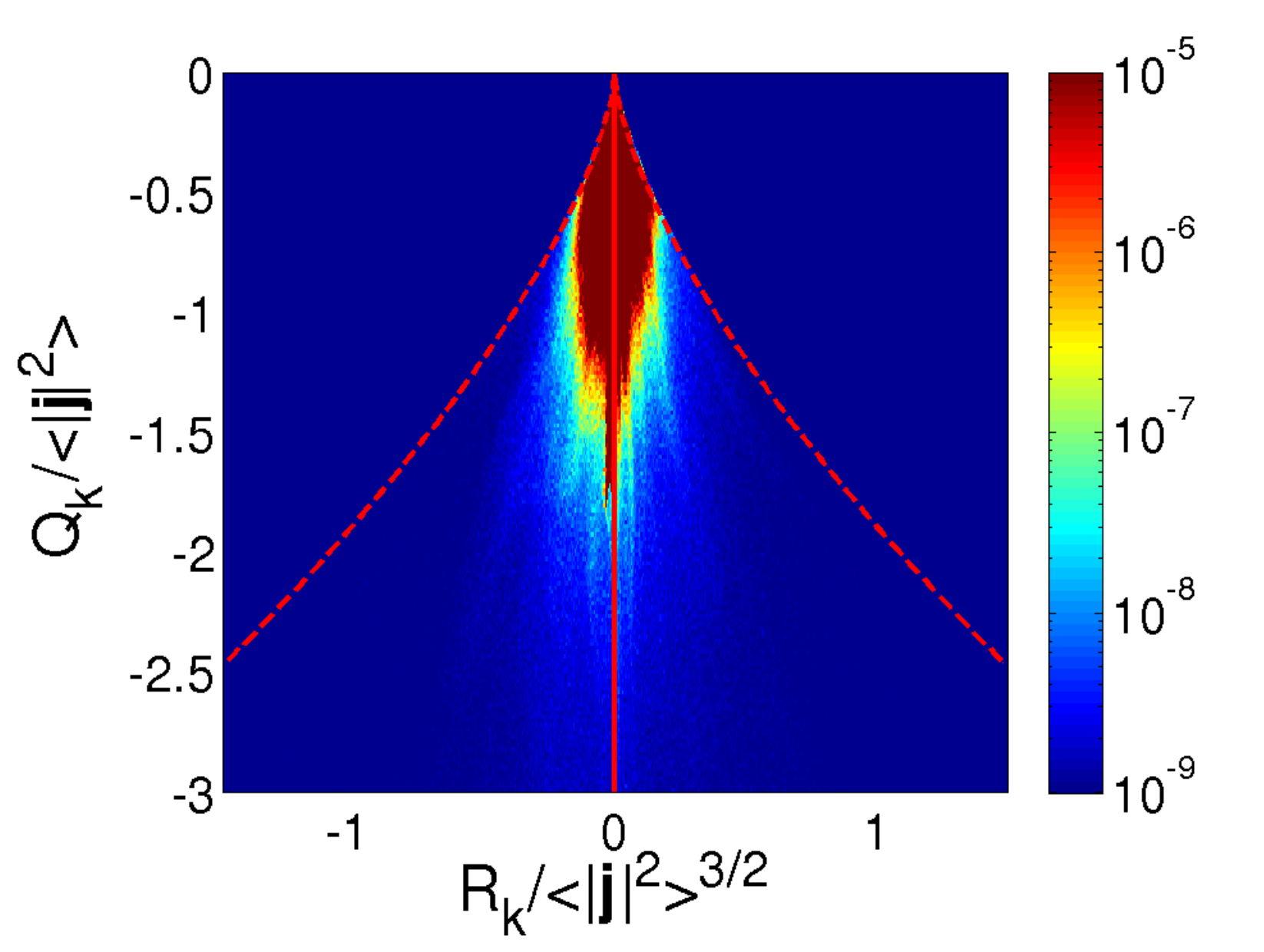}
  \caption{Run C}
 \end{subfigure}
 \begin{subfigure}{0.35\textwidth}
  \includegraphics[width=\textwidth]{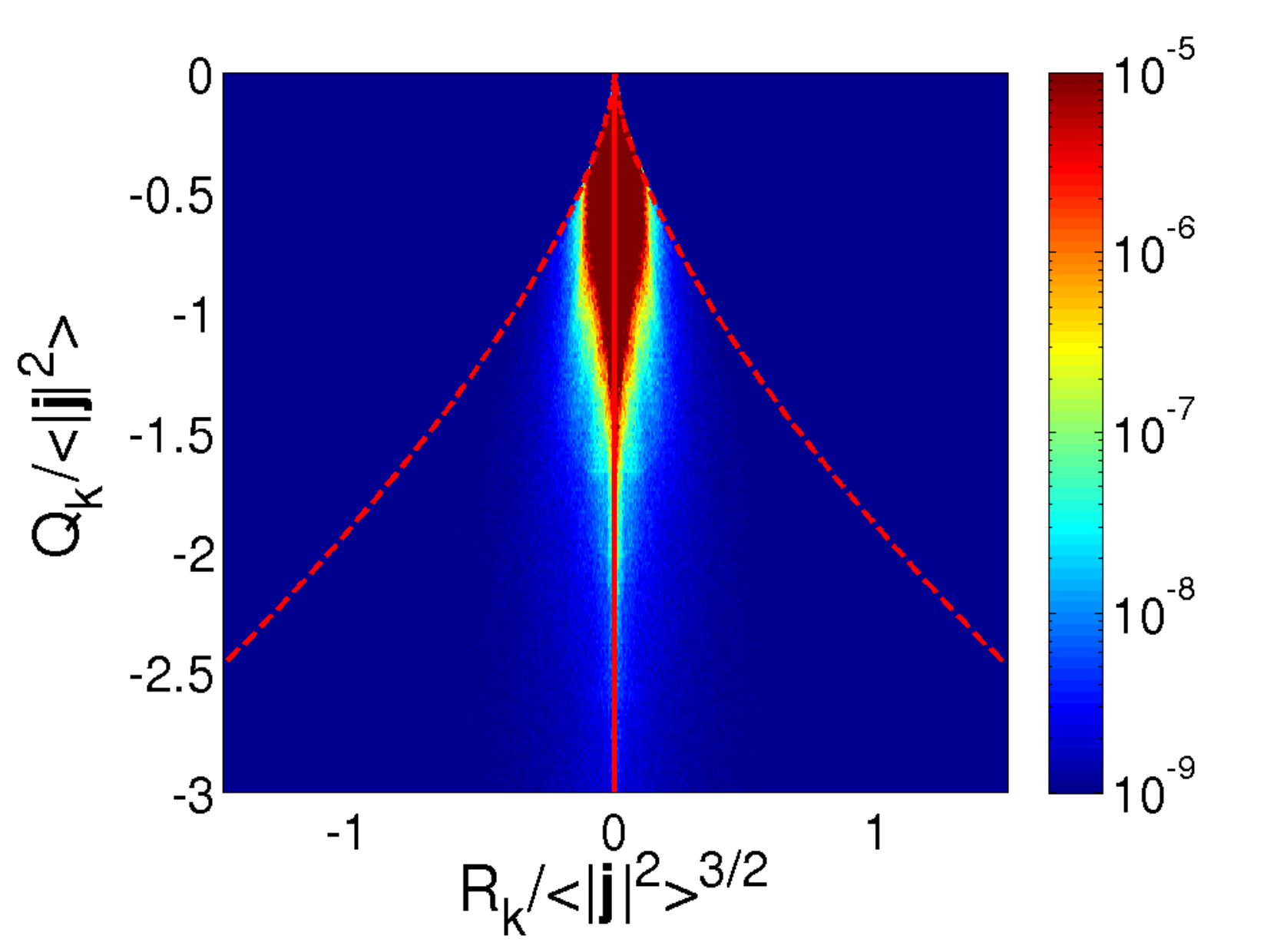}
  \caption{Run A}
 \end{subfigure}
  \caption{(Color online) Joint PDFs of the second invariant $Q_K$ and the third invariant $R_K$ of the magnetic strain rate tensor normalised appropriately by powers of the mean squared current density for a) run R, b) run I, c) run C and  d) run A of Table \ref{tbl:dnsparam}. The line $D_K = \tfrac{27}{4} R_K ^2 + Q_K^3 = 0$ is plotted for reference.}
  \label{fig:QkRk}
\end{figure}

Looking at the joint PDFs of the second and third invariants of $\bm K$ we can study the geometry of the local magnetic straining. The invariants of the magnetic strain rate tensor can be obtained by setting $\bm j = 0$ in Eqs. \eqref{eq:Qx} and \eqref{eq:Rx}, which reduce to
\begin{equation}
 Q_K = -\tfrac{1}{2}tr(\bm K^2) 
\end{equation}
and
\begin{equation}
 R_K = -\tfrac{1}{3}tr(\bm K^3),
\end{equation}
where $Q_K$ is negative definite due to the symmetric nature of $\bm K$. Note that $Q_K$ is not directly related to Ohmic dissipation in contrast to the $Q_s$ for viscous dissipation. Then, the physical interpretation of the ($R_K$,$Q_K$) invariant map is quite different from Fig. \ref{fig:QsRs_map} but similar in terms of flow topology. So, very low values of $Q_K$ in Fig. \ref{fig:QkRk} can be physically interpreted as regions of high magnetic-strain or regions where the Lorentz force is small. The third invariant $R_K$ can be written as the product of the eigenvalues of $K_{\alpha\beta}$ in analogy to $R_s$ (see Eq. \eqref{eq:rseig}). Then, the interpretation of $R_K$ in terms of sheetlike and tubelike structures is also determined by $\text{sgn}(R_K) = \text{sgn}(\lambda_2)$.

The joint PDFs between $R_K$ and $Q_K$, representing the local topology of the structures related to magnetic strain rate, appear to be symmetric along the $R_K = 0$ axis for most of the runs of Table \ref{tbl:dnsparam} (see Fig. \ref{fig:QkRk}).
In detail, the joint PDF of run R (Fig. \ref{fig:QkRk}a) illustrates an equipartition between tubelike and sheetlike structures associated with $\bm K$. The shapes of the ($R_K,Q_K$) invariant map for runs I and A (see Figs. \ref{fig:QkRk}b and \ref{fig:QkRk}d respectively) are also symmetric and they can be well approximated by a magnetic field gradient of the form of Eq. \eqref{eq:approxA} with $Q_K = -\tfrac{1}{4}[(\pd_yb_x)^2 + (\pd_yb_z)^2]$ and $R_K = 0$. In Fig. \ref{fig:QkRk}b there are very low values of $Q_K$ correlated with $R_K = 0$ in comparison to the rest of the flows. Therefore, this approximation for $\bm X$ is especially valid for the small scale structures that correspond to low values of $Q_K$ in this joint PDF.
Figure \ref{fig:QkRk}c (run C), on the other hand, is slightly asymmetric, showing a tangible inclination of the joint PDF towards $R_K < 0$. This implies that there is a preference for the intermediate eigenvalue of $K_{\alpha\beta}$ to be negative and hence a tendency for tubelike structures.

\begin{figure}[!ht]
 \includegraphics[width=0.5\textwidth]{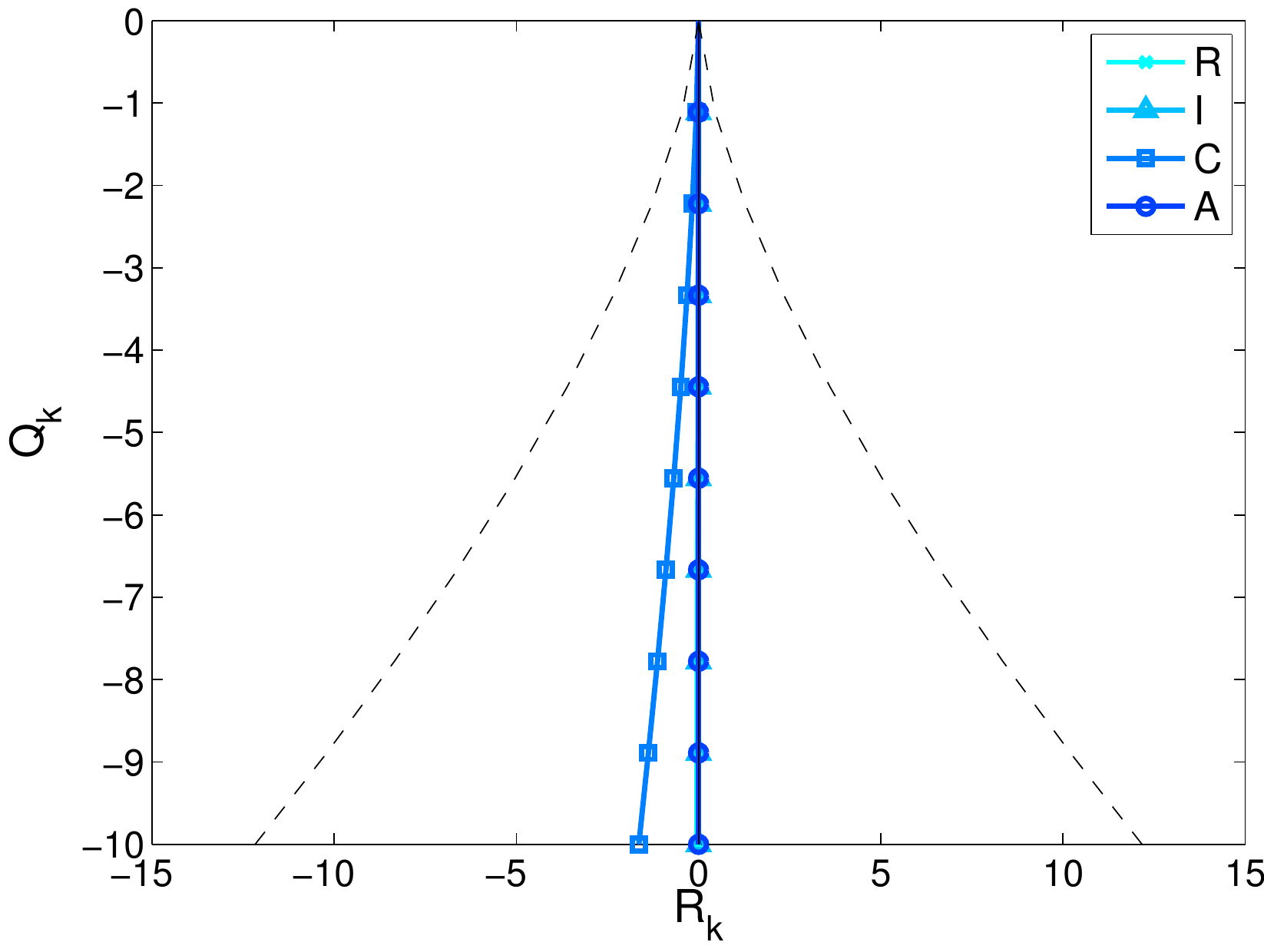}
 \caption{(Color online) Plots of Eq. \eqref{eq:schematic} using the mean eigenvalues of $K_{\alpha\beta}$ from Table \ref{tbl:Keig}. The dashed line $D_K = \tfrac{27}{4} R_K ^2 + Q_K^3 = 0$ is plotted for reference.}
 \label{fig:Keig}
\end{figure}

Now, we attempt to provide an outline of the joint PDFs of Fig. \ref{fig:QkRk} by tabulating the mean eigenvalues of the magnetic strain rate tensor (see Table \ref{tbl:Keig}) and by plotting the analogous expression to Eq. \eqref{eq:schematic} for $R_K$ and $Q_K$ using the mean eigenvalues of Table \ref{tbl:Keig} (see Fig. \ref{fig:Keig}).
\begin{table}[!ht]
 \caption{Mean eigenvalues of the magnetic strain rate tensor $K_{\alpha\beta}$ and their ratios.}
 \label{tbl:Keig}
 \begin{ruledtabular}
    \begin{tabular}{*{6}{c}} 
      \textbf{Run} & $\bm{\avg{\lambda_1}}$ & $\bm{\avg{\lambda_2}}$ & $\bm{\avg{\lambda_3}}$ & $\bm{\avg{\lambda_1}:\avg{\lambda_2}:\avg{\lambda_3}}$ \\
     \hline
      R & 0.26 &  0.00 & -0.26 & 1 : 0 : -1 \\
      I & 0.44 &  0.00 & -0.44 & 1 : 0 : -1 \\
      C & 0.35 & -0.02 & -0.33 & 18 : -1 : -17 \\
      A & 0.44 &  0.00 & -0.44 & 1 : 0 : -1 \\
    \end{tabular}
 \end{ruledtabular}
\end{table}
The values of the mean eigenvalue ratios tell us that on average run I and C are described by quasi two-dimensional structures in agreement with the joint PDF analysis. Moreover, the values 1:0:-1 that we obtain for run R agree with the argument that the joint PDF of Fig. \ref{fig:QkRk}a is symmetric but also express that in an average sense the flow topology is locally invariant in one direction. The only case that deviates from two-dimensionality is run C, which is on average characterised by biaxial contraction (i.e. $\avg{\lambda_2} < 0$) and thereby tubelike structures.

\subsection{Joint PDFs of the second invariants of the magnetic strain and current density tensors}
The skew-symmetric part of the magnetic field gradient tensor, $\bm J$ has only one invariant in analogy to the rotation rate tensor $\bm \Omega$. This can be obtained by letting $\bm K$ to be zero in Eqs. \eqref{eq:Qx} and \eqref{eq:Rx}, then
\begin{equation}
 Q_j = -\tfrac{1}{2}tr(\bm J^2) = \tfrac{1}{4}\bm j^2,
\end{equation}
which is also related to the second invariants of $\bm X$ and $\bm K$ through $Q_j = Q_X - Q_K$. 

The ($Q_j,-Q_K$) invariant map describes the relative importance between the straining and rotational part of the magnetic field gradient in analogy to ($Q_\omega,-Q_s$) map for the velocity gradient (see Fig. \ref{fig:QwQs_map}). However, the important difference in this case is that the rotational part of $\bm X$ is directly related to Ohmic dissipation and not the straining part. Hence, the points of the joint PDFs close to the $Q_j$ axis that are nearly in solid-body rotation are regions in the flow of strong Ohmic dissipation and negligible magnetic straining in contrast to the picture we get from Fig. \ref{fig:QwQs_map}. On the other side, points adjacent to the $-Q_K$ axis express nearly pure magnetic straining motions in regions of where the current is negligible and thereby Lorentz force is suppressed.

\begin{figure}[!ht]
  \begin{subfigure}{0.35\textwidth}
   \includegraphics[width=\textwidth]{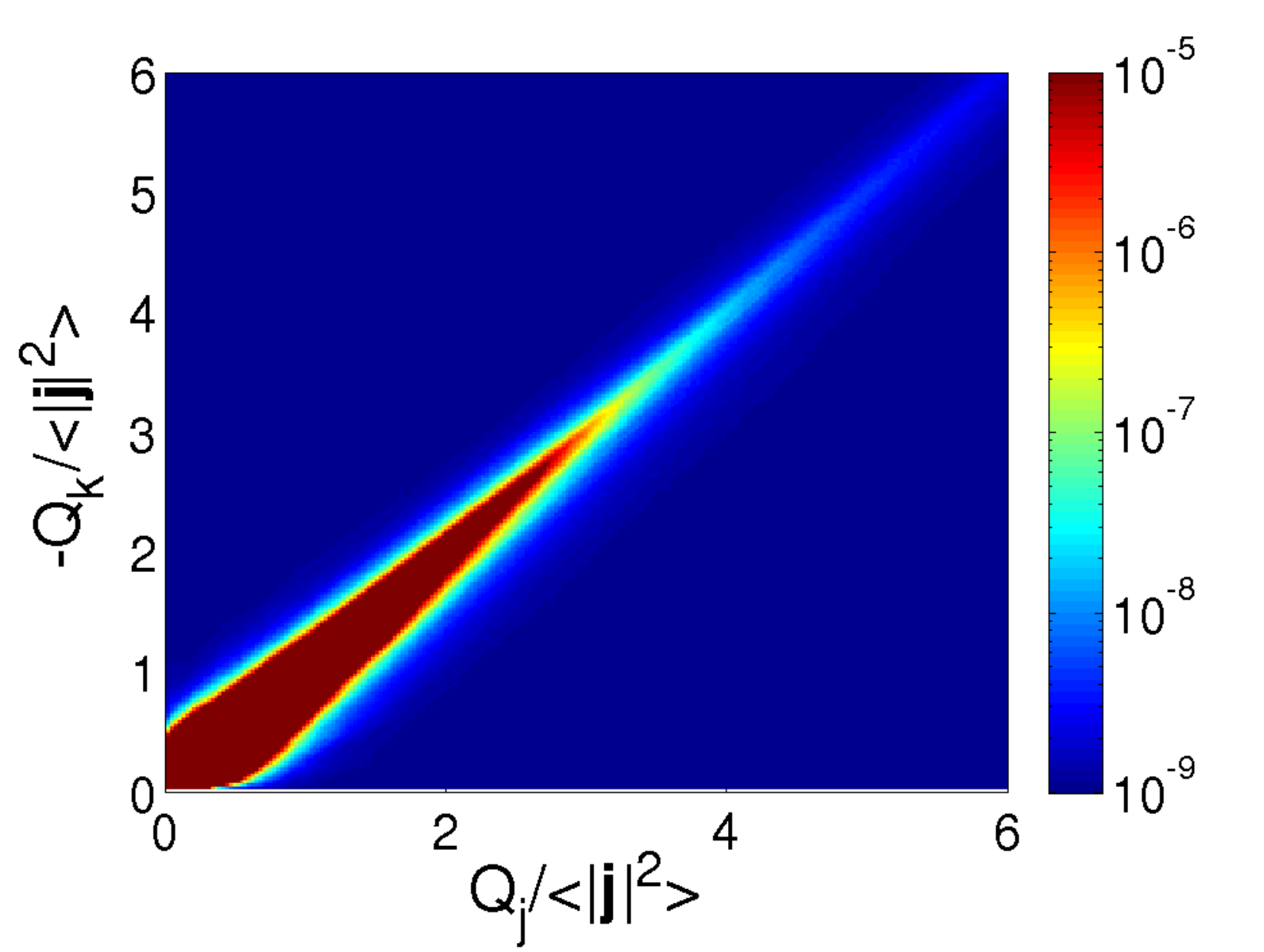}
   \caption{Run R}
  \end{subfigure}
  \begin{subfigure}{0.35\textwidth}
   \includegraphics[width=\textwidth]{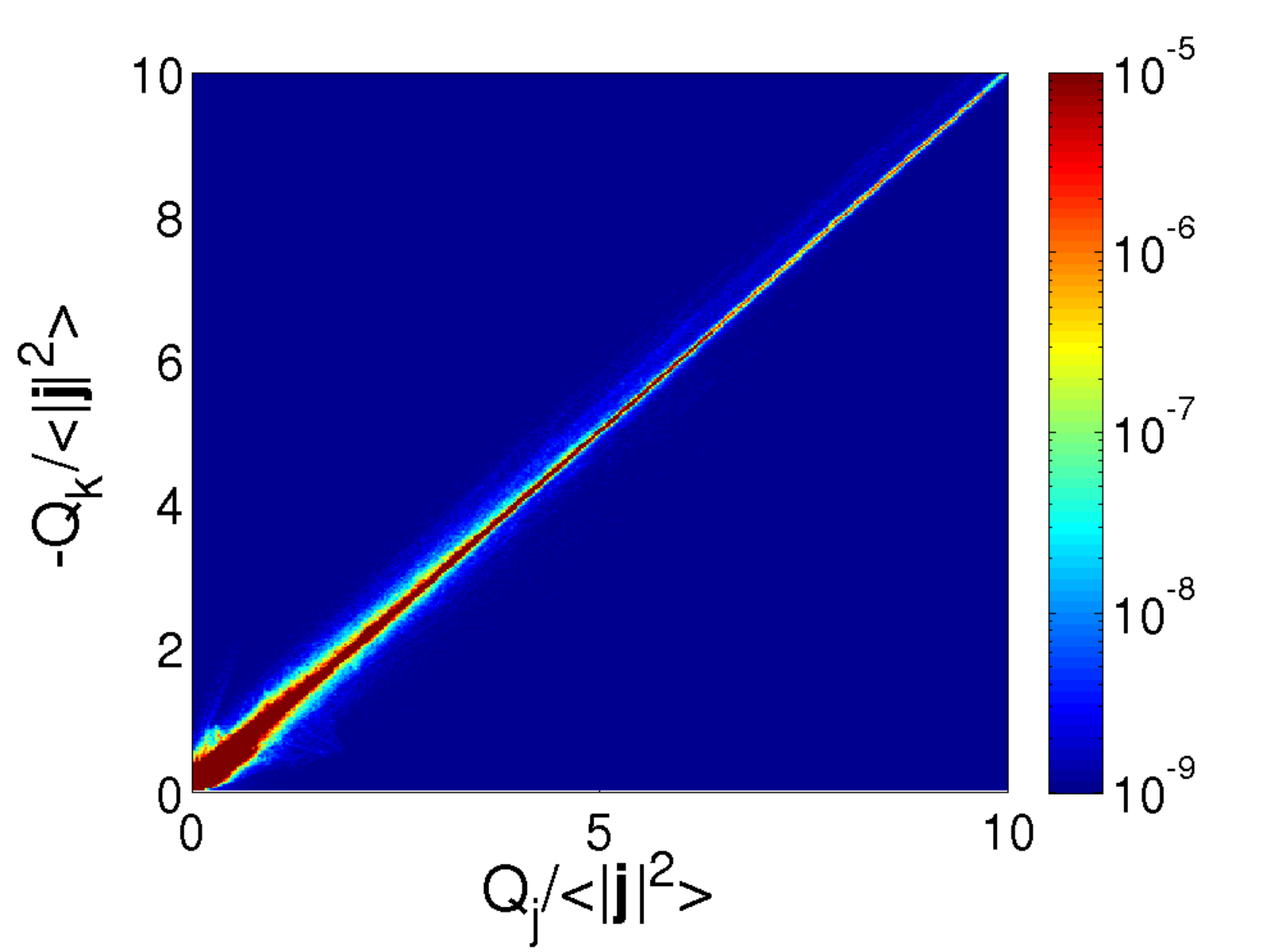}
   \caption{Run I}
  \end{subfigure} \\
  \begin{subfigure}{0.35\textwidth}
   \includegraphics[width=\textwidth]{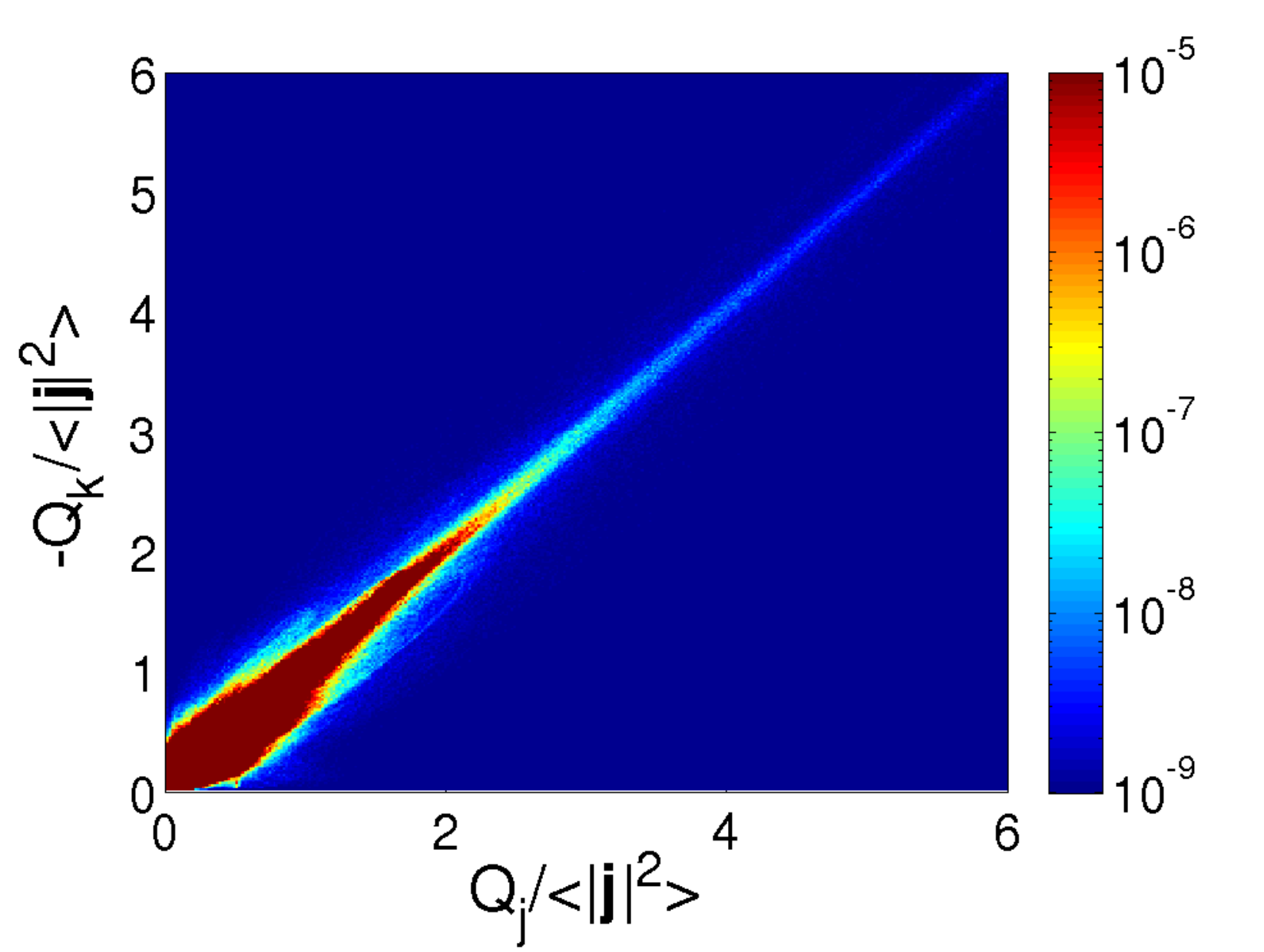}
   \caption{Run C}
  \end{subfigure}
  \begin{subfigure}{0.35\textwidth}
   \includegraphics[width=\textwidth]{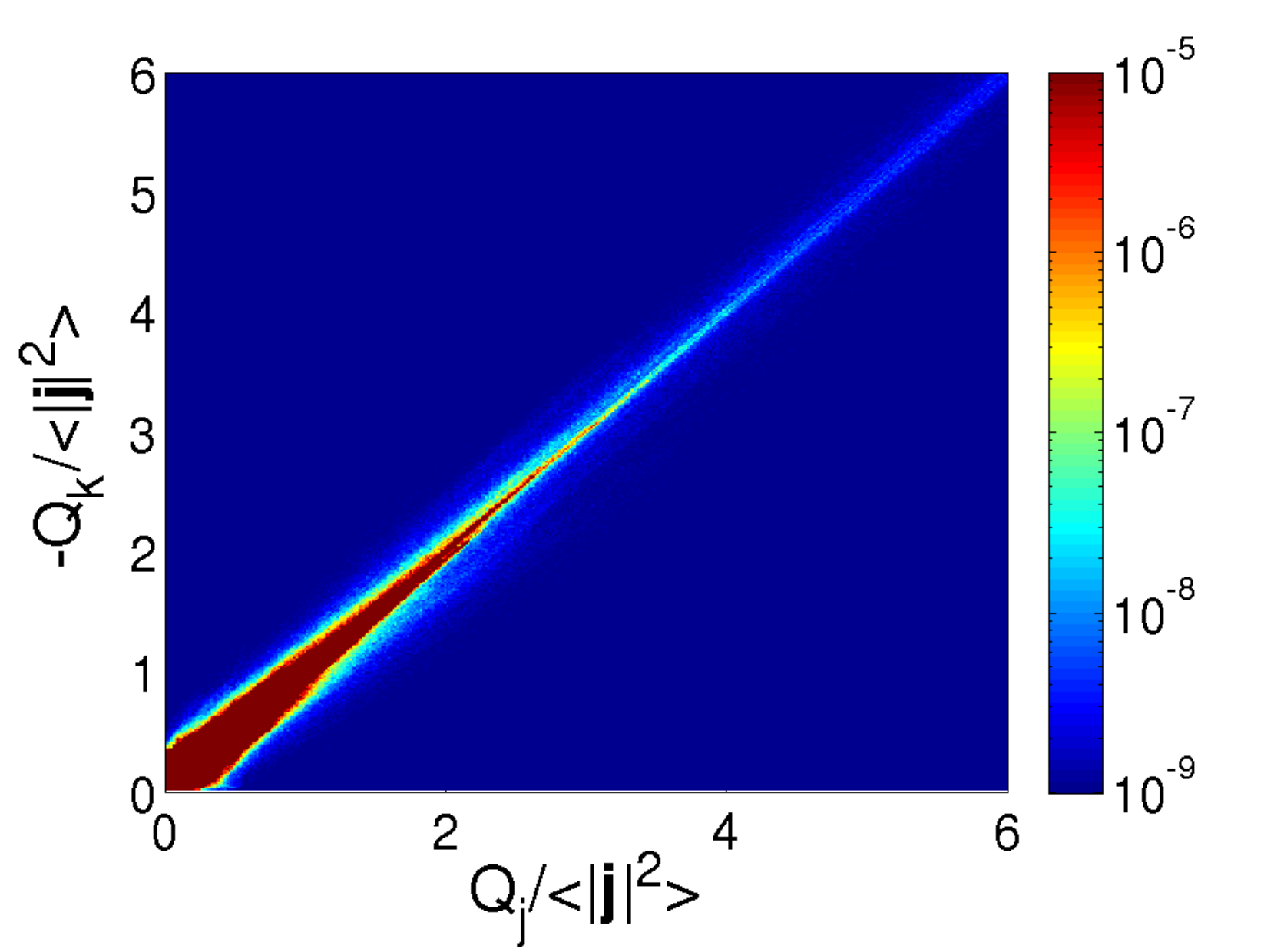}
   \caption{Run A}
  \end{subfigure}
  \caption{(Color online) Joint PDFs of the second invariants of magnetic strain rate and current density rate tensors normalised by the mean squared current density for a) run R, b) run I, c) run C and d) run A of Table \ref{tbl:dnsparam}.}
  \label{fig:QjQk}
 \end{figure}

The joint PDFs of Fig. \ref{fig:QjQk} show that points near the axes are rare in MHD turbulent flows and are related only to the large scales of the flows, where $Q_j$ and $-Q_K$ are small in comparison to $\avg{|\bm j|^2}$.
Most of the points in the plots of Fig. \ref{fig:QjQk} lie near the main diagonal, revealing that Ohmic dissipation occurs in current sheets. Here, the magnetic field gradient tensor can be well approximated by the form of Eq. \eqref{eq:approxA}, which gives $Q_j = -Q_K = \tfrac{1}{4}[(\pd_yb_x)^2 + (\pd_yb_z)^2]$. This is particularly a good approximation for runs I and A (see Figs. \ref{fig:QjQk}b and \ref{fig:QjQk}d respectively), where $Q_j$ and $-Q_K$ are strongly correlated for all scales. It can also be argued that this approximation is also valid for the small scales of runs R and C (see Figs. \ref{fig:QjQk}a and \ref{fig:QjQk}c respectively) that correspond to high values of $Q_j$ and $-Q_K$.

\subsection{Structures in the current density field}
\begin{figure}[!ht]
  \begin{subfigure}{0.3\textwidth}
   \includegraphics[width=\textwidth]{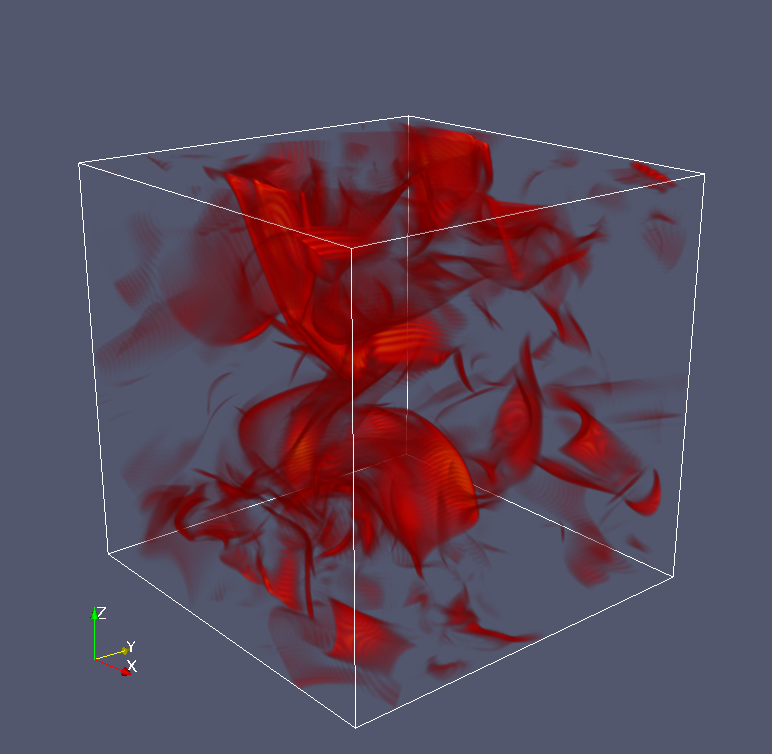}
   \caption{Run R}
  \end{subfigure}
  \begin{subfigure}{0.3\textwidth}
   \includegraphics[width=\textwidth]{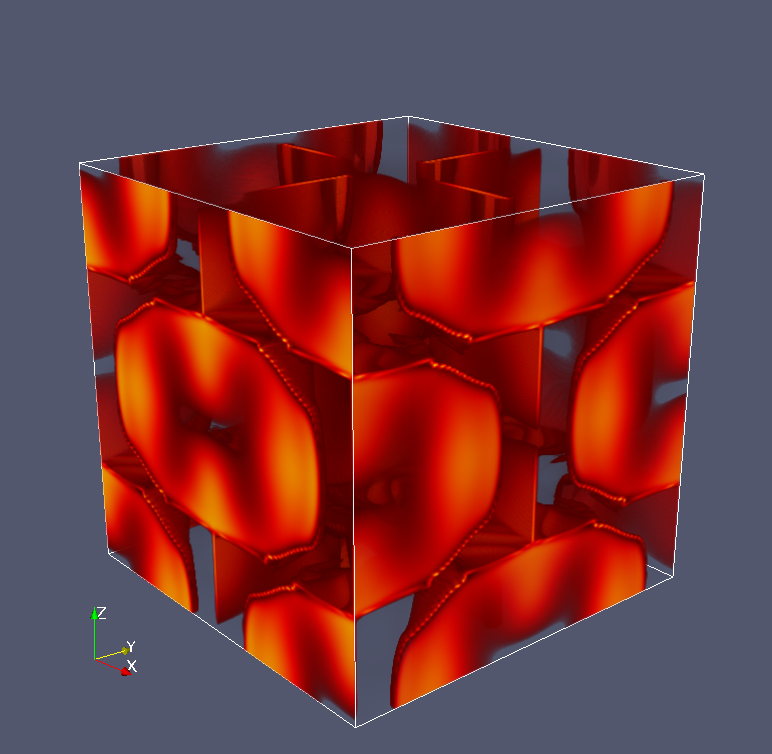}
   \caption{Run I}
  \end{subfigure} \\
  \begin{subfigure}{0.3\textwidth}
   \includegraphics[width=\textwidth]{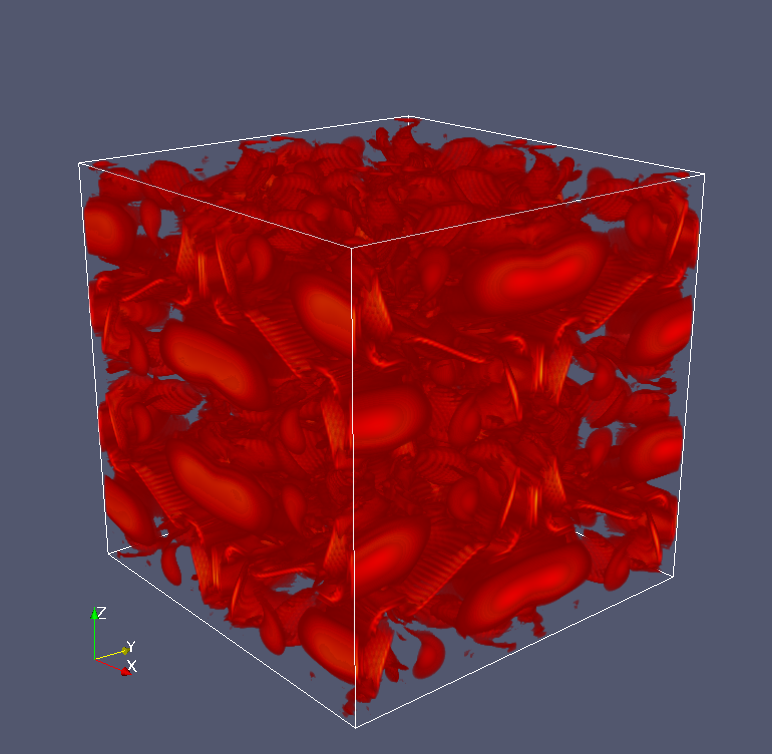}
   \caption{Run C}
  \end{subfigure}
  \begin{subfigure}{0.3\textwidth}
   \includegraphics[width=\textwidth]{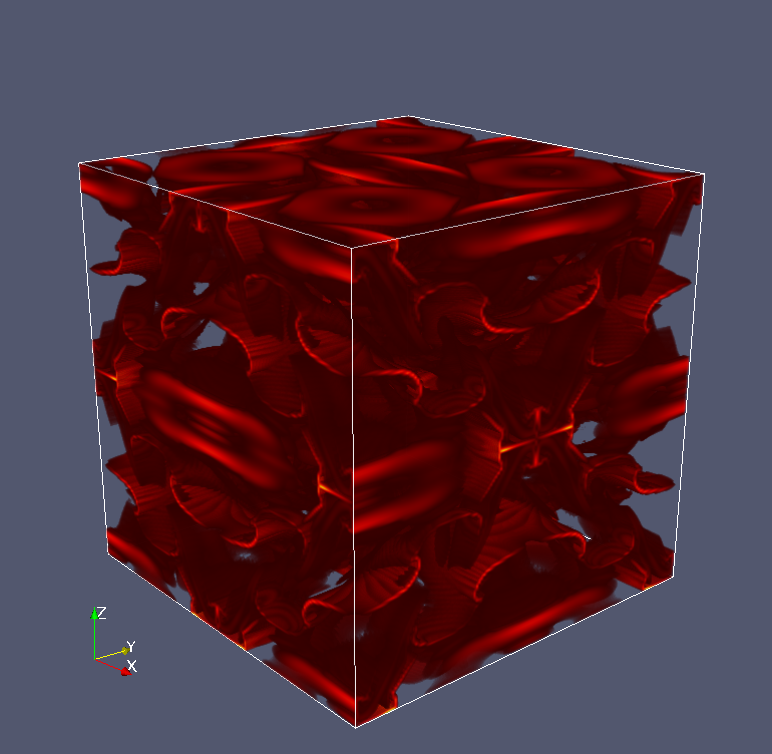}
   \caption{Run A}
  \end{subfigure}
  \caption{(Color online) Current density field iso-contours with $|j| \geq 6j'$ for a) run R, b) run I, c) run C and d) run A of Table \ref{tbl:dnsparam}.}
  \label{fig:visj}
 \end{figure}

To further validate our joint PDF approach, we present indicatively iso-contours of current density (Fig. \ref{fig:visj}) in our $[0,2\pi]^3$ periodic boxes at the moment of maximum dissipation for all the runs that we have considered (see Table \ref{tbl:dnsparam}). All the visualisations of Fig. \ref{fig:visj} display current density iso-contours with $|j| \ge 6j'$ where $j' \equiv (|\bm j|^2)^{1/2}$.
The field of current density for run R (Fig. \ref{fig:visj}a) is composed by randomly oriented sheetlike structures which seem to be extremely thin, supporting the fact that the values of the mean eigenvalue ratios for the magnetic strain rate tensor are $1:0:-1$. It is clear that locally quasi two-dimensional structures are the dominant structures in Figs. \ref{fig:visj}b and \ref{fig:visj}d (runs I and A respectively), validating the joint PDFs of ($R_K,Q_K$). These 2D current sheets are also the structures where most of the Ohmic dissipation occurs. For run I these dominant structures are formed at the faces of the $[0,\pi]^3$ boxes, whereas for run A these are randomly oriented. On the other hand, run C (Fig. \ref{fig:visj}c) seem to be dominated by tubelike structures but one can also observe the coexistence of isolated thin current sheets in agreement to our analysis. Finally, the TG vortex symmetries are clearly depicted in these visualisations with each TG flow having different degree of randomness. 
This raises again questions as to what degree these symmetries restrict the dynamics of the flows.

Here, we would like to emphasize that the structures related to the magnetic field gradient have different characteristics than those related to the velocity gradient. This might well be a reason that the energy spectra that we obtain, as well as other studies, for the kinetic and magnetic energy (not shown here) seem to obey different scaling exponents. Moreover, we conjecture that these quasi two-dimensional organised structures that appear in runs I and A both in the vorticity and the current density fields are the reason to obtain a $k^{-2}$ scaling that we observe in the total energy spectra in Figs. \ref{fig:et_spectra}b and \ref{fig:et_spectra}d. This, however, needs to be further investigated and it will be reported elsewhere.

\section{\label{sec:end} Conclusions}
The universality of the energy spectrum in MHD turbulence is in doubt by various studies. One aspect is the manifestation of different, dubious scaling exponents. In order to avoid ambiguity between scaling exponents, we explore various statistics based on the invariants of the velocity gradient and related tensors. Note that for a big family of hydrodynamic turbulent flows, the joint PDF of the invariants of the velocity gradient is generally considered to be universal. We further extend this analysis to the invariants of gradient statistics related to the magnetic field. In particular, we explore DNS data of decaying MHD turbulence with random initial conditions as well as a set of three different Taylor-Green type initial conditions without imposing any symmetry constrains in our flows during their evolution. The TG flows were chosen to be examined since recently, Lee et al. \cite{leeetal10} reported that the scaling of the energy spectrum at the peak of dissipation depends on the initial conditions.

Our study attempts to classify the structures of our MHD flows. The structures related to the strain rate tensor are predominantly sheetlike structures (i.e. $\avg{\lambda_2} > 0$) for all the flows apart from run I (see Fig. \ref{fig:Seig}), which is quasi two-dimensional (i.e. $\avg{\lambda_2} \simeq 0$). 
The biaxial stretching for our MHD flows is different in comparison to hydrodynamic turbulence, namely $\avg{\lambda_1}:\avg{\lambda_2}:\avg{\lambda_3} = 3:1:-4$ (see Table \ref{tbl:Seig}). Furthermore, the enstrophy dominated regions are well correlated with region of high viscous dissipation in contrast to hydrodynamic flows. We also find that viscous dissipation is an intrinsic element of vortex sheets.

On the other side, magnetic field consists of quasi two-dimensional structures, i.e. $\avg{\lambda_2} = 0$, for all the cases apart from run C, which is on average dominated by tubelike structures, i.e. $\avg{\lambda_2} < 0$ (see Table \ref{tbl:Keig}). The correlation between magnetic strain dominated regions and regions of high Ohmic dissipation is generally stronger than the correlation between enstrophy and viscous dissipation. We also obtain that Ohmic dissipation resides in current sheets, which are thinner than the vortex sheets in the same flow. Visualisations support further our joint PDFs analysis of the invariants. 

Our results also demonstrate that small scales depend on the initial conditions in decaying MHD turbulence. This is dramatically illustrated through the joint PDF of $R_A$ with $Q_A$ (see Fig. \ref{fig:QaRa}), which has a universal teardrop shape for hydrodynamic turbulence away from walls. Lack of small scale universality in decaying MHD turbulence will have important implications in modelling. The main idea of small-scale universality applied in LES (i.e. that although large scales may be dependent on boundary conditions or initial conditions, smaller scales are less flow dependent and more amenable to modelling) seems to fail for decaying MHD turbulent flows. Therefore, if MHD turbulence is non-universal, then the construction of subgrid-scale models for MHD flows might be doubtful.

However, there is an important element regarding the TG flows that one has to address before claiming that small scale universality is absent in these flows. This element is the self-preservation of TG vortex symmetries during the evolution of the flow, which seem to be a strong property of the MHD equations. Therefore, a natural question that emerges is: what happens if we perturb the TG flows in order to break these symmetries before the peak of dissipation? Will the joint PDFs converge to a single shape or/and the scaling of the energy spectra to a single value? What is the role of the symmetries imposed by the initial conditions in terms of the dynamics? Do we have classes of universality for these moderate Reynolds numbers or is there a universal power law in the high Reynolds number limit? We plan to address these questions in our future work.

\begin{acknowledgements}
The authors acknowledge interesting and stimulating discussions with Christos Vassilicos and Marc-Etienne Brachet. V.D. acknowledges the financial support from EU-funded Marie Curie Actions---Intra-European Fellowships (FP7-PEOPLE-2011-IEF, MHDTURB, Project No. 299973). All computations were performed using the HPC resources from
GENCI-CINES (Grant No. 2012026421).
\end{acknowledgements}

\bibliography{references}
\end{document}